\documentclass[twoside,english,final, 3p, smallcondensed, sort&compress, jhep]{svjour3}
\usepackage[T1]{fontenc}
\usepackage[latin9]{inputenc}
\usepackage{geometry}
\usepackage{array}
\usepackage{float}
\usepackage{mathtools}
\usepackage{url}
\usepackage{multirow}
\usepackage{amsmath}
\usepackage{amssymb}

\makeatletter

\newcommand{\noun}[1]{\textsc{#1}}
\newcommand{\lyxmathsym}[1]{\ifmmode\begingroup\def\b@ld{bold}
  \text{\ifx\math@version\b@ld\bfseries\fi#1}\endgroup\else#1\fi}

\providecommand{\tabularnewline}{\\}

\usepackage{nag}
\usepackage{bm}
\usepackage[final]{microtype}
\usepackage[unicode=true, bookmarks=true,bookmarksnumbered=false,
bookmarksopen=false, breaklinks=true,pdfborder={0 0 0},
pdfborderstyle={},backref=false,colorlinks=true,
citecolor={cyan},linkcolor={cyan}] {hyperref}
\allowdisplaybreaks
\usepackage{pifont}
\date{Final version published in The Journal of High Energy Physics, 110, 2018. \\ https://doi.org/10.1007/JHEP10(2018)110}

\makeatother

\usepackage{babel}
\begin{document}

\title{Solving and classifying the solutions of the Yang-Baxter equation
through a differential approach. }

\subtitle{Two-state systems }

\author{R. S. Vieira}

\institute{Universidade Federal de São Carlos (UFSCar), Departamento de Física,
C.P. 676, CEP. 13565-905, São Carlos, SP, Brasil (Currently at Universidade
Estadual Paulista (UNESP), Faculdade de Ciências e Tecnologia, Departamento
de Matemática Aplicada e Computacional, CEP. 19060-900, Presidente
Prudente, SP, Brasil) \email{rsvieira@df.ufscar.br}}
\maketitle
\begin{abstract}
The formal derivatives of the Yang-Baxter equation with respect to
its spectral parameters, evaluated at some fixed point of these parameters,
provide us with two systems of differential equations. The derivatives
of the $R$ matrix elements, however, can be regarded as independent
variables and eliminated from the systems, after which two systems
of polynomial equations are obtained in place. In general, these polynomial
systems have a non-zero Hilbert dimension, which means that not all
elements of the $R$ matrix can be fixed through them. Nonetheless,
the remaining unknowns can be found by solving a few number of simple
differential equations that arise as consistency conditions of the
method. The branches of the solutions can also be easily analyzed
by this method, which ensures the uniqueness and generality of the
solutions. In this work we considered the Yang-Baxter equation for
two-state systems, up to the eight-vertex model. This differential
approach allowed us to solve the Yang-Baxter equation in a systematic
way and also to completely classify its regular solutions.

\end{abstract}

\tableofcontents{}

\section{The Yang-Baxter equation\label{SecYBE}}

The Yang-Baxter equation (\noun{ybe}) is one of the most important
equations of contemporary mathematical-physics. It originally emerged
in two different contexts of theoretical physics: in quantum field
theory, the\noun{ ybe} appeared as a sufficient condition for the
many-body scattering amplitudes to factor into the product of pairwise
scattering amplitudes \cite{Yang1967,Yang1968,Zamolodchikov1979};
in statistical mechanics it represented a sufficient condition for
the transfer matrix of a given statistical model to commute for different
values of the spectral parameters \cite{Baxter1972,Baxter1978}.

Since the pioneer works in quantum integrable systems -- see \cite{KulishSklyanin1982,Jimbo1990,Kulish1996}
for a historical background --, the \noun{ybe} has become a cornerstone
in several fields of physics and mathematics: it is most known for
its fundamental role in the quantum inverse scattering method and
in the algebraic Bethe Ansatz \cite{SklyaninTakhtadzhyanFaddeev1979,TakhtadzhanFaddeev1979,Sklyanin1982B},
although it also revealed to be important in the formulation of Hopf
algebras and quantum groups \cite{Sklyanin1982A,Jimbo1985,Drinfeld1985,Drinfeld1988B,FaddeevReshetikhinTakhtajan1988},
in knot theory \cite{Turaev1988}, in quantum computation \cite{KauffmanLomonaco2004},
in\emph{ AdS-CFT} correspondence \cite{MinahanZarembo2003,BeisertEtal2012}
and, more recently, in gauge theory \cite{Witten1989,CostelloWittenYamazaki2017,CostelloWittenYamazaki2018}. 

The \noun{ybe} can be seen as a matrix relation defined in $\mathrm{End}\left(V\otimes V\otimes V\right)$,
where $V$ is an $n$th dimensional complex vector space. In the most
general case, it reads: 
\begin{equation}
R_{12}(x,y)R_{13}(x,z)R_{23}(y,z)=R_{23}(y,z)R_{13}(x,z)R_{12}(x,y),\label{YBE}
\end{equation}
where the arguments $x$, $y$ and $z$, called \emph{spectral parameters},
have values in $\mathbb{C}$. The solution of the \noun{ybe} is an
$R$ matrix defined in $\mathrm{End}\left(V\otimes V\right)$. The
indexed matrices $R_{ij}$ appearing in (\ref{YBE}) are defined in
$\mathrm{End}\left(V\otimes V\otimes V\right)$ through the formulas
\begin{equation}
R_{12}=R\otimes I,\qquad R_{23}=I\otimes R,\qquad\text{and}\qquad R_{13}=P_{23}R_{12}P_{23}=P_{12}R_{23}P_{12},
\end{equation}
where $I\in\mathrm{End}\left(V\right)$ is the identity matrix, $P\in\mathrm{End}\left(V\otimes V\right)$
is the \emph{permutator matrix} (defined by the relation $P\left(A\otimes B\right)P=B\otimes A$
for  $\forall\,A,B\in\mathrm{End}(V)$) and $P_{12}=P\otimes I$,
$P_{23}=I\otimes P$. 

For physical reasons, a particular form of the \noun{ybe} is usually
considered: this consists in assuming that the $R$ matrix depends
only on the difference of the spectral parameters $x$, $y$, and
$z$. In this case, the \noun{ybe} assumes the simpler form:
\begin{equation}
R_{12}(u)R_{13}(u+v)R_{23}(v)=R_{23}(v)R_{13}(u+v)R_{12}(u),\label{ybeA}
\end{equation}
where the new spectral parameters are related to the older ones by
$u=x-y$ and $v=y-z$. In this work, we shall consider only the ``additive''
\noun{ybe} (\ref{ybeA}).

For each solution of the \noun{ybe}, a given integrable system can
be associated. In fact, in statistical mechanics, the $R$ matrix
represents the Boltzmann weights of a given statistical model while,
in quantum field theory, the $R$ matrix is associated with factorizable
scattering amplitudes between relativistic particles. From the\noun{
ybe} we can prove that systems described by an $R$ matrix possess
infinitely many conserved quantities in involution -- the Hamiltonian
being one of them --, the reason why they are called \emph{integrable}
\cite{Korepin1997}. 

We say that a given solution $R(u)$ of the\noun{ ybe} (\ref{ybeA})
is regular if $R(0)=P$. Regular solutions of the\noun{ ybe} have
several important properties. We list below some of them \cite{Kulish1996}:
\begin{itemize}
\item \emph{Multiplicative property}: If $R(u)$ is a regular solution of
the \noun{ybe} (\ref{ybeA}), then the matrix $\bar{R}(u)=f(u)R(u)$,
where $f(u)$ is any complex function of the variable $u$ satisfying
$f(0)=1$, is also a regular solution of the \noun{ybe} (\ref{ybeA}). 
\item \emph{Similarity Property}: A similarity transformation of the form
$\bar{R}(u)=SR(u)S^{-1}$, where $S$ is any non-degenerated matrix
defined on $\mathrm{End}\left(V\otimes V\right)$, provides another
regular solution of the \noun{ybe} (\ref{ybeA}). 
\item \emph{Rescaling Property}: Replacing the spectral parameter $u$ by
$\omega u$, where $\omega$ is any non-null complex number, results
in another regular solution $R(\omega u)$ of the \noun{ybe} (\ref{ybeA}).
\end{itemize}
Two solutions that differ from each other only by these transformations
are said to be equivalent. Regular solutions of the \noun{ybe} can
also have several properties or symmetries; the most common are the
following \cite{Kulish1996}: 
\begin{itemize}
\item Unitarity (U) symmetry: $R(u)PR(-u)P=\rho(u)I$;
\item Permutation (P) symmetry: $PR(u)P=R(u)$;
\item Transposition (T) symmetry: $R(u)^{t}=R(u)$;
\item Permutation-Transposition (PT) symmetry: $PR(u)P=R(u)^{t}$;
\item Crossing (C) symmetry: $R(u)^{t_{1}}R(-u-2\zeta)^{t_{1}}=\sigma(u)I$.
\end{itemize}
Here, $t$ denotes transposition in $\mathrm{End}(V\otimes V)$, $t_{1}$
and $t_{2}$ denote the partial transposition in the first and second
vector spaces, respectively, $\zeta$ is called the crossing parameter
of the $R$ matrix, $I$ is the identity matrix and, finally, $\rho(u)$
and $\sigma(u)$ are two complex functions specific to each model.

We highlight that we shall not impose any of these symmetries in our
search for solutions of the \noun{ybe} (\ref{ybeA}). Nevertheless,
we indicate in Appendix \ref{AppendixS} which symmetries each solution
presents. 

\section{The differential Yang-Baxter equations }

The \noun{ybe} corresponds to a system of non-linear functional equations.
Several particular solutions of the \noun{ybe} are known \cite{KulishSklyanin1982,Jimbo1990,Kulish1996}.
The first solutions were found by a direct inspection of the functional
equations, which are in fact very simple because the $R$ matrix is
assumed to have many symmetries. Nevertheless, there are other more
advanced methods for solving the\noun{ ybe}: we can cite, for instance,
the Baxterization of braid relations \cite{Jones1990}, the use of
Lie algebras and superalgebras \cite{Jimbo1986A,Bazhanov1987,BazhanovShadrikov1987},
the construction of Hopf algebras and quantum groups \cite{Jimbo1985,Drinfeld1985,Drinfeld1988B},
and also techniques relying on algebraic geometry \cite{Krichever1981}. 

The methods mentioned above usually require that the $R$ matrix presents
one or more symmetries from the very start. From a mathematical point
of view, would be desirable to develop a method that requires in principle
as few as possible symmetries and, at the same time, that is powerful
enough in order to find and classify the solutions of the \noun{ybe.}
This paper is concerned with the development and extensive use of
such a method, which is based on a differential approach. To be more
precise, this method consists mainly of the following: if we take
the formal derivatives of the \noun{ybe} (\ref{ybeA}) with respect
to the spectral parameters $u$ and $v$ and then evaluate the derivatives
at some fixed point of those variables (say at zero), then we shall
get two systems of ordinary non-linear differential equations for
the elements of the $R$ matrix. The derivatives of the $R$ matrix
elements, however, can be regarded as independent variables, so that,
after they are eliminated, two systems of polynomial equations are
obtained in place. Thus, these polynomial systems can be analyzed
-- for instance, through techniques of the computational algebraic
geometry \cite{CoxLittleOshea2015} -- and eventually completely
solved. It happens, however, that these polynomial systems usually
have a positive Hilbert dimension, which means that the systems are
satisfied even when some of the $R$ matrix elements are still arbitrary.
The remaining unknowns, nonetheless, can be found by solving a few
number of differential equations that arise from the expressions for
the derivatives we had eliminated before. These auxiliary differential
equations, therefore, can be thought as consistency conditions of
the method. 

For example, if we take the formal derivative of (\ref{ybeA}) with
respect to $v$ and then evaluate the result at the point $v=0$,
then we shall get the equation, 
\begin{equation}
E\coloneqq R_{12}(u)D_{13}(u)P_{23}+R_{12}(u)R_{13}(u)H_{23}=H_{23}R_{13}(u)R_{12}(u)+P_{23}D_{13}(u)R_{12}(u),\label{DYB1}
\end{equation}
while, on the other hand, if we take the derivative of (\ref{ybeA})
with respect to $u$ and then evaluate the result at $u=0$, we shall
get,
\begin{equation}
F\coloneqq H_{12}R_{13}(v)R_{23}(v)+P_{12}D_{13}(v)R_{23}(v)=R_{23}(v)D_{13}(v)P_{12}+R_{23}(v)R_{13}(v)H_{12}.\label{DYB2}
\end{equation}
In (\ref{DYB1}) and (\ref{DYB2}), we introduced the quantities:
\begin{equation}
D(u)=\left.\frac{\mathrm{d}R(u+v)}{\mathrm{d}v}\right|_{v=0},\qquad D(v)=\left.\frac{\mathrm{d}R(u+v)}{\mathrm{d}u}\right|_{u=0},\label{D}
\end{equation}
where $P$ is the permutator matrix\footnote{Notice that the condition $R(0)=P$ means that we are considering
here \emph{only} regular solutions of the \noun{ybe}. If this requirement
is weakened then other equations similar to (\ref{DYB1}) and (\ref{DYB2})
can be derived and other classes of solutions can be found. Nonetheless,
the differential method \emph{per se} is not changed.} so that $R(0)=P$ and $H=D(0)$. We highlight that $H=\mathcal{H}_{L}P$,
where $\mathcal{H}_{L}$ is nothing but the local Hamiltonian associated
with the model -- see, for instance, \cite{KulishSklyanin1982,Kulish1996}.

The idea of transforming a functional equation into a differential
one goes back to the works of the Niels Henrik Abel, who solved several
functional equations in this way \cite{Abel1823}. Abel's method presents
many advantages when compared with other methods of solving functional
equations. For instance, it consists in a general method that can
be applied to a huge class of functional equations; it establishes
the generality and uniqueness of the solutions (which would be difficult,
if not impossible, to establish in other ways) by reducing the problem
to the theory of differential equations and so on -- see \cite{Aczel1966}
for more. Notice moreover that although Abel's method requires the
solutions to be differentiable (there can be non-differentiable solutions
of some functional equations), this restriction is not a problem when
dealing with the \noun{ybe}, as its solutions are always assumed to
be differentiable because of the connection between the $R$ matrix
and the corresponding local Hamiltonian. 

Concerning the theory of integrable systems, the differential method
is perhaps most known in connection with boundary \noun{ybe} \cite{Sklyanin1988,Mezincescu1991}:
\begin{equation}
R(u-v)K_{1}(u)PR(u+v)PK_{2}(v)=K_{2}(v)R(u+v)K_{1}(u)PR(u-v)P.\label{BYBE}
\end{equation}
This equation -- which is also known as the \emph{reflection equation
}-- is a generalization of the \noun{ybe} for non-periodic boundary
conditions. The fundamental unknown of the reflection equation is
the $K$ matrix -- also known as the \emph{reflection matrix }--,
while\emph{ }the $R$ matrix is assumed to be given. Notice that the
$K$ matrices figuring in (\ref{BYBE}) always depend only on a single
variable, which is why the differential method transforms (\ref{BYBE})
into a system of \emph{linear algebraic equations} instead of a non-linear
differential system. This particularity makes the differential method
as simple as powerful to study the reflection equation (\ref{BYBE}).
In fact, this method was extensively used by Lima-Santos and collaborators
in a series of papers, where solutions of (\ref{BYBE}) associated
with non-exceptional Lie algebras and superalgebras were found and
classified \cite{MalaraLima2006,Lima2009A,Lima2009B,Lima2009C,Lima2010,VieiraLima2013,VieiraLima2017A,VieiraLima2017C}.
Differential methods were also employed to solve the periodic \noun{ybe}
(\ref{ybeA}). In fact the first solutions of the \noun{ybe} were
found precisely in this way \cite{KulishSklyanin1982}. What distinguishes
the present approach from the previous ones is that here we regard
the derivatives of the $R$ matrix elements as independent variables,
so that the system of equations (\ref{DYB1}) and (\ref{DYB2}) can
be solved in an algebraic way -- it is only at the end of the calculations
that a few number of differential equations arise as consistency conditions.
This approach allowed us to make a systematic study of the \noun{ybe;}
it provided a simpler and comprehensive analysis of the possible branches
of the solutions which enabled us to make a complete classification
of the regular $R$ matrices for two-state quantum systems, up to
the eight-vertex model\footnote{A priori, the most general $R$ matrix associated with a two-state
system would be a four-by-four matrix with no zero elements -- in
this case, the \noun{ybe} (\ref{ybeA}) would represent a set of $64$
functional equations for $16$ unknowns. The Hamiltonian of such a
\emph{sixteen-vertex model} would describe a completely anisotropic
Heisenberg chain in the presence of external fields and with arbitrary
ionized configurations \cite{SuzukiFisher1971}. It is not known,
however, if the sixteen-vertex model is integrable: its $R$ matrix,
if exists, would have no symmetry at all -- not even the unitarity
one. For this reason (and because the problem of solving the \noun{ybe}
for the sixteen-vertex model is insurmountable at the present), we
shall restrict ourselves to the \emph{eight-vertex model}, whose Hamiltonian
is related to a Heisenberg chain in the presence of external fields
but with ionized configurations occurring only in pairs \cite{SuzukiFisher1971}. }. Our results agree with early classifications proposed by Sogo \&
al. in \cite{SogoEtal1982} and by Khachatryan and Sedrakyan in \cite{KhachatryanSedrakyan2013};
in fact, many of the solutions derived here are equivalent to the
ones presented in \cite{SogoEtal1982} and \cite{KhachatryanSedrakyan2013},
although a few of them seems to be new -- which is the case, for
instance, of some solutions for six-vertex models with unusual shapes,
among others. 

The eight-vertex model is the most general two-state vertex model
satisfying the $\mathbb{Z}_{2}$-symmetry \cite{SuzukiFisher1971,Baxter1985},
which means that the non-null elements $r_{i_{1},i_{2}}^{j_{1},j_{2}}(u)$
of the $R$ matrix must satisfy the relation $i_{1}-j_{1}+i_{2}-j_{2}\equiv0\text{ mod 2}$,
where the indexes $i_{1}$, $i_{2}$, $j_{1}$ and $j_{2}$ can assume
only the values $0$ or $1$. Therefore, let us write the most general
$R$ matrix associated with the eight-vertex model as follows: 
\begin{equation}
R(u)=\begin{pmatrix}a_{1}(u) & 0 & 0 & d_{1}(u)\\
0 & b_{1}(u) & c_{1}(u) & 0\\
0 & c_{2}(u) & b_{2}(u) & 0\\
d_{2}(u) & 0 & 0 & a_{2}(u)
\end{pmatrix}.\label{8VR0}
\end{equation}
Besides, let us denote the matrices $D(u)=R'(u)$ and $H=D(0)$ respectively
by
\begin{equation}
D(u)=\begin{pmatrix}a_{1}'(u) & 0 & 0 & d_{1}'(u)\\
0 & b_{1}'(u) & c_{1}'(u) & 0\\
0 & c_{2}'(u) & b_{2}'(u) & 0\\
d_{2}'(u) & 0 & 0 & a_{2}'(u)
\end{pmatrix},\qquad\text{and}\qquad H=\begin{pmatrix}\alpha_{1} & 0 & 0 & \delta_{1}\\
0 & \beta_{1} & \gamma_{1} & 0\\
0 & \gamma_{2} & \beta_{2} & 0\\
\delta_{2} & 0 & 0 & \alpha_{2}
\end{pmatrix}.\label{H}
\end{equation}
In this work, we shall look for regular solutions of the \noun{ybe}
(\ref{ybeA}) which means that $R(0)=P$, where $P$ is the permutator
matrix, explicitly given in this case by
\begin{equation}
P=\begin{pmatrix}1 & 0 & 0 & 0\\
0 & 0 & 1 & 0\\
0 & 1 & 0 & 0\\
0 & 0 & 0 & 1
\end{pmatrix}.
\end{equation}

In the next section we shall present a detailed analysis of the differential
\noun{ybe}'s (\ref{DYB1}) and (\ref{DYB2}), from which we derive
the possible solutions of the \noun{ybe} (\ref{ybeA}) for two-state
systems, up to eight-vertex model. Other important properties of these
solutions will be presented in the Appendixes.

\section{Solutions for the four-vertex model \label{Sec4V}}

The most simple regular solution of the \noun{ybe} occurs when the
$R$ matrix has the same shape as the permutator matrix. This means
that\footnote{For sake of clarity, we shall often hide the dependence of the $R$
matrix elements on the spectral parameter $u$.} $b_{1}=b_{2}=d_{1}=d_{2}=0$ in (\ref{8VR0}) and we are left with
a\emph{ four-vertex model}. In this case the $R$ matrix is,
\begin{equation}
R=\begin{pmatrix}a_{1} & 0 & 0 & 0\\
0 & 0 & c_{1} & 0\\
0 & c_{2} & 0 & 0\\
0 & 0 & 0 & a_{2}
\end{pmatrix}.
\end{equation}
Equations (\ref{DYB1}) and (\ref{DYB2}) become identically the same
in this case, and in fact there are only five linearly independent
equations, namely,\begin{subequations} 
\begin{align}
E_{2,5}=\frac{a_{1}'}{a_{1}}-\frac{c_{1}'}{c_{1}} & =\alpha_{1}-\gamma_{1}, & E_{4,7}=\frac{a_{2}'}{a_{2}}-\frac{c_{1}'}{c_{1}} & =\alpha_{2}-\gamma_{1}, & E_{3,3}=\frac{c_{2}'}{c_{2}}-\frac{c_{1}'}{c_{1}} & =\gamma_{2}-\gamma_{1},\\
E_{5,2}=\frac{a_{1}'}{a_{1}}-\frac{c_{2}'}{c_{2}} & =\alpha_{1}-\gamma_{2}, & E_{7,4}=\frac{a_{2}'}{a_{2}}-\frac{c_{2}'}{c_{2}} & =\alpha_{2}-\gamma_{2}.
\end{align}
\end{subequations} Each of the equations above has the general form
$f'/f-g'/g=\kappa$ for a given constant $\kappa$ and two functions
$f$ and $g$. The general solution of this differential equation
satisfying the initial condition $f(0)/g(0)=1$ is $f/g=\mathrm{e}^{\kappa u}$.
Therefore, the equations above fix the ratios of the $R$ matrix elements
as follows:
\begin{equation}
\frac{a_{1}}{c_{1}}=\mathrm{e}^{\left(\alpha_{1}-\gamma_{1}\right)u},\qquad\frac{a_{2}}{c_{1}}=\mathrm{e}^{\left(\alpha_{2}-\gamma_{1}\right)u},\qquad\frac{a_{1}}{c_{2}}=\mathrm{e}^{\left(\alpha_{1}-\gamma_{2}\right)u},\qquad\frac{a_{2}}{c_{2}}=\mathrm{e}^{\left(\alpha_{2}-\gamma_{2}\right)u},\qquad\frac{c_{2}}{c_{1}}=\mathrm{e}^{\left(\gamma_{2}-\gamma_{1}\right)u}.
\end{equation}
This means that we can write:
\begin{equation}
a_{1}=\mathrm{e}^{\alpha_{1}u},\qquad a_{2}=\mathrm{e}^{\alpha_{2}u},\qquad c_{1}=\mathrm{e}^{\gamma_{1}u},\qquad c_{2}=\mathrm{e}^{\gamma_{2}u},
\end{equation}
so that the general $R$ matrix of the four-vertex model is:
\begin{equation}
R(u)=\begin{pmatrix}\mathrm{e}^{\alpha_{1}u} & 0 & 0 & 0\\
0 & 0 & \mathrm{e}^{\gamma_{1}u} & 0\\
0 & \mathrm{e}^{\gamma_{2}u} & 0 & 0\\
0 & 0 & 0 & \mathrm{e}^{\alpha_{2}u}
\end{pmatrix}.\label{4V}
\end{equation}

The solution depends on four \emph{free-parameters} (namely, $\alpha_{1}$,
$\alpha_{2}$, $\gamma_{1}$ and $\gamma_{2}$). Notice, however,
that one of these parameters can be removed due to the multiplicative
property of the \noun{ybe}, which means that the four-vertex $R$
matrix above presents only three\emph{ bare free-parameters}\footnote{In general, one or more free-parameters can be removed from the solutions
thanks to the equivalence properties of regular $R$ matrices (see
Section \ref{SecYBE}).  For example, by multiplying the $R$ matrix
through a given regular function, redefining the spectral parameter,
performing a similarity transformation or by grouping together some
of the parameters and renaming them. The remaining parameters that
cannot be removed anymore from the $R$ matrix through these equivalence
transformations will be called \emph{bare free-parameters} of the
solutions. }.

\section{Solutions for the usual six-vertex model\label{Sec6V}}

Let us consider now the usual \emph{six-vertex model}. In this case
we require just that $d_{1}=d_{2}=0$ in (\ref{8VR0}) and, consequently,
the most general six-vertex $R$ matrix becomes: 
\begin{equation}
R=\begin{pmatrix}a_{1} & 0 & 0 & 0\\
0 & b_{1} & c_{1} & 0\\
0 & c_{2} & b_{2} & 0\\
0 & 0 & 0 & a_{2}
\end{pmatrix}.\label{6VR}
\end{equation}

We can verify that in this case the systems $E$ and $F$ -- respectively,
the equations (\ref{DYB1}) and (\ref{DYB2}) -- become different
each from the other, so that we get two complementary systems at our
disposal. In fact, several simple relations immediately follow from
these equations: for instance, subtracting $E_{2,5}$ from $F_{2,5}$
we are led to the relation,
\begin{equation}
\frac{b_{2}}{b_{1}}=\frac{\beta_{2}}{\beta_{1}}.\label{6Vb1b2}
\end{equation}
On the other hand, from $E_{3,3}$ or $F_{3,3}$ it follows as well
that, 
\begin{equation}
\frac{c_{2}}{c_{1}}=\mathrm{e}^{(\gamma_{2}-\gamma_{1})u}.\label{6Vc1c2}
\end{equation}
This means that we can write: 
\begin{equation}
c_{1}=\mathrm{e}^{\gamma_{1}u},\qquad c_{2}=\mathrm{e}^{\gamma_{2}u}.\label{6Vc1c2B}
\end{equation}

Now we can solve the equations $E_{2,5}$, $E_{4,7}$, $E_{4,6}$
and $E_{3,5}$ for $a_{1}'$, $a_{2}'$, $b_{1}'$ and $b_{2}'$,
respectively, which provide us with, \begin{subequations}\label{6Vderivatives}
\begin{align}
a_{1}' & =\alpha_{1}a_{1}-\beta_{1}b_{2}, & a_{2}' & =\alpha_{2}a_{2}-\beta_{2}b_{1},\\
b_{1}' & =\beta_{1}a_{2}+\left(\gamma_{1}+\gamma_{2}-\alpha_{2}\right)b_{1}, & b_{2}' & =\beta_{2}a_{1}+\left(\gamma_{1}+\gamma_{2}-\alpha_{1}\right)b_{2}.
\end{align}
 \end{subequations}

After these derivatives are eliminated, we can verify that the equations
$E_{2,3}$ and $E_{6,7}$ become, respectively, \begin{subequations}\label{6Vabc}
\begin{align}
\left(a_{1}a_{2}+b_{1}b_{2}-c_{1}c_{2}\right)\beta_{1} & =\left(\alpha_{1}+\alpha_{2}-\gamma_{1}-\gamma_{2}\right)a_{1}b_{1},\label{6Vabc1}\\
\left(a_{1}a_{2}+b_{1}b_{2}-c_{1}c_{2}\right)\beta_{2} & =\left(\alpha_{1}+\alpha_{2}-\gamma_{1}-\gamma_{2}\right)a_{2}b_{2}.\label{6Vabc2}
\end{align}
 \end{subequations} On the other hand, the equation $F_{2,3}$ gives
us a relation between $a_{2}$ and $a_{1}$: 
\begin{equation}
a_{2}-a_{1}=\left(\frac{\alpha_{2}-\alpha_{1}}{\beta_{1}}\right)b_{1}.\label{a2a1}
\end{equation}
Then, using (\ref{6Vb1b2}) and (\ref{a2a1}), it follows from (\ref{6Vabc1})
and (\ref{6Vabc2}), assuming that $b_{1}\neq0$ and $b_{2}\neq0$,
the following constraint:
\begin{equation}
\left(\alpha_{2}-\alpha_{1}\right)\left(\alpha_{1}+\alpha_{2}-\gamma_{1}-\gamma_{2}\right)=0.\label{Branch}
\end{equation}
This means that the solutions of the \noun{ybe} for the six-vertex
model admit two main branches.

\subsection{The first case $a_{2}=a_{1}$\label{Sec6VA}}

Let us first assume that $\alpha_{2}=\alpha_{1}$ which, according
to (\ref{a2a1}), implies $a_{2}=a_{1}$. Then we look for a solution
with the following properties: 
\begin{equation}
\frac{a_{2}}{a_{1}}=1,\qquad\frac{b_{1}}{b_{2}}=\frac{\beta_{1}}{\beta_{2}},\qquad\frac{c_{1}}{c_{2}}=\mathrm{e}^{\left(\gamma_{1}-\gamma_{2}\right)u}.
\end{equation}

Notice that, because $b_{2}$ is fixed through (\ref{6Vb1b2}) and
$c_{1}$, $c_{2}$ are given by (\ref{6Vc1c2B}), it only remains
to find $a_{1}$ and $b_{1}$. From (\ref{6Vderivatives}) it follows
that, 
\begin{equation}
a_{1}'=\alpha_{1}a_{1}-\beta_{2}b_{1},\qquad\text{and}\qquad b_{1}'=\beta_{1}a_{1}+\left(\gamma_{1}+\gamma_{2}-\alpha_{1}\right)b_{1}.
\end{equation}
This system of linear differential equations can be easily solved
with the initial conditions $a_{1}(0)=1$ and $b_{2}(0)=0$. The solution
is: 
\begin{align}
a_{1} & =\mathrm{e}^{\frac{1}{2}\left(\gamma_{1}+\gamma_{2}\right)u}\left[\cosh\left(\omega u\right)+\left(\tfrac{2\alpha_{1}-\gamma_{1}-\gamma_{2}}{2\omega}\right)\sinh\left(\omega u\right)\right], & b_{1} & =\beta_{1}\mathrm{e}^{\frac{1}{2}\left(\gamma_{1}+\gamma_{2}\right)u}\sinh\left(\omega u\right)/\omega,
\end{align}
where, 
\begin{equation}
\omega=\tfrac{1}{2}\sqrt{\left(2\alpha_{1}-\gamma_{1}-\gamma_{2}\right){}^{2}-4\beta_{1}\beta_{2}}.
\end{equation}

At this point, all equations of the systems $E$ and $F$ are satisfied.
Moreover, we indeed have $b_{2}'=\left(\beta_{2}/\beta_{1}\right)b_{1}'$,
so that the solution is consistent with the differential method. Therefore
we get the following solution: \begin{subequations}\label{6VFb1}
\begin{align}
a_{1} & =\mathrm{e}^{\frac{1}{2}\left(\gamma_{1}+\gamma_{2}\right)u}\left[\cosh\left(\omega u\right)+\left(\tfrac{2\alpha_{1}-\gamma_{1}-\gamma_{2}}{2\omega}\right)\sinh\left(\omega u\right)\right], & a_{2} & =\mathrm{e}^{\frac{1}{2}\left(\gamma_{1}+\gamma_{2}\right)u}\left[\cosh\left(\omega u\right)+\left(\tfrac{2\alpha_{1}-\gamma_{1}-\gamma_{2}}{2\omega}\right)\sinh\left(\omega u\right)\right],\\
b_{1} & =\left(\beta_{1}/\omega\right)\mathrm{e}^{\frac{1}{2}\left(\gamma_{1}+\gamma_{2}\right)u}\sinh\left(\omega u\right), & b_{2} & =\left(\beta_{2}/\omega\right)\mathrm{e}^{\frac{1}{2}\left(\gamma_{1}+\gamma_{2}\right)u}\sinh\left(\omega u\right),\\
c_{1} & =\mathrm{e}^{\gamma_{1}u}, & c_{2} & =\mathrm{e}^{\gamma_{2}u},
\end{align}
 \end{subequations} which depends on five free-parameters, namely,
$\alpha_{1}$, $\beta_{1}$, $\beta_{2}$, $\gamma_{1}$ and $\gamma_{2}$.
This solution can also be written in a more convenient form by introducing
the parameter $\eta$ defined by the relation
\begin{equation}
\coth\omega\eta=\frac{2\alpha_{1}-\gamma_{1}-\gamma_{2}}{2\omega},
\end{equation}
so that we get\footnote{In the whole paper, we shall use the notation $\epsilon=\pm1$. }
$\sinh\left(\omega\eta\right)=\epsilon\omega/\sqrt{\beta_{1}\beta_{2}}$,
where $\epsilon=\text{sign}\left(2\alpha_{1}-\gamma_{1}-\gamma_{2}\right)$.
Thus, in terms of this new parameter, the solution becomes:
\begin{equation}
R(u)=\begin{pmatrix}\mathrm{e}^{\frac{1}{2}(\gamma_{1}+\gamma_{2})u}\frac{\sinh\left[\omega(\epsilon u+\eta)\right]}{\sinh(\omega\eta)} & 0 & 0 & 0\\
0 & \epsilon\sqrt{\frac{\beta_{1}}{\beta_{2}}}\mathrm{e}^{\frac{1}{2}(\gamma_{1}+\gamma_{2})u}\frac{\sinh(\omega u)}{\sinh(\omega\eta)} & \mathrm{e}^{\gamma_{1}u} & 0\\
0 & \mathrm{e}^{\gamma_{2}u} & \epsilon\sqrt{\frac{\beta_{2}}{\beta_{1}}}\mathrm{e}^{\frac{1}{2}(\gamma_{1}+\gamma_{2})u}\frac{\sinh(\omega u)}{\sinh(\omega\eta)} & 0\\
0 & 0 & 0 & \mathrm{e}^{\frac{1}{2}(\gamma_{1}+\gamma_{2})u}\frac{\sinh\left[\omega(\epsilon u+\eta)\right]}{\sinh(\omega\eta)}
\end{pmatrix}.\label{6VR1}
\end{equation}
In order to count the number of bare free-parameters of this solution
we can proceed as follows: first, we can simplify the $R$ matrix
by dividing all its elements by $\mathrm{e}^{\frac{1}{2}(\gamma_{1}+\gamma_{2})u}$
(thanks to the multiplicative property of the \noun{ybe} this gives
another equivalent solution). After that, we may notice that only
the ratio between $\beta_{1}$ and $\beta_{2}$ appears in the solution,
so that we can set $\beta_{1}/\beta_{2}\rightarrow\beta$. In the
same fashion, $\eta$ always appears multiplied by $\omega$ so that
we can let $\omega\eta\rightarrow\lambda$. Finally, we can redefine
the spectral parameter $u$ through $\omega u\rightarrow u$, after
which $\gamma_{1}$ and $\gamma_{2}$ will appear only in the combination
$\left(\gamma_{1}-\gamma_{2}\right)/\omega$, which we may call $\gamma$.
This means that the solution has actually only three bare free-parameters,
namely, $\beta$, $\lambda$ and $\gamma$. It follows therefore that
the $R$ matrix above is equivalent to the solution named 6V(I) by
Sogo \& al. in \cite{SogoEtal1982} and that one given by equation
(2.15) in the work of Khachatryan \& Sedrakyan in \cite{KhachatryanSedrakyan2013}.

\subsection{The second case $a_{2}\protect\neq a_{1}$\label{Sec6VB}}

Now, let us suppose that 
\begin{equation}
\alpha_{1}+\alpha_{2}=\gamma_{1}+\gamma_{2}.\label{6VconstraintB}
\end{equation}
Taking (\ref{a2a1}) into account, this means that we are looking
for a solution with the properties: 

\begin{equation}
a_{2}-a_{1}=\left(\frac{\alpha_{2}-\alpha_{1}}{\beta_{1}}\right)b_{1},\qquad\frac{b_{1}}{b_{2}}=\frac{\beta_{1}}{\beta_{2}},\qquad\frac{c_{1}}{c_{2}}=\mathrm{e}^{\left(\gamma_{1}-\gamma_{2}\right)u}.
\end{equation}
In this case, the derivatives (\ref{6Vderivatives}) become,
\begin{align}
a_{1}' & =\alpha_{1}a_{1}-\beta_{1}b_{2}, & a_{2}' & =\alpha_{2}a_{2}-\beta_{2}b_{1}, & b_{1}' & =b_{1}\alpha_{1}+a_{2}\beta_{1}, & b_{2}' & =b_{2}\alpha_{2}+a_{1}\beta_{2},
\end{align}
which can be easily solved as we impose the initial conditions $a_{1}(0)=a_{2}(0)=1$
and $b_{1}(0)=b_{2}(0)=0$. The solution is: \begin{subequations}
\begin{align}
a_{1} & =\mathrm{e}^{\frac{1}{2}(\alpha_{1}+\alpha_{2})u}\left[\cosh\left(\omega u\right)+\left(\tfrac{\alpha_{1}-\alpha_{2}}{2\omega}\right)\sinh\left(\omega u\right)\right], & a_{2} & =\mathrm{e}^{\frac{1}{2}(\alpha_{1}+\alpha_{2})u}\left[\cosh\left(\omega u\right)-\left(\tfrac{\alpha_{1}-\alpha_{2}}{2\omega}\right)\sinh\left(\omega u\right)\right],\\
b_{1} & =\left(\beta_{1}/\omega\right)\mathrm{e}^{\frac{1}{2}(\alpha_{1}+\alpha_{2})u}\sinh\left(\omega u\right), & b_{2} & =\left(\beta_{2}/\omega\right)\mathrm{e}^{\frac{1}{2}(\alpha_{1}+\alpha_{2})u}\sinh\left(\omega u\right),
\end{align}
 \end{subequations} where, here, 
\begin{equation}
\omega=\tfrac{1}{2}\sqrt{\left(\alpha_{1}-\alpha_{2}\right){}^{2}-4\beta_{1}\beta_{2}}.
\end{equation}

We can verify that all the equations are satisfied when the constraint
$\alpha_{1}+\alpha_{2}=\gamma_{1}+\gamma_{2}$ is taken into account.
Therefore, together with the expressions for $c_{1}=\mathrm{e}^{\gamma_{1}u}$
and $c_{2}=\mathrm{e}^{\gamma_{2}u}$ given by (\ref{6Vc1c2B}) we
got a solution with five free-parameters. This solution can also be
written in a simpler form after we make the transformation 
\begin{equation}
\coth\omega\eta=\frac{\alpha_{1}-\alpha_{2}}{2\omega},
\end{equation}
so that $\sinh\left(\omega\eta\right)=\epsilon\omega/\sqrt{\beta_{1}\beta_{2}}$,
where $\epsilon=\text{sign}\left(\alpha_{1}-\alpha_{2}\right)$. In
fact, using (\ref{6VconstraintB}), the solution above becomes, 
\begin{equation}
R(u)=\begin{pmatrix}\mathrm{e}^{\frac{1}{2}(\gamma_{1}+\gamma_{2})u}\frac{\sinh\left[\omega(\epsilon u+\eta)\right]}{\sinh(\omega\eta)} & 0 & 0 & 0\\
0 & \epsilon\sqrt{\frac{\beta_{1}}{\beta_{2}}}\mathrm{e}^{\frac{1}{2}(\gamma_{1}+\gamma_{2})u}\frac{\sinh(\omega u)}{\sinh(\omega\eta)} & \mathrm{e}^{\gamma_{1}u} & 0\\
0 & \mathrm{e}^{\gamma_{2}u} & \epsilon\sqrt{\frac{\beta_{2}}{\beta_{1}}}\mathrm{e}^{\frac{1}{2}(\gamma_{1}+\gamma_{2})u}\frac{\sinh(\omega u)}{\sinh(\omega\eta)} & 0\\
0 & 0 & 0 & \mathrm{e}^{\frac{1}{2}(\gamma_{1}+\gamma_{2})u}\frac{\sinh\left[\omega(\eta-\epsilon u)\right]}{\sinh(\omega\eta)}
\end{pmatrix}.\label{6VR2}
\end{equation}
After simplifying this solution through the equivalence transformations
of the \noun{ybe} (see Section \ref{SecYBE}), we can verify that
it presents three bare free-parameters. This $R$ matrix is therefore
equivalent to the solution named 6V(II) in \cite{SogoEtal1982} and
that given by equation (2.15) in \cite{KhachatryanSedrakyan2013}.

\section{Solutions for unusual six-vertex models \label{Sec6VU}}

In the previous section, the six-vertex $R$ matrix (\ref{6VR}) was
obtained from the most general eight-vertex $R$ matrix (\ref{8VR0})
by zeroing the elements $d_{1}$ and $d_{2}$. This, however, is not
the only possibility for constructing six-vertex $R$ matrices that
are compatible with the regularity condition. Indeed, we might have
zeroed any two of the elements $b_{1}$, $b_{2}$, $d_{1}$ and $d_{2}$
instead. For instance, if we vanish the elements $b_{1}$ and $b_{2}$
then we would be led to a six-vertex model whose $R$ matrix has the
following unusual shape:
\begin{equation}
R=\begin{pmatrix}a_{1} & 0 & 0 & d_{1}\\
0 & 0 & c_{1} & 0\\
0 & c_{2} & 0 & 0\\
d_{2} & 0 & 0 & a_{2}
\end{pmatrix}.\label{RU}
\end{equation}
 In the other cases, we would get the following unusual six-vertex
models: 
\begin{equation}
R=\begin{pmatrix}a_{1} & 0 & 0 & 0\\
0 & 0 & c_{1} & 0\\
0 & c_{2} & b_{2} & 0\\
d_{2} & 0 & 0 & a_{2}
\end{pmatrix},\qquad R=\begin{pmatrix}a_{1} & 0 & 0 & d_{1}\\
0 & 0 & c_{1} & 0\\
0 & c_{2} & b_{2} & 0\\
0 & 0 & 0 & a_{2}
\end{pmatrix},\qquad R=\begin{pmatrix}a_{1} & 0 & 0 & 0\\
0 & b_{1} & c_{1} & 0\\
0 & c_{2} & 0 & 0\\
d_{2} & 0 & 0 & a_{2}
\end{pmatrix},\qquad R=\begin{pmatrix}a_{1} & 0 & 0 & d_{1}\\
0 & b_{1} & c_{1} & 0\\
0 & c_{2} & 0 & 0\\
0 & 0 & 0 & a_{2}
\end{pmatrix}.\label{R6U}
\end{equation}
In this section we shall show that the \noun{ybe} (\ref{ybeA}) admits
solutions for such unusual six-vertex $R$ matrices.

\subsection{The first case $b_{1}=0$ and $b_{2}=0$}

Let us first consider the $R$ matrix given by (\ref{RU}). Our starting
point here is again the analysis of equations $E_{3,3}$ and $F_{3,3}$.
As in the previous cases, these equations fix the ratio between $c_{2}$
and $c_{1}$ through the simple relation
\begin{equation}
\frac{c_{2}}{c_{1}}=\mathrm{e}^{\gamma_{2}-\gamma_{1}}.
\end{equation}
Here, however, equations $E_{1,1}$ and $F_{1,1}$ are not null, and
difference of them provide us with a nice relation between $d_{2}$
and $d_{1}$: 
\begin{equation}
\frac{d_{2}}{d_{1}}=\frac{\delta_{2}}{\delta_{1}}.\label{6VUd1d2}
\end{equation}
Returning to equations $E_{3,3}$ and $F_{3,3}$, we realize that
$c_{2}$ must equal $c_{1}$. Thanks to the multiplicative property
of the \noun{ybe} (see Section \ref{SecYBE}), this means that we
can write\footnote{\label{FootnoteGamma}Notice that this is the same as setting $\gamma_{2}=\gamma_{1}=0$.
If we wish, we can recover the parameter $\gamma_{1}$ by renormalyzing
the solution (e.g., by multiplying the $R$ matrix by $\mathrm{e}^{\gamma_{1}u}$).
With that the \noun{ybe} will still be satisfied, although differential
\noun{ybe}'s will be satisfied only if we redefine the other parameters
of the solution so that the renormalized $R$ matrix satisfies the
consistency condition $R'(0)=H$. The number of bare free-parameters
of the solutions are not altered by this choice, of course.}:
\begin{equation}
c_{2}=c_{1}=1.\label{c2c1=00003D1}
\end{equation}
Thus, we can go on by eliminating the derivatives $a_{1}'$, $a_{2}'$,
$d_{1}'$ and $d_{2}'$ through the equations $E_{2,5}$, $E_{4,7}$,
$E_{1,4}$ and $E_{2,2}$, respectively, from which we get the relations:
\begin{equation}
a_{1}'=\alpha_{1}a_{1}+\delta_{2}a_{2}d_{1},\qquad a_{2}'=\alpha_{2}a_{2}+\delta_{2}a_{1}d_{1},\qquad d_{1}'=(\alpha_{2}-\alpha_{1})d_{1}+\delta_{1}a_{1}^{2},\qquad d_{2}'=(\delta_{2}/\delta_{1})d_{1}.\label{6VUderivatives}
\end{equation}

Now, from $E_{1,7}$, it follows that 
\begin{equation}
a_{2}=a_{1}\frac{\left(\alpha_{2}-3\alpha_{1}\right)d_{1}+\delta_{1}a_{1}^{2}}{\delta_{2}d_{1}^{2}+\delta_{1}},\label{6VUa2}
\end{equation}
after which $E_{2,8}$ becomes a quadratic equation for $a_{1}^{2}$:

\begin{equation}
\delta_{1}^{2}a_{1}^{4}-\left(4\alpha_{1}\delta_{1}d_{1}\right)a_{1}^{2}-\left\{ \delta_{2}^{2}d_{1}^{4}+\left[2\delta_{1}\delta_{2}-\left(3\alpha_{1}-\alpha_{2}\right)\left(\alpha_{1}+\alpha_{2}\right)\right]d_{1}^{2}+\delta_{1}^{2}\right\} =0.
\end{equation}
Assuming $\delta_{1}$ positive, the only solution of this equation
that satisfies the initial conditions $a_{1}(0)=a_{2}(0)=1$ and $d_{1}(0)=0$
is
\begin{equation}
a_{1}=\sqrt{2\left(\frac{\alpha_{1}}{\delta_{1}}\right)d_{1}+\sqrt{\left(\frac{\delta_{2}}{\delta_{1}}\right)^{2}d_{1}^{4}+\left[\left(\frac{\alpha_{1}-\alpha_{2}}{\delta_{1}}\right)^{2}+2\left(\frac{\delta_{2}}{\delta_{1}}\right)\right]d_{1}^{2}+1}}.\label{6VUa1}
\end{equation}
The other solutions differ from the above one by negative signs in
front of the square roots, but they do not satisfy the required initial
conditions for $\delta_{1}$ positive\footnote{\label{FootNoteBranches}In general, for some complexes values of
$\delta_{1}$, the initial conditions can also be satisfied when the
signs in front the square roots are negative. A detailed study of
these cases shows, however, that this leads to the same solution as
the one presented in the text. This can be explained from the fact
that $\delta_{1}$ is a free-parameter of the solution, so that the
negative signs can be absorbed into its definition. For this reason,
in similar situations, we shall assume that the free-parameters of
the solutions are positive, although the corresponding solution may
be valid as well for negative, or even complex, values of these parameters.}.

After $a_{1}$ is fixed by (\ref{6VUa1}), we can verify that the
remaining equations imply the constraint $\alpha_{2}^{2}=\alpha_{1}^{2}$,
so that the solutions branches into two ways.

\subsubsection{The subcase $a_{2}=a_{1}$}

Let us consider first the branch in which $\alpha_{2}=\alpha_{1}$.
In this case, from (\ref{6VUa2}) and (\ref{6VUa1}) we get that
\begin{equation}
a_{2}=a_{1}=\sqrt{\frac{\delta_{2}d_{1}^{2}+2\alpha_{1}d_{1}+\delta_{1}}{\delta_{1}}}.\label{6Ua1}
\end{equation}
Therefore, the solution is characterized by 
\begin{equation}
\frac{a_{2}}{a_{1}}=1,\qquad\frac{c_{2}}{c_{1}}=1,\qquad\frac{d_{2}}{d_{1}}=\frac{\delta_{2}}{\delta_{1}}.
\end{equation}

As $d_{2}$ is already fixed by (\ref{6VUd1d2}), it remains to find
$d_{1}$. This follows from the third equation in (\ref{6VUderivatives}),
which provides the following non-linear differential equation: 
\begin{equation}
d_{1}'=\delta_{2}d_{1}^{2}+2\alpha_{1}d_{1}+\delta_{1}.\label{6VUedo}
\end{equation}
This is a Riccati differential equation with constant coefficients
(see \cite{NIST2010}). To solve it, let us rewrite it in the form
(we have written $y$ in place of $d_{1}$ for convenience):
\begin{equation}
y'=Ay^{2}+By+C=A(y-\xi_{1})(y-\xi_{2}),\label{6VUedoY}
\end{equation}
 where $A=\delta_{2}$, $B=2\alpha_{1}$, $C=\delta_{1}$ and $\xi_{1}$,
$\xi_{2}$ are the roots of the quadratic equation $Ay^{2}+By+C=0$. 

Let us first assume that $\xi_{2}\neq\xi_{1}$ (the degenerated case
$\xi_{2}=\xi_{1}$ will be presented in Appendix \ref{AppendixReduced}).
In this case, the differential equation (\ref{6VUedoY}) can be reduced
to an integral, which, by its turn, can be easily solved through partial
fractions method: 
\begin{equation}
u(y)=\frac{1}{A}\int\frac{\mathrm{d}y}{(y-\xi_{1})(y-\xi_{2})}=\frac{\log\left(y-\xi_{1}\right)-\log\left(y-\xi_{2}\right)}{A(\xi_{1}-\xi_{2})}+c,
\end{equation}
where $c$ is the constant of integration. Inverting this relation,
we get the desired general solution of (\ref{6VUedoY}): 
\begin{equation}
y(u)=\frac{\xi_{2}-\xi_{1}}{\mathrm{e}^{A\left(\xi_{2}-\xi_{1}\right)(c-u)}-1}+\xi_{2}.
\end{equation}
Thereby, the corresponding solution of (\ref{6VUedo}) can be found
by imposing the initial condition $d_{1}(0)=y(0)=0$, which implies
the value $c=\log\left(\xi_{2}/\xi_{1}\right)/\left[A(\xi_{1}-\xi_{2})\right]$
for the constant of integration. Thus, after we replace back the values
of $A$, $\xi_{1}$ and $\xi_{2}$, we shall get that, 
\begin{equation}
d_{1}=\frac{\delta_{1}\tanh\left(u\sqrt{\alpha_{1}^{2}-\delta_{1}\delta_{2}}\right)}{\sqrt{\alpha_{1}^{2}-\delta_{1}\delta_{2}}-\alpha_{1}\tanh\left(u\sqrt{\alpha_{1}^{2}-\delta_{1}\delta_{2}}\right)},
\end{equation}
from what follows the desired $R$ matrix: 
\begin{equation}
R(u)=\begin{pmatrix}\dfrac{\omega\,\text{sech}\left(\omega u\right)}{\omega-\alpha_{1}\tanh\left(\omega u\right)} & 0 & 0 & \dfrac{\delta_{1}\tanh\left(\omega u\right)}{\omega-\alpha_{1}\tanh\left(\omega u\right)}\\
0 & 0 & 1 & 0\\
0 & 1 & 0 & 0\\
\dfrac{\delta_{2}\tanh\left(\omega u\right)}{\omega-\alpha_{1}\tanh\left(\omega u\right)} & 0 & 0 & \dfrac{\omega\,\text{sech}\left(\omega u\right)}{\omega-\alpha_{1}\tanh\left(\omega u\right)}
\end{pmatrix},\qquad\omega=\sqrt{\alpha_{1}^{2}-\delta_{1}\delta_{2}},\label{6UR1}
\end{equation}
which depends on three free-parameters (namely, $\alpha_{1}$, $\delta_{1}$
and $\delta_{2}$). Introducing the parameter $\eta$ through the
relation $\coth\left(\omega\eta\right)=\alpha_{1}/\omega$ so that
$\sinh\left(\omega\eta\right)=\omega/\sqrt{\delta_{1}\delta_{2}}$,
the solution above can be rewritten as 
\begin{equation}
R(u)=\begin{pmatrix}\dfrac{\sinh\left(\omega\eta\right)}{\sinh\left[\omega(\eta-u)\right]} & 0 & 0 & \sqrt{\dfrac{\delta_{1}}{\delta_{2}}}\dfrac{\sinh\left(\omega u\right)}{\sinh\left[\omega(\eta-u)\right]}\\
0 & 0 & 1 & 0\\
0 & 1 & 0 & 0\\
\sqrt{\dfrac{\delta_{2}}{\delta_{1}}}\dfrac{\sinh\left(\omega u\right)}{\sinh\left[\omega(\eta-u)\right]} & 0 & 0 & \dfrac{\sinh\left(\omega\eta\right)}{\sinh\left[\omega(\eta-u)\right]}
\end{pmatrix},\qquad\omega=\sqrt{\alpha_{1}^{2}-\delta_{1}\delta_{2}}.
\end{equation}
from which we see that the number of bare free-parameters is two ($\omega$
can be removed by redefining $u$ and $\eta$). Therefore, it can
be verified that this solution is equivalent to that given by equation
(5.25) in \cite{KhachatryanSedrakyan2013}.

\subsubsection{The subcase $a_{2}\protect\neq a_{1}$}

Now, let us consider the second possibility in which $\alpha_{2}=-\alpha_{1}$.
From (\ref{6VUa2}) and (\ref{6VUa1}) it follows that we have, in
this case, 
\begin{equation}
a_{1}=\sqrt{2\left(\frac{\alpha_{1}}{\delta_{1}}\right)d_{1}+\sqrt{\left(\frac{\delta_{2}}{\delta_{1}}\right)^{2}d_{1}^{4}+\left[4\left(\frac{\alpha_{1}}{\delta_{1}}\right)^{2}+2\left(\frac{\delta_{2}}{\delta_{1}}\right)\right]d_{1}^{2}+1}},\qquad\text{and}\qquad a_{2}=\frac{1}{a_{1}}\left(1+\frac{\delta_{2}}{\delta_{1}}d_{1}^{2}\right).
\end{equation}
Therefore, this branch is characterized by 
\begin{equation}
\frac{a_{2}}{a_{1}}\neq1,\qquad\frac{c_{2}}{c_{1}}=1,\qquad\frac{d_{2}}{d_{1}}=\frac{\delta_{2}}{\delta_{1}}.
\end{equation}

As in the previous case, $d_{1}$ can be found through the differential
equation provided by the third equation in (\ref{6VUderivatives}).
Here, however, we get the following non-linear differential equation\footnote{We highlight that the non-linear differential equation $\left(\mathrm{d}y/\mathrm{d}x\right)^{2}=Ay^{4}+By^{2}+C$
appears in several fields of mathematics and physics. In fact, in
the last decade it becomes a cornerstone in some expansion methods
applied to non-linear wave equations, providing in this way new elliptic
solutions for several non-linear partial differential equations. For
instance, this expansion method provided a countless number of solutions
for the Klein-Gordon-Schrödinger \cite{Wang2003}, Kronig-Penny \cite{HuaiTangHongQing2003},
Boussinesq \cite{HuaiTangHongQing2004}, Korteweg-de-Vries \cite{Yan2004},
Burgers \cite{Kudryashov2009}, Ostrovsky \cite{EbaidAly2012} equations
and their generalizations -- only to cite a few. It is interesting
to notice that all the elliptic solutions of the \noun{ybe} derived
here follows from the solutions of this differential equation.}: 
\begin{equation}
d_{1}'=\sqrt{Ad_{1}^{4}+Bd_{1}^{2}+C},\label{6VUedoD}
\end{equation}
where 
\begin{equation}
A=\delta_{2}^{2},\qquad B=4\alpha_{1}^{2}+2\delta_{1}\delta_{2},\qquad C=\delta_{1}^{2}.\label{6VUabc2}
\end{equation}

In the what follows, we shall show that the general solution of (\ref{6VUedoD})
is given in terms of \emph{Jacobian elliptic functions}. To see why,
notice first that (\ref{6VUedoD}) can be reduced to the following
integral: 
\begin{equation}
u(y)=\int\frac{\mathrm{d}y}{\sqrt{Ay^{4}+By^{2}+C}}=\int\frac{\mathrm{d}y}{\sqrt{A\left(y^{2}-\xi_{1}^{2}\right)\left(y^{2}-\xi_{2}^{2}\right)}},\label{IntJ}
\end{equation}
where we have written $y$ in place of $d_{1}$ for convenience and
$\xi_{1}^{2}$ and $\xi_{2}^{2}$ are the two roots of the quadratic
equation $Ay^{2}+By^{2}+C=0$ with $A$, $B$ and $C$ given by (\ref{6VUabc2}).
Now, assuming $\xi_{2}\neq\xi_{1}$ (the degenerated case $\xi_{2}=\xi_{1}$
will be presented in Appendix \ref{AppendixReduced}) and making the
change of variable $y\rightarrow\xi_{1}\sin\phi$, it follows that
(\ref{IntJ}) can be rewritten as: 
\begin{equation}
u(\phi)=\frac{1}{\sqrt{\xi_{2}^{2}A}}\int\frac{\mathrm{d}\phi}{\sqrt{1-k^{2}\sin^{2}\phi}},\label{Int}
\end{equation}
where $k=\xi_{1}/\xi_{2}$. The integral above is the definition of
the \emph{trigonometric elliptic integral of first kind of modulus
$k$} (see \cite{NIST2010}), denoted here by $F(\phi|k)$, so that
we get,
\begin{equation}
u(\phi)=\frac{F(\phi|k)}{\sqrt{\xi_{2}^{2}A}}+c,
\end{equation}
where $c$ is the constant of integration. Now, the inverse of $F(\phi|k)$
is the \emph{Jacobi amplitude function}, $\phi(u)=\text{am}(u|k)$,
from which we get the general solution for the function $\phi$:
\begin{equation}
\phi(u)=\mathrm{\text{am}}\left(\left(u-c\right)\sqrt{\xi_{2}^{2}A}\left|\frac{\xi_{1}}{\xi_{2}}\right.\right).\label{AM}
\end{equation}

The solution for $d_{1}$ follows from the relation $d_{1}=\xi_{1}\sin\phi$,
after we use the identity $\sin(\text{am}(u,k))=\text{sn}(u,k)$ and
impose the correct initial conditions. In fact, the condition $d_{1}(0)=0$
implies $c=0$, and from $A=\delta_{2}^{2}$, we get that\footnote{\label{FootnoteInversion}From the identity $\mathrm{sn}\left(x|1/k\right)=k\,\mathrm{sn}\left(x/k|k\right)$
for the inversion of the elliptic modulus $k$ (see \cite{NIST2010}),
we also have that $d_{1}=\xi_{2}\,\text{sn}\left(\delta_{2}\xi_{1}u|\xi_{2}/\xi_{1}\right)$.
This property shows us that for every elliptic solution of the \noun{ybe,}
another one can be obtained by exchanging $\xi_{1}$ and $\xi_{2}$
(i.e., by inverting the modulus $k$). In the present case of the
$R$ matrix (\ref{6VUR2A}), the inversion of the modulus has the
effect of exchanging $a_{1}$ and $a_{2}$. },
\begin{equation}
d_{1}=\epsilon\,\xi_{1}\text{sn}\left(\xi_{2}\delta_{2}u\left|\frac{\xi_{1}}{\xi_{2}}\right.\right),\label{d1}
\end{equation}
where $\epsilon=\text{cosign}\left(\xi_{2}\delta_{2}\right)$ . Therefore,
we have determined all elements of the $R$ matrix. After using the
following well-known identities for the elliptic functions (see \cite{NIST2010}),
\begin{equation}
\mathrm{cn}^{2}\left(x|k\right)=1-\mathrm{sn}^{2}\left(x|k\right),\qquad\mathrm{dn}^{2}\left(x|k\right)=1-k^{2}\mathrm{sn}^{2}\left(x|k\right),\label{Identity1}
\end{equation}
to simplify the elements of the $R$ matrix, we shall get the solution,
\begin{subequations}\label{6VUR2A}
\begin{align}
a_{1} & =\sqrt{\text{cn}\left(\omega u|k\right)\text{dn}\left(\omega u|k\right)+2\epsilon\frac{\alpha_{1}\xi_{1}}{\delta_{1}}\text{sn}\left(\omega u|k\right)}, & a_{2} & =\dfrac{1+\left(\dfrac{\delta_{2}}{\delta_{1}}\right)\xi_{1}^{2}\text{sn}^{2}\left(\omega u|k\right)}{\sqrt{\text{cn}\left(\omega u|k\right)\text{dn}\left(\omega u|k\right)+2\epsilon\dfrac{\alpha_{1}\xi_{1}}{\delta_{1}}\text{sn}\left(\omega u|k\right)}},\\
b_{1} & =0, & b_{2} & =0,\\
c_{1} & =1, & c_{2} & =1,\\
d_{1} & =\epsilon\xi_{1}\text{sn}\left(\omega u|k\right), & d_{2} & =\epsilon\left(\frac{\delta_{2}}{\delta_{1}}\right)\xi_{1}\text{sn}\left(\omega u|k\right),
\end{align}
\end{subequations} where $\omega=i\alpha_{1}+\epsilon i\sqrt{\alpha_{1}^{2}+\delta_{1}\delta_{2}}$,
$\xi_{1}=\delta_{1}/\omega$, $\xi_{2}=\omega/\delta_{2}$ and $k=\xi_{1}/\xi_{2}=\delta_{1}\delta_{2}/\omega^{2}$.

This solution can be simplified further by introducing a parameter
$\eta$ through the relation $i\epsilon\sqrt{\delta_{2}/\delta_{1}}\text{sn}\left(\omega\eta|k\right)=1/\xi_{1}$.
Thus, we can verify from the relations (\ref{6VUabc2}) and the identities
(\ref{Identity1}) that $\text{cn}\left(\omega\eta|k\right)\text{dn}\left(\omega\eta|k\right)=2\alpha_{1}\xi_{1}/\delta_{1}$.
Then, from the addition formula of the elliptic sinus (see \cite{NIST2010}),
\begin{equation}
\text{sn}\left(x+y|k\right)=\frac{\text{sn}\left(x|k\right)\text{cn}\left(y|k\right)\text{dn}\left(y|k\right)+\text{sn}\left(y|k\right)\text{cn}\left(x|k\right)\text{dn}\left(x|k\right)}{1-k^{2}\text{sn}^{2}\left(x|k\right)\text{sn}^{2}\left(y|k\right)},\label{Identity2}
\end{equation}
it follows that (\ref{6VUR2A}) can be rewritten as 
\begin{equation}
R(u)=\begin{pmatrix}\sqrt{\varLambda\left(u,\eta\right)\dfrac{\text{sn}\left(\omega\left(\epsilon u+\eta\right)|k\right)}{\text{sn}\left(\omega\eta|k\right)}} & 0 & 0 & i\epsilon\sqrt{\dfrac{\delta_{1}}{\delta_{2}}}\dfrac{\text{sn}(\omega u|k)}{\text{sn}\left(\omega\eta|k\right)}\\
0 & 0 & 1 & 0\\
0 & 1 & 0 & 0\\
i\epsilon\sqrt{\dfrac{\delta_{2}}{\delta_{1}}}\dfrac{\text{sn}(\omega u|k)}{\text{sn}\left(\omega\eta|k\right)} & 0 & 0 & \dfrac{1-\dfrac{\text{sn}^{2}\left(\omega u|k\right)}{\text{sn}^{2}\left(\omega\eta|k\right)}}{\sqrt{\varLambda\left(u,\eta\right)\dfrac{\text{sn}\left(\omega\left(\epsilon u+\eta\right)|k\right)}{\text{sn}\left(\omega\eta|k\right)}}}
\end{pmatrix},\label{6VUR2B}
\end{equation}
where $\varLambda\left(u,\eta\right)=1-k^{2}\text{sn}^{2}\left(\omega u|k\right)\text{sn}^{2}\left(\omega\eta|k\right)$.
Alternatively, from the identity (see \cite{NIST2010}), 
\begin{equation}
{\it \text{sn}}\left(x-y|k\right)=\frac{\text{sn}^{2}\left(x|k\right)-\text{sn}^{2}\left(y|k\right)}{\text{sn}\left(x|k\right)\text{cn}\left(y|k\right)\text{dn}\left(y|k\right)+\text{sn}\left(y|k\right)\text{cn}\left(x|k\right)\text{dn}\left(x|k\right)},
\end{equation}
we can also rewrite (\ref{6VUR2A}) as follows\footnote{Notice that (\ref{Identity1}) and (\ref{Identity2}) imply the remarkable
identity: ${\it \text{sn}}\left(x+y|k\right){\it \text{sn}}\left(x-y|k\right)=\frac{\text{sn}^{2}\left(x|k\right)-\text{sn}^{2}\left(y|k\right)}{1-k^{2}\text{sn}^{2}\left(x|k\right)\text{sn}^{2}\left(y|k\right)}$.
This provides another way of written the $R$ matrix (\ref{6VUR2A}).}: 
\begin{equation}
R(u)=\begin{pmatrix}\sqrt{\left[1-\dfrac{\text{sn}^{2}\left(\omega u|k\right)}{\text{sn}^{2}\left(\omega\eta|k\right)}\right]\dfrac{\text{sn}\left(\omega\eta|k\right)}{{\it \text{sn}}\left(\omega(\eta-\epsilon u)|k\right)}} & 0 & 0 & i\epsilon\sqrt{\dfrac{\delta_{1}}{\delta_{2}}}\dfrac{\text{sn}(\omega u|k)}{\text{sn}\left(\omega\eta|k\right)}\\
0 & 0 & 1 & 0\\
0 & 1 & 0 & 0\\
i\epsilon\sqrt{\dfrac{\delta_{2}}{\delta_{1}}}\dfrac{\text{sn}(\omega u|k)}{\text{sn}\left(\omega\eta|k\right)} & 0 & 0 & \sqrt{\left[1-\dfrac{\text{sn}^{2}\left(\omega u|k\right)}{\text{sn}^{2}\left(\omega\eta|k\right)}\right]\dfrac{{\it \text{sn}}\left(\omega(\eta-\epsilon u)|k\right)}{\text{sn}\left(\omega\eta|k\right)}}
\end{pmatrix}.\label{6VUR2}
\end{equation}

This solution is characterized by two bare free-parameters (e.g.,
$\delta_{1}/\delta_{2}$ and $k$, as $\omega\eta$ is a function
of $k$). We did not find such $R$ matrices in the literature; we
mention, however, that an elliptic $R$ matrix with this same shape
was already discussed in \cite{KhachatryanSedrakyan2013}, which corresponds
to a specific limit of two asymmetric eight-vertex $R$ matrices (which
are equivalent to the solutions discussed in Section \ref{Sec8Vdifferent}
of this paper), after several elliptic transformations are performed.
We were not able to verify if the $R$ matrix (\ref{6VUR2}) can be
mapped to that one reported in \cite{KhachatryanSedrakyan2013} by
performing some combinations of elliptic transformations and identities
(e.g., Jacobi's imaginary modulus transformation, the double-argument
identities or others, see \cite{NIST2010}). Nonetheless, we did check
that these solutions have essentially the same trigonometric limits
-- see Appendix \ref{AppendixReduced}.

\subsection{The remaining cases }

For the remaining cases of the unusual six-vertex $R$ matrices given
by (\ref{R6U}), the functional equations are quite simple, reason
for which we shall report only the final results here. They are:

\begin{equation}
R(u)=\begin{pmatrix}\mathrm{e}^{\alpha_{1}u} & 0 & 0 & 0\\
0 & 0 & \mathrm{e}^{\alpha_{1}u} & 0\\
0 & \mathrm{e}^{\alpha_{2}u} & \epsilon\left(\mathrm{e}^{\alpha_{2}u}-\mathrm{e}^{\alpha_{1}u}\right) & 0\\
\delta_{2}\frac{\mathrm{e}^{\alpha_{2}u}-\mathrm{e}^{\alpha_{1}u}}{\alpha_{2}-\alpha_{1}} & 0 & 0 & \mathrm{e}^{\alpha_{2}u}
\end{pmatrix},\qquad R(u)=\begin{pmatrix}\mathrm{e}^{\alpha_{1}u} & 0 & 0 & \delta_{1}\frac{\mathrm{e}^{\alpha_{2}u}-\mathrm{e}^{\alpha_{1}u}}{\alpha_{2}-\alpha_{1}}\\
0 & 0 & \mathrm{e}^{\alpha_{2}u} & 0\\
0 & \mathrm{e}^{\alpha_{1}u} & \epsilon\left(\mathrm{e}^{\alpha_{2}u}-\mathrm{e}^{\alpha_{1}u}\right) & 0\\
0 & 0 & 0 & \mathrm{e}^{\alpha_{2}u}
\end{pmatrix},\label{R6U34}
\end{equation}
 and
\begin{equation}
R(u)=\begin{pmatrix}\mathrm{e}^{\alpha_{1}u} & 0 & 0 & 0\\
0 & \epsilon\left(\mathrm{e}^{\alpha_{2}u}-\mathrm{e}^{\alpha_{1}u}\right) & \mathrm{e}^{\alpha_{2}u} & 0\\
0 & \mathrm{e}^{\alpha_{1}u} & 0 & 0\\
\delta_{2}\frac{\mathrm{e}^{\alpha_{2}u}-\mathrm{e}^{\alpha_{1}u}}{\alpha_{2}-\alpha_{1}} & 0 & 0 & \mathrm{e}^{\alpha_{2}u}
\end{pmatrix},\qquad R(u)=\begin{pmatrix}\mathrm{e}^{\alpha_{1}u} & 0 & 0 & \delta_{1}\frac{\mathrm{e}^{\alpha_{2}u}-\mathrm{e}^{\alpha_{1}u}}{\alpha_{2}-\alpha_{1}}\\
0 & \epsilon\left(\mathrm{e}^{\alpha_{2}u}-\mathrm{e}^{\alpha_{1}u}\right) & \mathrm{e}^{\alpha_{1}u} & 0\\
0 & \mathrm{e}^{\alpha_{2}u} & 0 & 0\\
0 & 0 & 0 & \mathrm{e}^{\alpha_{2}u}
\end{pmatrix}.\label{R6U56}
\end{equation}

\section{Solutions for the eight-vertex model \label{Sec8V}}

The most general $R$ matrix considered in this work belongs to the
eight-vertex model and it has the following shape: 
\begin{equation}
R=\begin{pmatrix}a_{1} & 0 & 0 & d_{1}\\
0 & b_{1} & c_{1} & 0\\
0 & c_{2} & b_{2} & 0\\
d_{2} & 0 & 0 & a_{2}
\end{pmatrix}.\label{8VR}
\end{equation}
Here again, we can verify that the equations $E_{1,1}$, $F_{1,1}$,
$E_{3,3}$ and $F_{3,3}$ imply the relations 
\begin{equation}
\frac{c_{2}}{c_{1}}=1,\qquad\text{and}\qquad\frac{d_{2}}{d_{1}}=\frac{\delta_{2}}{\delta_{1}}.
\end{equation}
Therefore, without loss, we can assume henceforward that
\begin{equation}
c_{2}=c_{1}=1.\label{8Vc2c1}
\end{equation}
The difference of $E_{2,5}$ with $F_{2,5}$ also provides the relation:
\begin{equation}
\frac{b_{2}}{b_{1}}=\frac{\beta_{2}}{\beta_{1}}.
\end{equation}

Now we can eliminate the derivatives $a_{1}'$, $a_{2}'$, $b_{1}'$
and $b_{2}'$ from the equations $E_{2,5}$, $E_{4,7}$, $E_{4,6}$
and $E_{3,5}$, respectively, which become, after simplification,
\begin{subequations}\label{8Vderivatives}
\begin{align}
a_{1}' & =\alpha_{1}a_{1}-b_{2}\beta_{1}+\delta_{2}a_{2}d_{1}, & a_{2}' & =\alpha_{2}a_{2}-\beta_{2}b_{1}+\delta_{2}a_{1}d_{1},\\
b_{1}' & =\beta_{1}a_{2}-\alpha_{2}b_{1}-\delta_{2}b_{2}d_{1}, & b_{2}' & =\beta_{2}a_{1}-\alpha_{1}b_{2}-\delta_{2}b_{1}d_{1}.
\end{align}
 \end{subequations} Besides, equation $F_{2,3}$ also provides,
\begin{equation}
a_{2}=a_{1}+\left(\frac{\alpha_{2}-\alpha_{1}}{\beta_{1}}\right)b_{1},\label{8Va2a1}
\end{equation}
in the same fashion as in the usual six-vertex model. 

At this point, if we take the difference between $E_{1,4}$ and $F_{2,8}$
and assume that $\beta_{1}\neq0$, then we shall get the relation
\begin{equation}
\left(\alpha_{2}-\alpha_{1}\right)\left\{ 2\beta_{1}^{2}d_{1}-\delta_{1}b_{1}\left[\left(\alpha_{2}-\alpha_{1}\right)b_{1}+2\beta_{1}a_{1}\right]\right\} =0,\label{8vBranch1}
\end{equation}
which evince how the solution branches. 

\subsection{The first case $a_{2}=a_{1}$\label{Sec8Vequal}}

The first branch occurs when $\alpha_{2}=\alpha_{1}$ which, according
to (\ref{8Va2a1}), implies as well that 
\begin{equation}
a_{2}=a_{1}.
\end{equation}

Now, multiplying $E_{2,3}$ by $\beta_{2}$ and $E_{6,7}$ by $\beta_{1}$
and taking the difference of them, we are led to the equation, 
\begin{equation}
\left(\beta_{1}^{2}-\beta_{2}^{2}\right)\left(\beta_{1}d_{1}-a_{1}b_{1}\delta_{1}\right)=0,\label{8VBranch2}
\end{equation}
for $\beta_{1}$, $\delta_{1}$ and $\delta_{2}$ different from zero.
It seems, therefore, that we have other three possible branches for
the solutions. It can be verified, however, that the constraint $\beta_{1}d_{1}-a_{1}b_{1}\delta_{1}=0$
is equivalent to the condition $\beta_{2}=\beta_{1}$, so that there
are actually only two possibilities to consider, namely, either $\beta_{2}=\beta_{1}$
or $\beta_{2}=-\beta_{1}$.

\subsubsection{The subcase $b_{2}=b_{1}$ }

Let us consider first the possibility $\beta_{2}=\beta_{1}$, which
implies $b_{2}=b_{1}$. Therefore, this solution is characterized
by the ratios:
\begin{equation}
\frac{a_{2}}{a_{1}}=1,\qquad\frac{b_{2}}{b_{1}}=1,\qquad\frac{c_{2}}{c_{1}}=1,\qquad\text{and}\qquad d_{2}=\frac{\delta_{2}}{\delta_{1}}d_{1}.
\end{equation}

Now, from equation $E_{1,4}$ we can eliminate $d_{1}'$: 
\begin{equation}
d_{1}'=\delta_{1}\left(a_{1}^{2}-b_{1}^{2}\right).
\end{equation}
Besides, from $E_{1,6}$ and $E_{1,7}$ we can fix $a_{1}$ and $d_{1}$:
\begin{subequations}
\begin{align}
a_{1} & =\frac{\beta_{1}}{\beta_{1}^{2}-\delta_{1}\delta_{2}b_{1}^{2}}\left(\alpha_{1}b_{1}+\sqrt{\delta_{1}\delta_{2}b_{1}^{4}+\left(\alpha_{1}^{2}-\beta_{1}^{2}-\delta_{1}\delta_{2}\right)b_{1}^{2}+\beta_{1}^{2}}\right),\\
d_{1} & =\frac{\delta_{1}b_{1}}{\beta_{1}^{2}-\delta_{1}\delta_{2}b_{1}^{2}}\left(\alpha_{1}b_{1}+\sqrt{\delta_{1}\delta_{2}b_{1}^{4}+\left(\alpha_{1}^{2}-\beta_{1}^{2}-\delta_{1}\delta_{2}\right)b_{1}^{2}+\beta_{1}^{2}}\right),
\end{align}
 \end{subequations} where we assumed $\beta_{1}$ positive (see Footnote
\ref{FootNoteBranches}) and $a_{1}^{2}\neq b_{1}^{2}$, since otherwise
the solution would not be regular.

At this point, we can verify that all the equations of the systems
$E$ and $F$ are satisfied. It remains, however, to find $b_{1}$.
This can be achieved from the third equation in (\ref{8Vderivatives}),
which provides the following non-linear differential equation:
\begin{equation}
b_{1}'=\sqrt{Ab_{1}^{4}+Bb_{1}^{2}+C},\label{EDO1}
\end{equation}
 with
\begin{equation}
A=\delta_{1}\delta_{2},\qquad B=\alpha_{1}^{2}-\beta_{1}^{2}-\delta_{1}\delta_{2},\qquad C=\beta_{1}^{2}.\label{8VABC1}
\end{equation}
The differential equation (\ref{EDO1}) has the same form as (\ref{6VUedoD}),
provided we replace $b_{1}$ with $d_{1}$. Moreover, since both $b_{1}$
and $d_{1}$ should satisfy the same initial condition (namely, $b_{1}(0)=d_{1}(0)=0$),
it follows that the required solution for $b_{1}$ has same form as
that for $d_{1}$. Therefore, the desired solution for $b_{1}$ is
\begin{equation}
b_{1}=\xi_{1}\,\mathrm{sn}\left(u\sqrt{\xi_{2}^{2}A}\left|\frac{\xi_{1}}{\xi_{2}}\right.\right)=\xi_{1}\,\mathrm{sn}\left(u\sqrt{\xi_{2}^{2}\delta_{1}\delta_{2}}\left|\frac{\xi_{1}}{\xi_{2}}\right.\right),
\end{equation}
 where $\xi_{1}^{2}$ and $\xi_{2}^{2}$ are the roots of the quadratic
equation $Ax^{2}+Bx+C=0$, with $A$, $B$ and $C$ given by (\ref{8VABC1})
and we assumed $\xi_{2}^{2}\neq\xi_{1}^{2}$ (for the degenerated
case $\xi_{2}^{2}=\xi_{1}^{2}$, see Appendix \ref{AppendixReduced}).

From the identities (\ref{Identity1}), we can simplify all the elements
of the $R$ matrix, from which we obtain the following solution: \begin{subequations}\label{8VRSymmetric}
\begin{align}
a_{1} & =\frac{\xi_{1}^{2}}{\xi_{1}^{2}-k^{2}\mathrm{sn}^{2}\left(\omega u|k\right)}\left[\frac{\alpha_{1}\xi_{1}}{\beta_{1}}\mathrm{sn}\left(\omega u|k\right)+\mathrm{cn}\left(\omega u|k\right)\mathrm{dn}\left(\omega u|k\right)\right],\\
a_{2} & =\frac{\xi_{1}^{2}}{\xi_{1}^{2}-k^{2}\mathrm{sn}^{2}\left(\omega u|k\right)}\left[\frac{\alpha_{1}\xi_{1}}{\beta_{1}}\mathrm{sn}\left(\omega u|k\right)+\mathrm{cn}\left(\omega u|k\right)\mathrm{dn}\left(\omega u|k\right)\right],\\
b_{1} & =\xi_{1}\mathrm{sn}\left(\omega u|k\right),\\
b_{2} & =\xi_{1}\mathrm{sn}\left(\omega u|k\right),\\
c_{1} & =1,\\
c_{2} & =1,\\
d_{1} & =\frac{\delta_{1}\xi_{1}}{\beta_{1}}\mathrm{sn}\left(\omega u|k\right)\left\{ \frac{\xi_{1}^{2}}{\xi_{1}^{2}-k^{2}\mathrm{sn}^{2}\left(\omega u|k\right)}\left[\frac{\alpha_{1}\xi_{1}}{\beta_{1}}\mathrm{sn}\left(\omega u|k\right)+\mathrm{cn}\left(\omega u|k\right)\mathrm{dn}\left(\omega u|k\right)\right]\right\} ,\\
d_{2} & =\frac{\delta_{2}\xi_{1}}{\beta_{1}}\mathrm{sn}\left(\omega u|k\right)\left\{ \frac{\xi_{1}^{2}}{\xi_{1}^{2}-k^{2}\mathrm{sn}^{2}\left(\omega u|k\right)}\left[\frac{\alpha_{1}\xi_{1}}{\beta_{1}}\mathrm{sn}\left(\omega u|k\right)+\mathrm{cn}\left(\omega u|k\right)\mathrm{dn}\left(\omega u|k\right)\right]\right\} ,
\end{align}
\end{subequations} where $\omega=\sqrt{\xi_{2}^{2}\delta_{1}\delta_{2}}=\sqrt{\beta_{1}^{2}/\xi_{1}^{2}}$
and $k=\xi_{1}/\xi_{2}$ is the modulus of the elliptic functions.
Notice that $\omega$, $k$, $\xi_{1}$ and $\xi_{2}$ are functions
of the parameters $\alpha_{1}$, $\beta_{1}$, $\delta_{1}$ and $\delta_{2}$.

This solution (\ref{8VRSymmetric}) can be simplified further if we
introduce a new parameter $\eta$ through the relation $\mathrm{sn}\left(\omega\eta|k\right)=1/\xi_{1}$,
so that $\mathrm{cn}\left(\omega\eta|k\right)\mathrm{dn}\left(\omega\eta|k\right)=\epsilon\alpha_{1}/\beta_{1}$.
Then, from the addition formula of the elliptic sinus (\ref{Identity2})
and using the relations (\ref{8VABC1}), it follows that (\ref{8VRSymmetric})
can be rewritten as 
\begin{equation}
R(u)=\begin{pmatrix}\dfrac{\mathrm{sn}\left(\omega(\epsilon u+\eta)|k\right)}{\mathrm{sn}\left(\omega\eta|k\right)} & 0 & 0 & k\sqrt{\dfrac{\delta_{1}}{\delta_{2}}}\mathrm{sn}\left(\omega u|k\right)\mathrm{sn}\left(\omega(\epsilon u+\eta)|k\right)\\
0 & \dfrac{\mathrm{sn}\left(\omega u|k\right)}{\mathrm{sn}\left(\omega\eta|k\right)} & 1 & 0\\
0 & 1 & \dfrac{\mathrm{sn}\left(\omega u|k\right)}{\mathrm{sn}\left(\omega\eta|k\right)} & 0\\
k\sqrt{\dfrac{\delta_{2}}{\delta_{1}}}\mathrm{sn}\left(\omega u|k\right)\mathrm{sn}\left(\omega(\epsilon u+\eta)|k\right) & 0 & 0 & \dfrac{\mathrm{sn}\left(\omega(\epsilon u+\eta)|k\right)}{\mathrm{sn}\left(\omega\eta|k\right)}
\end{pmatrix}.\label{8VR1}
\end{equation}
This solutions corresponds to a generalization of the eight-vertex
$R$ matrix found originally by Baxter in \cite{Baxter1972,Baxter1978}
and by Zamolodchikov in \cite{Zamolodchikov1979B}. It contains three
bare free-parameters (e.g., $k$, $\omega\eta$ and $\delta_{1}/\delta_{2}$)
and it is equivalent to the solution named 8V(I) in \cite{SogoEtal1982}
and that given by equation (3.7) in \cite{KhachatryanSedrakyan2013}.

\subsubsection{The subcase $b_{2}=-b_{1}$ }

Let us now to consider the possibility $\beta_{2}=-\beta_{1}$ in
(\ref{8VBranch2}), which implies $b_{2}=-b_{1}$. This means that
we are looking for a solution with the following properties: 
\begin{equation}
\frac{a_{2}}{a_{1}}=1,\qquad\frac{b_{2}}{b_{1}}=-1,\qquad\frac{c_{2}}{c_{1}}=1,\qquad\text{and}\qquad d_{2}=\frac{\delta_{2}}{\delta_{1}}d_{1}.\label{8VABC2}
\end{equation}

As in the previous case, we can eliminate $d_{1}'$ from the equation
$E_{1,4}$: 
\begin{equation}
d_{1}'=\delta_{1}\left(a_{1}^{2}-b_{1}^{2}\right).
\end{equation}
 Then, multiplying $E_{2,8}$ by $b_{1}$ and $E_{3,8}$ by $a_{1}$
and taking the difference of them, we are led to the equation 
\begin{equation}
\alpha_{1}a_{1}b_{1}d_{1}=0.
\end{equation}
The only possibility for $a_{1}$, $b_{1}$ and $d_{1}$ different
from zero is $\alpha_{1}=0$, which, of course, \emph{does not} means
that $a_{1}=0$. 

Now, multiplying $E_{1,6}$ by $b_{1}$ and $F_{1,7}$ by $a_{1}$
and taking again the difference, we get the equation 
\begin{equation}
\left(a_{1}^{2}-b_{1}^{2}\right)\left[\delta_{2}d_{1}^{2}-\delta_{1}\left(a_{1}^{2}-b_{1}^{2}-1\right)\right]=0.\label{8VE16F17}
\end{equation}
 Clearly $a_{1}^{2}\neq b_{1}^{2}$ if the solution is regular, hence,
the second factor in the equation above should vanish. Solving (\ref{8VE16F17})
for $d_{1}$ we get, 
\begin{equation}
d_{1}=\epsilon\sqrt{\left(\delta_{1}/\delta_{2}\right)\left(a_{1}^{2}-b_{1}^{2}-1\right)}.
\end{equation}

At this point, all equations of the systems $E$ and $F$ are satisfied.
It remains however to find $a_{1}$ and $b_{1}$. These remained unknowns
can be found through the respective differential equations in (\ref{8Vderivatives}),
which become, 
\begin{equation}
a_{1}'=b_{1}\beta_{1}+\epsilon a_{1}\sqrt{\delta_{1}\delta_{2}\left(a_{1}^{2}-b_{1}^{2}-1\right)},\qquad b_{1}'=a_{1}\beta_{1}+\epsilon b_{1}\sqrt{\delta_{1}\delta_{2}\left(a_{1}^{2}-b_{1}^{2}-1\right)}.
\end{equation}
Here we remark that $a_{1}^{2}-b_{1}^{2}-1=0$ implies $d_{1}=0$,
which would lead us to a particular solution of the usual six-vertex
model. Therefore, let us assume that $a_{1}^{2}-b_{1}^{2}-1\neq0$.
In this case, the system of differential equations above, with the
initial conditions $a_{1}(0)=1$ and $b_{1}(0)=0$, has the solution:
\begin{equation}
a_{1}=\cosh\left(\beta_{1}u\right)\sec\left(u\sqrt{\delta_{1}\delta_{2}}\right),\qquad b_{1}=\epsilon\sinh\left(\beta_{1}u\right)\sec\left(u\sqrt{\delta_{1}\delta_{2}}\right).
\end{equation}

Now, taking the derivative of $b_{1}$ and $b_{2}$ at $u=0$ we get
that $b_{1}'(0)=\epsilon\beta_{1}$ and $b_{2}'(0)=\epsilon\beta_{2}$.
This means that we must set $\epsilon=1$ for consistency with the
differential method\footnote{The case $\epsilon=-1$ leads to the same solution after a redefinition
of $\beta_{1}$ and $\beta_{2}$.}. Therefore, after simplify the previous expressions, we find the
solution we were looking for: 
\begin{equation}
R(u)=\begin{pmatrix}\dfrac{\cosh\left(\beta_{1}u\right)}{\cos\left(u\sqrt{\delta_{1}\delta_{2}}\right)} & 0 & 0 & \sqrt{\dfrac{\delta_{1}}{\delta_{2}}}\tan\left(u\sqrt{\delta_{1}\delta_{2}}\right)\\
0 & \dfrac{\sinh\left(\beta_{1}u\right)}{\cos\left(u\sqrt{\delta_{1}\delta_{2}}\right)} & 1 & 0\\
0 & 1 & -\dfrac{\sinh\left(\beta_{1}u\right)}{\cos\left(u\sqrt{\delta_{1}\delta_{2}}\right)} & 0\\
\sqrt{\dfrac{\delta_{2}}{\delta_{1}}}\tan\left(u\sqrt{\delta_{1}\delta_{2}}\right) & 0 & 0 & \dfrac{\cosh\left(\beta_{1}u\right)}{\cos\left(u\sqrt{\delta_{1}\delta_{2}}\right)}
\end{pmatrix}.\label{8VR2}
\end{equation}
This $R$ matrix depends on three free-parameters ($\beta_{1}$, $\delta_{1}$
and $\delta_{2}$) but one of them can be removed by redefining $u$,
so that the number of bare free-parameters is two (say, $\text{\ensuremath{\beta_{1}/\sqrt{\delta_{1}\delta_{2}}} and \ensuremath{\delta_{1}/\delta_{2}}}$).
It is equivalent to the solution named 8V(III) in \cite{SogoEtal1982}
and that given by equation (3.11) in \cite{KhachatryanSedrakyan2013}.

\subsection{The second case $a_{2}\protect\neq a_{1}$\label{Sec8Vdifferent}}

Let us back now to equation (\ref{8vBranch1}) and assume $\alpha_{2}\neq\alpha_{1}$.
Thus, we are looking for a solution in which 
\begin{equation}
a_{2}=a_{1}+\left(\frac{\alpha_{2}-\alpha_{1}}{\beta_{1}}\right)b_{1},\qquad\frac{b_{2}}{b_{1}}=\frac{\beta_{2}}{\beta_{1}},\qquad\frac{c_{2}}{c_{1}}=1,\qquad\text{and}\qquad\frac{d_{2}}{d_{1}}=\frac{\delta_{2}}{\delta_{1}}.\label{8V3abcd}
\end{equation}
Before solving (\ref{8vBranch1}), let us eliminate $d_{1}'$ from
$E_{1,4}$: 
\begin{equation}
d_{1}'=\left(\alpha_{2}-\alpha_{1}\right)d_{1}+\delta_{1}\left(a_{1}^{2}-b_{1}^{2}\right).
\end{equation}
Then we can solve (\ref{8vBranch1}), say for $a_{1}$, which provides
\begin{equation}
a_{1}=\frac{\beta_{1}d_{1}}{\delta_{1}b_{1}}-\frac{\alpha_{2}-\alpha_{1}}{2\beta_{1}}b_{1}.\label{8V3a1}
\end{equation}
Besides, after $d_{1}'$ is eliminated, we can verify that $F_{1,7}$
reduces to 
\begin{equation}
\delta_{1}\left(\beta_{2}^{2}-\beta_{1}^{2}\right)\left(\frac{b_{1}}{\beta_{1}}\right)^{2}=0.
\end{equation}

It seems therefore that we have two additional branches to consider,
depending on whether $\beta_{2}=\beta_{1}$ or $\beta_{2}=-\beta_{1}$.
The second possibility, however, implies $\alpha_{2}=\alpha_{1}$,
which lead us to the previous case discussed above (in fact, assuming
$\beta_{2}=-\beta_{1}$ we can verify that the difference between
$E_{1,6}$ and $E_{8,3}$ gives the equation $\left(\alpha_{1}-\alpha_{2}\right)b_{1}d_{1}=0$).
Therefore, let us consider the case $\beta_{2}=\beta_{1}$, that is,
\begin{equation}
b_{2}=b_{1}.
\end{equation}

We continue in this way by solving $E_{1,6}$ for $d_{1}$: 
\begin{equation}
d_{1}=\frac{\delta_{1}b_{1}}{\beta_{1}^{2}-\delta_{1}\delta_{2}b_{1}^{2}}\left[\sqrt{\delta_{1}\delta_{2}\left(1-\frac{\alpha_{1}^{2}}{\beta_{1}^{2}}\right)b_{1}^{4}+\left(\alpha_{1}^{2}-\beta_{1}^{2}-\delta_{1}\delta_{2}\right)b_{1}^{2}+\beta_{1}^{2}}+\left(\frac{\alpha_{1}+\alpha_{2}}{2}\right)b_{1}\right],\label{8V3d1}
\end{equation}
where we assumed $\beta_{1}$ positive (see Footnote \ref{FootNoteBranches}).
Now, $E_{2,3}$ reduces to 
\begin{equation}
\left(\frac{\alpha_{2}^{2}-\alpha_{1}^{2}}{2\beta_{1}}\right)b_{1}^{2}=0.
\end{equation}
The first possibility $\alpha_{2}=\alpha_{1}$ lead us again to the
previous considered case in which $a_{2}=a_{1}$. Therefore, let us
consider that 
\begin{equation}
\alpha_{2}=-\alpha_{1}.
\end{equation}

Now all the functional equations are satisfied. It remains, however,
to found $b_{1}$. As in the previous cases, $b_{1}$ can be determined
through the differential equation 
\begin{equation}
b_{1}'=\sqrt{Ab_{1}^{4}+Bb_{1}^{2}+C},
\end{equation}
which is provided by the third equation in (\ref{8Vderivatives}).
In this case, however, 
\begin{equation}
A=\delta_{1}\delta_{2}\left(1-\frac{\alpha_{1}^{2}}{\beta_{1}^{2}}\right),\qquad B=\alpha_{1}^{2}-\beta_{1}^{2}-\delta_{1}\delta_{2},\qquad C=\beta_{1}^{2}.\label{8VABC3}
\end{equation}
Therefore, the desired solution of the differential equation above
satisfying the initial value $b_{1}(0)=0$ is, 
\begin{equation}
b_{1}=\xi_{1}\,\mathrm{sn}\left(u\sqrt{\xi_{2}^{2}A}\left|\frac{\xi_{1}}{\xi_{2}}\right.\right)=\xi_{1}\,\mathrm{sn}\left(\left.u\sqrt{\xi_{2}^{2}\delta_{1}\delta_{2}\left(1-\frac{\alpha_{1}^{2}}{\beta_{1}^{2}}\right)}\right|\frac{\xi_{1}}{\xi_{2}}\right),
\end{equation}
where $\xi_{1}^{2}$ and $\xi_{2}^{2}$ are the roots of the quadratic
equation $Ax^{2}+Bx+C=0$ with $A$, $B$ and $C$ given by (\ref{8VABC3})
and we have assumed $\xi_{1}^{2}\neq\xi_{2}^{2}$ (the case $\xi_{1}^{2}=\xi_{2}^{2}$
will be discussed in Appendix \ref{AppendixReduced}). 

Therefore, from the equations (\ref{8V3abcd}), (\ref{8V3a1}) and
(\ref{8V3d1}), we can write down all the elements of the $R$ matrix,
which are: \begin{subequations}\label{8VRNotSymmetric}
\begin{align}
a_{1} & =\frac{\beta_{1}^{2}\mathrm{cn}\left(\omega u|k\right)\mathrm{dn}\left(\omega u|k\right)}{\beta_{1}^{2}-\delta_{1}\delta_{2}\xi_{1}^{2}\mathrm{sn}^{2}\left(\omega u|k\right)}+\frac{\alpha_{1}\xi_{1}}{\beta_{1}}\mathrm{sn}\left(\omega u|k\right), & a_{2} & =\frac{\beta_{1}^{2}\mathrm{cn}\left(\omega u|k\right)\mathrm{dn}\left(\omega u|k\right)}{\beta_{1}^{2}-\delta_{1}\delta_{2}\xi_{1}^{2}\mathrm{sn}^{2}\left(\omega u|k\right)}-\frac{\alpha_{1}\xi_{1}}{\beta_{1}}\mathrm{sn}\left(\omega u|k\right),\\
b_{1} & =\xi_{1}\mathrm{sn}\left(\omega u|k\right), & b_{2} & =\xi_{1}\mathrm{sn}\left(\omega u|k\right),\\
c_{1} & =1, & c_{2} & =1,\\
d_{1} & =\frac{\delta_{1}\beta_{1}\xi_{1}\mathrm{sn}\left(\omega u|k\right)\mathrm{cn}\left(\omega u|k\right)\mathrm{dn}\left(\omega u|k\right)}{\beta_{1}^{2}-\delta_{1}\delta_{2}\xi_{1}^{2}\mathrm{sn}^{2}\left(\omega u|k\right)}, & d_{2} & =\frac{\delta_{2}\beta_{1}\xi_{1}\mathrm{sn}\left(\omega u|k\right)\mathrm{cn}\left(\omega u|k\right)\mathrm{dn}\left(\omega u|k\right)}{\beta_{1}^{2}-\delta_{1}\delta_{2}\xi_{1}^{2}\mathrm{sn}^{2}\left(\omega u|k\right)},
\end{align}
\end{subequations} where $k=\xi_{1}/\xi_{2}$, $\omega=\sqrt{\xi_{2}^{2}\delta_{1}\delta_{2}\left(1-\alpha_{1}^{2}/\beta_{1}^{2}\right)}$
and we have used the identities (\ref{Identity1}) to simplify the
square root in (\ref{8V3d1}). Notice that $\omega$, $k$, $\xi_{1}$
and $\xi_{2}$ are functions of the free-parameters $\alpha_{1}$,
$\beta_{1}$, $\delta_{1}$ and $\delta_{2}$.

This solution can still be simplified by introducing the parameter
$\eta$ through the relation $\text{sn}(\omega\eta|k)=1/\xi_{1}$
and noticing that, remarkably, the expressions for $\xi_{1}^{2}$
and $\xi_{2}^{2}$ are quite simple in this case. In fact, we can
set either 
\begin{equation}
\xi_{1}^{2}=\frac{\beta_{1}^{2}}{\beta_{1}^{2}-\alpha_{1}^{2}},\qquad\text{and}\qquad\xi_{2}^{2}=\frac{\beta_{1}^{2}}{\delta_{1}\delta_{2}},\label{xi1xi2}
\end{equation}
 or, conversely, 
\begin{equation}
\xi_{1}^{2}=\frac{\beta_{1}^{2}}{\delta_{1}\delta_{2}},\qquad\text{and}\qquad\xi_{2}^{2}=\frac{\beta_{1}^{2}}{\beta_{1}^{2}-\alpha_{1}^{2}}.\label{xi2xi1}
\end{equation}

The first case implies the identity $\text{cn}\left(\omega\eta|k\right)=\epsilon\alpha_{1}/\beta_{1}$
where $\epsilon=\text{cosign}\left(\alpha_{1}/\beta_{1}\right)$,
which leads to the following $R$ matrix:
\begin{equation}
R(u)=\begin{pmatrix}\dfrac{\mathrm{cn}\left(\omega u|k\right)}{\mathrm{dn}\left(\omega u|k\right)}+\epsilon\dfrac{\mathrm{sn}\left(\omega u|k\right)\mathrm{cn}\left(\omega\eta|k\right)}{\mathrm{sn}\left(\omega\eta|k\right)} & 0 & 0 & \dfrac{\delta_{1}}{\beta_{1}}\dfrac{\mathrm{cn}\left(\omega u|k\right)}{\mathrm{dn}\left(\omega u|k\right)}\dfrac{\mathrm{sn}\left(\omega u|k\right)}{\mathrm{sn}\left(\omega\eta|k\right)}\\
0 & \dfrac{\mathrm{sn}\left(\omega u|k\right)}{\mathrm{sn}\left(\omega\eta|k\right)} & 1 & 0\\
0 & 1 & \dfrac{\mathrm{sn}\left(\omega u|k\right)}{\mathrm{sn}\left(\omega\eta|k\right)} & 0\\
\dfrac{\delta_{2}}{\beta_{1}}\dfrac{\mathrm{cn}\left(\omega u|k\right)}{\mathrm{dn}\left(\omega u|k\right)}\dfrac{\mathrm{sn}\left(\omega u|k\right)}{\mathrm{sn}\left(\omega\eta|k\right)} & 0 & 0 & \dfrac{\mathrm{cn}\left(\omega u|k\right)}{\mathrm{dn}\left(\omega u|k\right)}-\epsilon\dfrac{\mathrm{sn}\left(\omega u|k\right)\mathrm{cn}\left(\omega\eta|k\right)}{\mathrm{sn}\left(\omega\eta|k\right)}
\end{pmatrix},\label{8VR3A}
\end{equation}
where $\omega=\sqrt{\beta_{1}^{2}-\alpha_{1}^{2}}$ and $k=\sqrt{\left(\delta_{1}\delta_{2}\right)/\left(\beta_{1}^{2}-\alpha_{1}^{2}\right)}$. 

The second case implies $\text{dn}\left(\omega\eta|k\right)=\epsilon\alpha_{1}/\beta_{1}$
where $\epsilon=\text{cosign}\left(\alpha_{1}/\beta_{1}\right)$,
from which we get the solution
\begin{equation}
R(u)=\begin{pmatrix}\dfrac{\mathrm{dn}\left(\omega u|k\right)}{\mathrm{cn}\left(\omega u|k\right)}+\epsilon\dfrac{\mathrm{sn}\left(\omega u|k\right)\mathrm{dn}\left(\omega\eta|k\right)}{\mathrm{sn}\left(\omega\eta|k\right)} & 0 & 0 & \dfrac{\delta_{1}}{\beta_{1}}\dfrac{\mathrm{dn}\left(\omega u|k\right)}{\mathrm{cn}\left(\omega u|k\right)}\dfrac{\mathrm{sn}\left(\omega u|k\right)}{\mathrm{sn}\left(\omega\eta|k\right)}\\
0 & \dfrac{\mathrm{sn}\left(\omega u|k\right)}{\mathrm{sn}\left(\omega\eta|k\right)} & 1 & 0\\
0 & 1 & \dfrac{\mathrm{sn}\left(\omega u|k\right)}{\mathrm{sn}\left(\omega\eta|k\right)} & 0\\
\dfrac{\delta_{2}}{\beta_{1}}\dfrac{\mathrm{dn}\left(\omega u|k\right)}{\mathrm{cn}\left(\omega u|k\right)}\dfrac{\mathrm{sn}\left(\omega u|k\right)}{\mathrm{sn}\left(\omega\eta|k\right)} & 0 & 0 & \dfrac{\mathrm{dn}\left(\omega u|k\right)}{\mathrm{cn}\left(\omega u|k\right)}-\epsilon\dfrac{\mathrm{sn}\left(\omega u|k\right)\mathrm{dn}\left(\omega\eta|k\right)}{\mathrm{sn}\left(\omega\eta|k\right)}
\end{pmatrix},\label{8VR3B}
\end{equation}
where, now, $\omega=\sqrt{\delta_{1}\delta_{2}}$ and $k=\sqrt{\left(\beta_{1}^{2}-\alpha_{1}^{2}\right)/\left(\delta_{1}\delta_{2}\right)}$. 

To count the number of bare free-parameters of the solutions (\ref{8VR3A})
and (\ref{8VR3B}), we should use the relations (\ref{8VABC3}) and
(\ref{xi1xi2}) or (\ref{xi2xi1}). In this way, we can verify that
$\beta_{1}$ can be removed and the solution will depend only on $\delta_{1}/\delta_{2}$,
$k$ and $\omega\eta$, so that these solutions are characterized
by three bare free-parameters. The $R$ matrices (\ref{8VR3A}) and
(\ref{8VR3B}) are related to the solutions originally found by Felderhof
in \cite{Felderhof1973} and by Bazhanov \& Stroganov in \cite{BazhanovStroganov1985}
and they are also equivalent to the solutions reported in \cite{SogoEtal1982}
(solution named 8V(II) in Table I) and in \cite{KhachatryanSedrakyan2013}
(eqs. (3.23) and (3.19), respectively). We remark that the $R$ matrices
(\ref{8VR3A}) and (\ref{8VR3B}) are related to each other by the
inversion of the modulus $k$ -- see Footnote \ref{FootnoteInversion}.

\section{Solutions for the five-vertex models }

In the previous sections, we obtained the general regular solutions
of the \noun{ybe} for the case where the $R$ matrix had an even number
of non-null elements. There exist, however, regular $R$ matrices
with an odd number of non-null elements, which correspond to the five
and seven-vertex models. In this section we shall concern with the
solutions for the five-vertex models. The solutions for the seven-vertex
models will be treated in the next section.

Five-vertex models can be obtained in four different ways. Two types
of five-vertex models can be obtained by zeroing one of the elements
$b_{1}$ or $b_{2}$ in the six-vertex $R$ matrix (\ref{6VR}): 
\begin{equation}
R=\begin{pmatrix}a_{1} & 0 & 0 & 0\\
0 & 0 & c_{1} & 0\\
0 & c_{2} & b_{2} & 0\\
0 & 0 & 0 & a_{2}
\end{pmatrix},\qquad R=\begin{pmatrix}a_{1} & 0 & 0 & 0\\
0 & b_{1} & c_{1} & 0\\
0 & c_{2} & 0 & 0\\
0 & 0 & 0 & a_{2}
\end{pmatrix}.\label{5VR}
\end{equation}
These five-vertex models are related to the so-called \emph{Totally
Asymmetric Simple Exclusion Process} (\noun{tasep}) -- see \cite{MacdonaldEtal1968,MotegiSakai2013}.

Other two types of five-vertex models can be obtained as well by zeroing
the elements $d_{1}$ or $d_{2}$ in the unusual six-vertex $R$ matrix
(\ref{RU}): 
\begin{equation}
R=\begin{pmatrix}a_{1} & 0 & 0 & 0\\
0 & 0 & c_{1} & 0\\
0 & c_{2} & 0 & 0\\
d_{2} & 0 & 0 & a_{2}
\end{pmatrix},\qquad R=\begin{pmatrix}a_{1} & 0 & 0 & d_{1}\\
0 & 0 & c_{1} & 0\\
0 & c_{2} & 0 & 0\\
0 & 0 & 0 & a_{2}
\end{pmatrix}.\label{5VUR}
\end{equation}
Because the main steps to solve the \noun{ybe} for the five-vertex
models are similar to the previous cases, in the what follows we shall
only comment on the possible branches and present the final results.

\subsection{The first case $b_{1}=0$ or $b_{2}=0$}

This is the case corresponding to the $R$ matrices given in (\ref{5VR}).
The solutions branch into two classes regarding on whether $\alpha_{2}=\alpha_{1}$
or $\gamma_{1}+\gamma_{2}=\alpha_{1}+\alpha_{2}$. In the first case
where $\alpha_{2}=\alpha_{1}$, we obtain the solutions, 
\begin{equation}
R(u)=\begin{pmatrix}\mathrm{e}^{\alpha_{1}u} & 0 & 0 & 0\\
0 & 0 & \mathrm{e}^{\gamma_{1}u} & 0\\
0 & \mathrm{e}^{u\gamma_{2}} & \dfrac{\mathrm{e}^{\alpha_{1}u}-\mathrm{e}^{\left(\gamma_{1}+\gamma_{2}-\alpha_{1}\right)u}}{2\alpha_{1}-\gamma_{1}-\gamma_{2}}\beta_{2} & 0\\
0 & 0 & 0 & \mathrm{e}^{\alpha_{1}u}
\end{pmatrix},\qquad R(u)=\begin{pmatrix}\mathrm{e}^{u\alpha_{1}} & 0 & 0 & 0\\
0 & \dfrac{\mathrm{e}^{\alpha_{1}u}-\mathrm{e}^{\left(\gamma_{1}+\gamma_{2}-\alpha_{1}\right)u}}{2\alpha_{1}-\gamma_{1}-\gamma_{2}}\beta_{1} & \mathrm{e}^{u\gamma_{1}} & 0\\
0 & \mathrm{e}^{u\gamma_{2}} & 0 & 0\\
0 & 0 & 0 & \mathrm{e}^{u\alpha_{1}}
\end{pmatrix}.\label{5V12}
\end{equation}
If we set $\gamma_{2}=\gamma_{1}$ in the solutions above and make
the change of variable $u\rightarrow\log\left(u\right)/(\alpha_{1}-\gamma_{1})$
then we shall obtain the same $R$ matrices presented by Motegi \&
Sakai in \cite{MotegiSakai2013}, which are related to the \noun{tasep}
models.

For the second case where $\gamma_{2}=\alpha_{1}+\alpha_{2}-\gamma_{1}$,
we get the solutions, 

\begin{equation}
R(u)=\begin{pmatrix}\mathrm{e}^{\alpha_{1}u} & 0 & 0 & 0\\
0 & 0 & \mathrm{e}^{\gamma_{1}u} & 0\\
0 & \mathrm{e}^{\left(\alpha_{1}+\alpha_{2}-\gamma_{1}\right)u} & \dfrac{\mathrm{e}^{\alpha_{1}u}-\mathrm{e}^{\alpha_{2}u}}{\alpha_{1}-\alpha_{2}}\beta_{2} & 0\\
0 & 0 & 0 & \mathrm{e}^{\alpha_{2}u}
\end{pmatrix},\qquad R(u)=\begin{pmatrix}\mathrm{e}^{\alpha_{1}u} & 0 & 0 & 0\\
0 & \dfrac{\mathrm{e}^{\alpha_{1}u}-\mathrm{e}^{\alpha_{2}u}}{\alpha_{1}-\alpha_{2}}\beta_{1} & \mathrm{e}^{\gamma_{1}u} & 0\\
0 & \mathrm{e}^{\left(\alpha_{1}+\alpha_{2}-\gamma_{1}\right)u} & 0 & 0\\
0 & 0 & 0 & \mathrm{e}^{\alpha_{2}u}
\end{pmatrix}.\label{5V34}
\end{equation}
These solutions present four free-parameters ($\alpha_{1}$, $\gamma_{1}$,
$\gamma_{2}$ and $\beta_{1}$ or $\beta_{1}$); however, one of them
can be removed by renormalization and another one by redefining the
spectral parameter. This means that the solutions above contain two
bare free-parameters only. Finally, we remark that these $R$ matrices
can also be obtained from the six-vertex $R$ matrices (\ref{6VR1})
and (\ref{6VR2}), respectively, by taking the limits $\beta_{1}\rightarrow0$
or $\beta_{2}\rightarrow0$ (it is necessary to eliminate $\eta$
before taking the limit to avoid discontinuities). 

\subsection{The second case $d_{1}=0$ or $d_{2}=0$}

This case corresponds to the $R$ matrices given by (\ref{5VUR}).
Here the solution branches into three classes regarding on whether
we have: $\alpha_{2}=\alpha_{1}$ and $\gamma_{2}\neq\gamma_{1}$,
$\alpha_{2}\neq\alpha_{1}$ and $\gamma_{2}=\gamma_{1}$ or, finally,
$\alpha_{2}=\alpha_{1}$ and $\gamma_{2}=\gamma_{1}$. 

The first branch, we have $\alpha_{2}=\alpha_{1}=\tfrac{1}{2}(\gamma_{1}+\gamma_{2})$
and $\gamma_{2}\neq\gamma_{1}$. The solutions are:
\begin{equation}
R(u)=\left(\begin{array}{cccc}
\mathrm{e}^{\frac{1}{2}(\gamma_{1}+\gamma_{2})u} & 0 & 0 & 0\\
0 & 0 & \mathrm{e}^{u\gamma_{1}} & 0\\
0 & \mathrm{e}^{u\gamma_{2}} & 0 & 0\\
\delta_{2}\dfrac{\mathrm{e}^{\gamma_{1}u}-\mathrm{e}^{\gamma_{2}u}}{\gamma_{1}-\gamma_{2}} & 0 & 0 & \mathrm{e}^{\frac{1}{2}(\gamma_{1}+\gamma_{2})u}
\end{array}\right),\qquad R(u)=\left(\begin{array}{cccc}
\mathrm{e}^{\frac{1}{2}(\gamma_{1}+\gamma_{2})u} & 0 & 0 & \delta_{1}\dfrac{\mathrm{e}^{\gamma_{1}u}-\mathrm{e}^{\gamma_{2}u}}{\gamma_{1}-\gamma_{2}}\\
0 & 0 & \mathrm{e}^{u\gamma_{1}} & 0\\
0 & \mathrm{e}^{u\gamma_{2}} & 0 & 0\\
0 & 0 & 0 & \mathrm{e}^{\frac{1}{2}(\gamma_{1}+\gamma_{2})u}
\end{array}\right).\label{5U12}
\end{equation}

For the second branch, we have $\alpha_{2}\neq\alpha_{1}$ and $\gamma_{2}=\gamma_{1}=\tfrac{1}{2}(\alpha_{1}+\alpha_{2})$,
from which we get the solutions
\begin{equation}
R(u)=\left(\begin{array}{cccc}
\mathrm{e}^{\alpha_{1}u} & 0 & 0 & 0\\
0 & 0 & \mathrm{e}^{\frac{1}{2}(\alpha_{1}+\alpha_{2})u} & 0\\
0 & \mathrm{e}^{\frac{1}{2}(\alpha_{1}+\alpha_{2})u} & 0 & 0\\
\delta_{2}\left(\dfrac{\mathrm{e}^{\frac{1}{2}(\alpha_{1}+\alpha_{2})u}}{\alpha_{1}-\alpha_{2}}\right)\sinh\left[\left(\alpha_{1}-\alpha_{2}\right)u\right] & 0 & 0 & \mathrm{e}^{\alpha_{2}u}
\end{array}\right),\label{5U3}
\end{equation}
and 
\begin{equation}
R(u)=\left(\begin{array}{cccc}
\mathrm{e}^{\alpha_{1}u} & 0 & 0 & \delta_{1}\left(\dfrac{\mathrm{e}^{\frac{1}{2}(\alpha_{1}+\alpha_{2})u}}{\alpha_{1}-\alpha_{2}}\right)\sinh\left[\left(\alpha_{1}-\alpha_{2}\right)u\right]\\
0 & 0 & \mathrm{e}^{\frac{1}{2}(\alpha_{1}+\alpha_{2})u} & 0\\
0 & \mathrm{e}^{\frac{1}{2}(\alpha_{1}+\alpha_{2})u} & 0 & 0\\
0 & 0 & 0 & \mathrm{e}^{\alpha_{2}u}
\end{array}\right).\label{5U4}
\end{equation}

Finally, for the third branch we have $\alpha_{2}=\alpha_{1}$ and
$\gamma_{2}=\gamma_{1}$. In this case, the solutions are:
\begin{equation}
R(u)=\left(\begin{array}{cccc}
\mathrm{e}^{\alpha_{1}u} & 0 & 0 & 0\\
0 & 0 & \mathrm{e}^{\gamma_{1}u} & 0\\
0 & \mathrm{e}^{\gamma_{1}u} & 0 & 0\\
\delta_{2}\mathrm{e}^{u\alpha_{1}}\frac{\sinh\left[\left(\alpha_{1}-\gamma_{1}\right)u\right]}{\alpha_{1}-\gamma_{1}} & 0 & 0 & \mathrm{e}^{\alpha_{1}u}
\end{array}\right),\qquad R(u)=\left(\begin{array}{cccc}
\mathrm{e}^{\alpha_{1}u} & 0 & 0 & \delta_{1}\mathrm{e}^{u\alpha_{1}}\frac{\sinh\left[\left(\alpha_{1}-\gamma_{1}\right)u\right]}{\alpha_{1}-\gamma_{1}}\\
0 & 0 & \mathrm{e}^{\gamma_{1}u} & 0\\
0 & \mathrm{e}^{\gamma_{1}u} & 0 & 0\\
0 & 0 & 0 & \mathrm{e}^{\alpha_{1}u}
\end{array}\right).\label{5U56}
\end{equation}

These solutions are characterized by two bare free-parameters (because
one of the parameters can be removed by renormalization). 

\section{Solutions for the seven-vertex models }

Finally, let us consider the case in which the $R$ matrix has seven
non-null elements. In principle, a seven-vertex model can be obtained
by zeroing one of the elements $b_{1}$, $b_{2}$, $d_{1}$ or $d_{2}$
in the eight-vertex $R$ matrix (\ref{8VR}). This would provide four
possible initial shapes for the seven-vertex $R$ matrices. It happens,
however, that $\beta_{1}=0$ implies $\beta_{2}=0$ and vice-versa
whenever $d_{1}$ and $d_{2}$ are both different from zero, which
is a consequence of the constraint $\beta_{2}^{2}=\beta_{1}^{2}$
remaining from the eight-vertex model case. Therefore, we must regard
both $b_{1}$ and $b_{2}$ different from zero here, which means that
there are only two possibilities for the initial shapes of the seven-vertex
$R$ matrices, namely,

\begin{equation}
R=\begin{pmatrix}a_{1} & 0 & 0 & 0\\
0 & b_{1} & c_{1} & 0\\
0 & c_{2} & b_{2} & 0\\
d_{2} & 0 & 0 & a_{2}
\end{pmatrix},\qquad\text{and}\qquad R=\begin{pmatrix}a_{1} & 0 & 0 & d_{1}\\
0 & b_{1} & c_{1} & 0\\
0 & c_{2} & b_{2} & 0\\
0 & 0 & 0 & a_{2}
\end{pmatrix}.\label{7VR}
\end{equation}

The \noun{ybe} for these seven-vertex models can be solved in the
same fashion as the usual six-vertex models. In fact, after the elements
$a_{1}$, $a_{2}$, $b_{1}$, $b_{2}$, $c_{1}$ and $c_{2}$ are
found by the same equations as before, other simple equations fix
$d_{1}$ or $d_{2}$ and their derivatives, while the remaining equations
provide some constraints and determine the branches of the solutions.
Relation (\ref{Branch}) still holds true here, which means that we
have two main branches depending on whether we set $\alpha_{2}=\alpha_{1}$
or $\alpha_{1}+\alpha_{2}=\gamma_{1}+\gamma_{2}$. In the first case,
we get a solution in which $a_{2}=a_{1}$; in the second case, we
have $a_{2}\neq a_{1}$. We shall present these solutions in the what
follows.

\subsection{The first case $a_{2}=a_{1}$}

In this case, we have $\alpha_{2}=\alpha_{1}$, which implies $a_{2}=a_{1}$.
The solution can be found by following the same steps presented Section
\ref{Sec6VA}. Some other simple equations determine $d_{1}$ (or
$d_{2}$). However, differently from what happens in the six-vertex
case, some of the remaining equations implies the constraint $\beta_{2}^{2}=\beta_{1}^{2}$,
which means that we have two subcases to work on. 

\subsubsection{The subcase $b_{2}=b_{1}$}

In the case where $\alpha_{2}=\alpha_{1}$ and $\beta_{2}=\beta_{1}$,
other equations also imply the constraint $\gamma_{2}=\gamma_{1}$,
which means that we can set $c_{2}=c_{1}=1$ without loss. In this
case we are led to the following solutions:
\begin{equation}
R(u)=\left(\begin{array}{cccc}
\dfrac{\sinh\left[\omega(\eta+u)\right]}{\sinh\left(\omega\eta\right)} & 0 & 0 & 0\\
0 & \epsilon\dfrac{\sinh\left(\omega u\right)}{\sinh\left(\omega\eta\right)} & 1 & 0\\
0 & 1 & \epsilon\dfrac{\sinh\left(\omega u\right)}{\sinh\left(\omega\eta\right)} & 0\\
\epsilon\dfrac{\delta_{2}}{\beta_{1}}\dfrac{\sinh\left[\omega(\eta+u)\right]\sinh\left(\omega u\right)}{\sinh^{2}\left(\omega\eta\right)} & 0 & 0 & \dfrac{\sinh\left[\omega(\eta+u)\right]}{\sinh\left(\omega\eta\right)}
\end{array}\right),\label{7V1}
\end{equation}
and 
\begin{equation}
R(u)=\left(\begin{array}{cccc}
\dfrac{\sinh\left[\omega(\eta+u)\right]}{\sinh\left(\omega\eta\right)} & 0 & 0 & \epsilon\dfrac{\delta_{1}}{\beta_{1}}\dfrac{\sinh\left[\omega(u+\eta)\right]\sinh\left(\omega u\right)}{\sinh^{2}\left(\omega\eta\right)}\\
0 & \epsilon\dfrac{\sinh\left(\omega u\right)}{\sinh\left(\omega\eta\right)} & 1 & 0\\
0 & 1 & \epsilon\dfrac{\sinh\left(\omega u\right)}{\sinh\left(\omega\eta\right)} & 0\\
0 & 0 & 0 & \dfrac{\sinh\left[\omega(\eta+u)\right]}{\sinh\left(\omega\eta\right)}
\end{array}\right),\label{7V2}
\end{equation}
where $\omega=\sqrt{\alpha_{1}^{2}-\beta_{1}^{2}}$ and $\coth\left(\omega\eta\right)=\alpha_{1}/\omega$,
so that $\sinh\left(\omega\eta\right)=\epsilon\omega/\beta_{1}$ with
$\epsilon=\text{sign}(\alpha_{1})$. We remark that these $R$ matrices
are can be obtained from the limits $\delta_{1}\rightarrow0$ or $\delta_{2}\rightarrow0$,
respectively, of the eight-vertex $R$ matrix (\ref{8VR1}) -- see
Appendix \ref{AppendixReduced}. The $R$ matrices (\ref{7V1}) and
(\ref{7V2}) depend on three parameters ($\alpha_{1}$, $\beta_{1}$
and $\delta_{1}$ or $\delta_{2}$), however, after we redefine the
spectral parameter $u$ and also $\eta$, we realize that the solutions
will present only two bare free-parameters (e.g., $\omega\eta$ and
$\delta_{1}/\beta_{1}$ or $\delta_{2}/\beta_{1}$). These $R$ matrices
are generalizations of the solution originally found by Cherednik
in \cite{Cherednik1980} and they are equivalent to those reported
as solution 7V(I) in \cite{SogoEtal1982} and that given by equation
(5.33) in \cite{KhachatryanSedrakyan2013}.

\subsubsection{The subcase $b_{2}=-b_{1}$}

Now, let us consider the case $\alpha_{2}=\alpha_{1}$ and $\beta_{2}=-\beta_{1}$.
Here $\gamma_{1}$ and $\gamma_{2}$ remain arbitrary and we are led
to the following solutions:
\begin{equation}
R(u)=\left(\begin{array}{cccc}
\mathrm{e}^{\frac{1}{2}\left(\gamma_{1}+\gamma_{2}\right)u}\cosh\left(\beta_{1}u\right) & 0 & 0 & 0\\
0 & \mathrm{e}^{\frac{1}{2}\left(\gamma_{1}+\gamma_{2}\right)u}\sinh\left(\beta_{1}u\right) & \mathrm{e}^{\gamma_{1}u} & 0\\
0 & \mathrm{e}^{\gamma_{2}u} & -\mathrm{e}^{\frac{1}{2}\left(\gamma_{1}+\gamma_{2}\right)u}\sinh\left(\beta_{1}u\right) & 0\\
\dfrac{\mathrm{e}^{\gamma_{1}u}-\mathrm{e}^{\gamma_{2}u}}{\gamma_{1}-\gamma_{2}}\delta_{2} & 0 & 0 & \mathrm{e}^{\frac{1}{2}\left(\gamma_{1}+\gamma_{2}\right)u}\cosh\left(\beta_{1}u\right)
\end{array}\right),\label{7V3}
\end{equation}
and
\begin{equation}
R(u)=\left(\begin{array}{cccc}
\mathrm{e}^{\frac{1}{2}\left(\gamma_{1}+\gamma_{2}\right)u}\cosh\left(\beta_{1}u\right) & 0 & 0 & \dfrac{\mathrm{e}^{\gamma_{1}u}-\mathrm{e}^{\gamma_{2}u}}{\gamma_{1}-\gamma_{2}}\delta_{1}\\
0 & \mathrm{e}^{\frac{1}{2}\left(\gamma_{1}+\gamma_{2}\right)u}\sinh\left(\beta_{1}u\right) & \mathrm{e}^{\gamma_{1}u} & 0\\
0 & \mathrm{e}^{u\gamma_{2}} & -\mathrm{e}^{\frac{1}{2}\left(\gamma_{1}+\gamma_{2}\right)u}\sinh\left(\beta_{1}u\right) & 0\\
0 & 0 & 0 & \mathrm{e}^{\frac{1}{2}\left(\gamma_{1}+\gamma_{2}\right)u}\cosh\left(\beta_{1}u\right)
\end{array}\right).\label{7V4}
\end{equation}
The $R$ matrices (\ref{7V3}) and (\ref{7V4}) depend on four free-parameters,
namely, $\beta_{1}$, $\gamma_{1}$, $\gamma_{2}$ and $\delta_{1}$
or $\delta_{2}$. However, $\beta_{1}$ can be removed by redefining
the spectral parameter and, if we divide everything by $\mathrm{e}^{\gamma_{1}u}$,
then we can set $\gamma=$$\left(\gamma_{2}-\gamma_{1}\right)/\beta_{1}$
as a bare free-parameter, so that the solutions have actually only
two bare free-parameters. These solutions are, therefore, equivalent
to the seven-vertex $R$ matrices presented as solution 7V(III) in
\cite{SogoEtal1982} and that given by equation (5.36) in \cite{KhachatryanSedrakyan2013}.
We remark as well that the limits $\beta_{1}\rightarrow0$ or $\beta_{2}\rightarrow0$
of (\ref{7V3}) and (\ref{7V4}) reproduces the five-vertex $R$ matrices
given in (\ref{5U12}).

\subsection{The second case $a_{2}\protect\neq a_{1}$}

Now, let us consider the case in which $\alpha_{1}+\alpha_{2}=\gamma_{1}+\gamma_{2}$.
The solution can be found by following the same steps presented in
Section \ref{Sec6VB}, among with other simple equations that determine
$d_{1}$ (or $d_{2}$). After all elements are eliminated from the
systems of equations, we shall come across with following constraint:
\begin{equation}
\left(\beta_{2}-\beta_{1}\right)\left(\alpha_{1}-\alpha_{2}+\beta_{1}+\beta_{2}\right)\left(\alpha_{1}-\alpha_{2}-\beta_{1}-\beta_{2}\right)=0.
\end{equation}
 We have therefore three cases to consider. In the first case we get
a solution in which $b_{2}=b_{1}$; in the other two cases we get
solutions in which $b_{2}\neq b_{1}$ (we kindly thank the anonymous
referees of this paper for drawing our attention for this possibility).

\subsubsection{The subcase $b_{2}=b_{1}$}

In the first case where $\beta_{2}=\beta_{1}$ we get a solution with
$b_{2}=b_{1}$. Here as well we have $\gamma_{2}=\gamma_{1}$, so
that we can set $c_{1}=c_{2}=1$, which also implies that $\alpha_{2}=-\alpha_{1}$.
In this way we are led to the following solutions: 
\begin{equation}
R(u)=\left(\begin{array}{cccc}
\dfrac{\sinh\left[\omega(\eta+u)\right]}{\sinh\left(\omega\eta\right)} & 0 & 0 & 0\\
0 & \epsilon\dfrac{\sinh\left(\omega u\right)}{\sinh\left(\omega\eta\right)} & 1 & 0\\
0 & 1 & \epsilon\dfrac{\sinh\left(\omega u\right)}{\sinh\left(\omega\eta\right)} & 0\\
\epsilon\dfrac{1}{2}\dfrac{\delta_{2}}{\beta_{1}}\dfrac{\sinh\left(2\omega u\right)}{\sinh\left(\omega\eta\right)} & 0 & 0 & \dfrac{\sinh\left[\omega(\eta-u)\right]}{\sinh\left(\omega\eta\right)}
\end{array}\right),\label{7V5}
\end{equation}
and 
\begin{equation}
R(u)=\left(\begin{array}{cccc}
\dfrac{\sinh\left[\omega(\eta+u)\right]}{\sinh\left(\omega\eta\right)} & 0 & 0 & \epsilon\dfrac{1}{2}\dfrac{\delta_{1}}{\beta_{1}}\dfrac{\sinh\left(2\omega u\right)}{\sinh\left(\omega\eta\right)}\\
0 & \epsilon\dfrac{\sinh\left(\omega u\right)}{\sinh\left(\omega\eta\right)} & 1 & 0\\
0 & 1 & \epsilon\dfrac{\sinh\left(\omega u\right)}{\sinh\left(\omega\eta\right)} & 0\\
0 & 0 & 0 & \dfrac{\sinh\left[\omega(\eta-u)\right]}{\sinh\left(\omega\eta\right)}
\end{array}\right),\label{7V6}
\end{equation}
where $\omega=\sqrt{\alpha_{1}^{2}-\beta_{1}^{2}}$ and $\coth\left(\omega\eta\right)=\alpha_{1}/\omega$
so that $\sinh\left(\omega\eta\right)=\epsilon\omega/\beta_{1}$ with
$\epsilon=\text{sign}(\alpha_{1})$. We remark that the $R$ matrices
above can also be obtained from the limits $\delta_{1}\rightarrow0$
or $\delta_{2}\rightarrow0$, respectively, of the eight-vertex $R$
matrix (\ref{8VR3A}) -- see Appendix \ref{AppendixReduced}. The
$R$ matrices (\ref{7V5}) and (\ref{7V6}) contain two bare free-parameters
(e.g. $\omega\eta$ and $\delta_{1}/\beta_{1}$or $\delta_{2}/\beta_{1}$)
and they are equivalent to the seven-vertex $R$ matrices reported
in \cite{SogoEtal1982} as the solution 7V(II) and those given by
equation (5.38) in \cite{KhachatryanSedrakyan2013}.

\subsubsection{The subcase $b_{2}\protect\neq b_{1}$}

Finally, the two remaining cases can be handled at once by setting
$\beta_{2}=\epsilon\left(\alpha_{2}-\alpha_{1}\right)-\beta_{1}$.
This leads to solutions in which $b_{2}\neq b_{1}$. Here as well,
we have $\alpha_{2}=-\alpha_{1}$, however, contrary to the previous
case, we have now $\gamma_{2}=-\gamma_{1}$. This leads to the following
solutions: 
\begin{equation}
R(u)=\left(\begin{array}{cccc}
\dfrac{\sinh\left[\omega(\eta+u)\right]}{\sinh\left(\omega\eta\right)} & 0 & 0 & 0\\
0 & -\epsilon\,\mathrm{e}^{-\omega\eta}\dfrac{\sinh\left(\omega u\right)}{\sinh\left(\omega\eta\right)} & \mathrm{e}^{\omega u} & 0\\
0 & \mathrm{e}^{-\omega u} & -\epsilon\,\mathrm{e}^{\omega\eta}\dfrac{\sinh\left(\omega u\right)}{\sinh\left(\omega\eta\right)} & 0\\
\dfrac{\delta_{2}}{\alpha_{1}}\dfrac{\sinh\left(\omega u\right)\cosh\left(\omega\eta\right)}{\sinh\left(\omega\eta\right)} & 0 & 0 & \dfrac{\sinh\left[\omega(\eta-u)\right]}{\sinh\left(\omega\eta\right)}
\end{array}\right),\label{7V7}
\end{equation}
 and 
\begin{equation}
R(u)=\left(\begin{array}{cccc}
\dfrac{\sinh\left[\omega(\eta+u)\right]}{\sinh\left(\omega\eta\right)} & 0 & 0 & \dfrac{\delta_{1}}{\alpha_{1}}\dfrac{\sinh\left(\omega u\right)\cosh\left(\omega\eta\right)}{\sinh\left(\omega\eta\right)}\\
0 & -\epsilon\,\mathrm{e}^{-\omega\eta}\dfrac{\sinh\left(\omega u\right)}{\sinh\left(\omega\eta\right)} & \mathrm{e}^{-\omega u} & 0\\
0 & \mathrm{e}^{\omega u} & -\epsilon\,\mathrm{e}^{\omega\eta}\dfrac{\sinh\left(\omega u\right)}{\sinh\left(\omega\eta\right)} & 0\\
0 & 0 & 0 & \dfrac{\sinh\left[\omega(\eta-u)\right]}{\sinh\left(\omega\eta\right)}
\end{array}\right),\label{7V8}
\end{equation}
where $\omega=\alpha_{1}+\epsilon\beta_{1}$ and $\coth\left(\omega\eta\right)=\alpha_{1}/\omega$
so that $\sinh\left(\omega\eta\right)=-\epsilon\omega/\sqrt{\beta_{1}\beta_{2}}$.
These solutions contain two bare free-parameters (we can redefine
the spectral parameter) and they were obtained first in \cite{KhachatryanSedrakyan2013}
as the $R$ matrix presented by equation (5.39). 

\section{Conclusions and generalizations}

In this work we developed a differential method for solving the \noun{ybe}
and to classify its solutions. This method allowed a systematic analysis
of the functional equations arising from the \noun{ybe} for two-state
systems: it revealed in a simple way how the solutions branch and
allowed a complete classification of their regular $R$ matrices.
In total we found thirty-one families of solutions that are associated
with the four, five, six, seven and eight-vertex models -- see Table
\ref{TableR}.

\begin{table}
\begin{center}%
\begin{tabular}{lllllll}
\hline 
Model & $a_{1}/a_{2}$ & $b_{1}/b_{2}$ & $c_{1}/c_{2}$ & $d_{1}/d_{2}$ & Free-Fermion & Bare free-parameters\tabularnewline
\hline 
\noun{4V} & $\mathrm{e}^{\left(\alpha_{1}-\alpha_{2}\right)u}$ & $\lyxmathsym{\textendash}$ & $\mathrm{e}^{\left(\gamma_{1}-\gamma_{2}\right)u}$ & $\lyxmathsym{\textendash}$ & \ding{55} & $3$\tabularnewline
5V1 & $1$ & $0$ & $\mathrm{e}^{\left(\gamma_{1}-\gamma_{2}\right)u}$ & $\lyxmathsym{\textendash}$ & \ding{55} & $2$\tabularnewline
5V2 & $1$ & $\infty$ & $\mathrm{e}^{\left(\gamma_{1}-\gamma_{2}\right)u}$ & $\lyxmathsym{\textendash}$ & \ding{55} & $2$\tabularnewline
5V3 & $\mathrm{e}^{\left(\alpha_{1}-\alpha_{2}\right)u}$ & $0$ & $\mathrm{e}^{\left(2\gamma_{1}-\alpha_{1}-\alpha_{2}\right)u}$ & $\lyxmathsym{\textendash}$ & \ding{51} & $2$\tabularnewline
5V4 & $\mathrm{e}^{\left(\alpha_{1}-\alpha_{2}\right)u}$ & $\infty$ & $\mathrm{e}^{\left(2\gamma_{1}-\alpha_{1}-\alpha_{2}\right)u}$ & $\lyxmathsym{\textendash}$ & \ding{51} & $2$\tabularnewline
5U1 & $1$ & $-$ & $\mathrm{e}^{\left(\gamma_{1}-\gamma_{2}\right)u}$ & $0$ & \ding{51} & $2$\tabularnewline
5U2 & $1$ & $-$ & $\mathrm{e}^{\left(\gamma_{1}-\gamma_{2}\right)u}$ & $\infty$ & \ding{51} & $2$\tabularnewline
5U3 & $\mathrm{e}^{\left(\alpha_{1}-\alpha_{2}\right)u}$ & $-$ & $1$ & $0$ & \ding{51} & $2$\tabularnewline
5U4 & $\mathrm{e}^{\left(\alpha_{1}-\alpha_{2}\right)u}$ & $-$ & $1$ & $\infty$ & \ding{51} & $2$\tabularnewline
5U5 & $1$ & $-$ & $1$ & $0$ & \ding{55} & $2$\tabularnewline
5U6 & $1$ & $-$ & $1$ & $\infty$ & \ding{55} & $2$\tabularnewline
6V1 & $1$ & $\beta_{1}/\beta_{2}$ & $\mathrm{e}^{\left(\gamma_{1}-\gamma_{2}\right)u}$ & $\lyxmathsym{\textendash}$ & \ding{55} & $3$\tabularnewline
6V2 & $\dfrac{\sinh\left[\omega(\eta+u)\right]}{\sinh\left[\omega(\eta-u)\right]}$ & $\beta_{1}/\beta_{2}$ & $\mathrm{e}^{\left(\gamma_{1}-\gamma_{2}\right)u}$ & $\lyxmathsym{\textendash}$ & \ding{51} & $3$\tabularnewline
6U1 & $1$ & $\lyxmathsym{\textendash}$ & $1$ & $\delta_{1}/\delta_{2}$ & \ding{55} & $2$\tabularnewline
6U2 & $\dfrac{\text{sn}\left(\omega\eta|k\right)}{\text{sn}\left(\omega(\eta-u)|k\right)}$ & $\lyxmathsym{\textendash}$ & $1$ & $\delta_{1}/\delta_{2}$ & \ding{51} & $2$\tabularnewline
6U3 & $\mathrm{e}^{\left(\alpha_{1}-\alpha_{2}\right)u}$ & $0$ & $\mathrm{e}^{\left(\alpha_{1}-\alpha_{2}\right)u}$ & $0$ & \ding{51} & $2$\tabularnewline
6U4 & $\mathrm{e}^{\left(\alpha_{1}-\alpha_{2}\right)u}$ & $0$ & $\mathrm{e}^{\left(\alpha_{2}-\alpha_{1}\right)u}$ & $\infty$ & \ding{51} & $2$\tabularnewline
6U5 & $\mathrm{e}^{\left(\alpha_{1}-\alpha_{2}\right)u}$ & $\infty$ & $\mathrm{e}^{\left(\alpha_{2}-\alpha_{1}\right)u}$ & $0$ & \ding{51} & $2$\tabularnewline
6U6 & $\mathrm{e}^{\left(\alpha_{1}-\alpha_{2}\right)u}$ & $\infty$ & $\mathrm{e}^{\left(\alpha_{1}-\alpha_{2}\right)u}$ & $\infty$ & \ding{51} & $2$\tabularnewline
7V1 & $1$ & $1$ & $1$ & $0$ & \ding{55} & $2$\tabularnewline
7V2 & $1$ & $1$ & $1$ & $\infty$ & \ding{55} & $2$\tabularnewline
7V3 & $1$ & $-1$ & $\mathrm{e}^{\left(\gamma_{1}-\gamma_{2}\right)u}$ & $0$ & \ding{51} & $2$\tabularnewline
7V4 & $1$ & $-1$ & $\mathrm{e}^{\left(\gamma_{1}-\gamma_{2}\right)u}$ & $\infty$ & \ding{51} & $2$\tabularnewline
7V5 & $\dfrac{\sinh\left[\omega(\eta+u)\right]}{\sinh\left[\omega(\eta-u)\right]}$ & $1$ & $1$ & $0$ & \ding{51} & $2$\tabularnewline
7V6 & $\dfrac{\sinh\left[\omega(\eta+u)\right]}{\sinh\left[\omega(\eta-u)\right]}$ & $1$ & $1$ & $\infty$ & \ding{51} & $2$\tabularnewline
7V7 & $\dfrac{\sinh\left[\omega(\eta+u)\right]}{\sinh\left[\omega(\eta-u)\right]}$ & $\beta_{1}/\beta_{2}$ & $\mathrm{e}^{\left(\gamma_{1}-\gamma_{2}\right)u}$ & $0$ & \ding{51} & $2$\tabularnewline
7V8 & $\dfrac{\sinh\left[\omega(\eta+u)\right]}{\sinh\left[\omega(\eta-u)\right]}$ & $\beta_{1}/\beta_{2}$ & $\mathrm{e}^{\left(\gamma_{1}-\gamma_{2}\right)u}$ & $\infty$ & \ding{51} & $2$\tabularnewline
8V1 & $1$ & $1$ & $1$ & $\delta_{1}/\delta_{2}$ & \ding{55} & $3$\tabularnewline
8V2 & $1$ & $-1$ & $1$ & $\delta_{1}/\delta_{2}$ & \ding{51} & $2$\tabularnewline
8V3 & $1+\dfrac{2\epsilon}{a_{2}}\dfrac{\text{sn}\left(\omega u|k\right)\text{cn}\left(\omega\eta|k\right)}{\text{sn}\left(\omega\eta|k\right)}$ & $1$ & $1$ & $\delta_{1}/\delta_{2}$ & \ding{51} & $3$\tabularnewline
8V4 & $1+\dfrac{2\epsilon}{a_{2}}\dfrac{\text{sn}\left(\omega u|k\right)\text{dn}\left(\omega\eta|k\right)}{\text{sn}\left(\omega\eta|k\right)}$ & $1$ & $1$ & $\delta_{1}/\delta_{2}$ & \ding{51} & $3$\tabularnewline
\hline 
\end{tabular}\end{center}

\caption{Classification of the solutions of the \noun{ybe} for two-state systems
according to the ratios of the $R$ matrix elements. We also indicate
whether or not the respective solution is of the free-fermion type
(i.e., if the quantity $\varPhi=a_{1}a_{2}+b_{1}b_{2}-c_{1}c_{2}-d_{1}d_{2}$
is zero or not), and we give as well the number of bare free-parameters
of the solutions. In total we found thirty-one families of solutions.}

\label{TableR}
\end{table}

In the Appendices, we also report interesting reduced solutions, which
are obtained from the general ones by fixing some of the free-parameters
of the $R$ matrices. In this way, trigonometric solutions are obtained
from the elliptic $R$ matrices as well as rational solutions are
derived from the trigonometric ones. The symmetries of the solutions,
their geometric invariants and manifolds, the corresponding Hamiltonians
and the respective classical limits (when they exist) are also discussed
in the Appendices.

This work can be generalized in several ways. The most obvious generalization
is the classification of the $R$ matrices associated with three-state
systems, which includes important models as the fifteen-vertex $R$
matrix of Cherednik and the nineteen-vertex $R$ matrices of Zamolodchikov-Fateev
and Izergin-Korepin. Such classification is already in preparation
and it will be communicated in the future. 

Other generalizations may include the computation of the reflection
matrices (solutions of the boundary \noun{ybe}) associated with the
$R$ matrices presented here, as well as the study of the statistical
mechanics of the respective integrable models. It would be interesting,
for instance, to present the Bethe Ansatz of these integrable models.
Finally, we believe that the differential method can also be useful
in the study of the non-additive \noun{ybe}, the tetrahedral equation
and also to find and classify the solutions of the classical \noun{ybe}
without make reference to the quantum \noun{ybe} -- this analysis
could provide classical $r$ matrices that fall outside the Belavin-Drinfel'd
classification given in \cite{BelavinDrinfeld1982}. 
\begin{acknowledgements}
The author kindly thanks Professor A. Lima-Santos for his comments
and also the referees for their valuable remarks and suggestions.
This work was fomented by Coordination for the Improvement of Higher
Level Personnel (CAPES).
\end{acknowledgements}

\section*{Appendices}
\appendix

\section{Reduced Solutions\label{AppendixReduced}}

The general solutions presented in this work contain several free-parameters.
When giving particular values to these parameters, particular solutions
are obtained. For instance, if we set $\omega=1$ and $\gamma_{2}=\gamma_{1}=0$
in the six-vertex $R$ matrices (\ref{6VR1}) and (\ref{6VR2}), respectively,
then we shall obtain the well-known simplest trigonometric $R$ matrices
of the six-vertex model \cite{Baxter1985}, namely,
\begin{equation}
R(u)=\begin{pmatrix}\sinh(u+\eta) & 0 & 0 & 0\\
0 & \sinh(u) & \sinh(\eta) & 0\\
0 & \sinh(\eta) & \sinh(u) & 0\\
0 & 0 & 0 & \sinh(\eta+u)
\end{pmatrix},\qquad\text{and}\qquad R(u)=\begin{pmatrix}\sinh(u+\eta) & 0 & 0 & 0\\
0 & \sinh(u) & \sinh(\eta) & 0\\
0 & \sinh(\eta) & \sinh(u) & 0\\
0 & 0 & 0 & \sinh(\eta-u)
\end{pmatrix}.
\end{equation}
In a similar way, the well-known $R$ matrices of the eight-vertex
model, for instance Baxter's $R$ matrix \cite{Baxter1972,Baxter1978,Baxter1985},
\begin{equation}
R(u)=\begin{pmatrix}\text{sn}(u+\eta|k) & 0 & 0 & k\,\text{sn}(\eta|k)\text{sn}(u|k)\text{sn}(u+\eta|k)\\
0 & \text{sn}(u|k) & \text{sn}(\eta|k) & 0\\
0 & \text{sn}(\eta|k) & \text{sn}(u|k) & 0\\
k\,\text{sn}(\eta|k)\text{sn}(u|k)\text{sn}(u+\eta|k) & 0 & 0 & \text{sn}(u+\eta|k)
\end{pmatrix},
\end{equation}
can be obtained from (\ref{8VR1}) after we set $\omega=1$ and $\delta_{1}=\delta_{2}$,
so that we get $\xi_{1}\xi_{2}=\beta_{1}/\delta_{1}$ with $k=\xi_{1}/\xi_{2}$. 

Another important example was suggested by one of the referees of
this paper. It corresponds to a reduced solution obtained from the
eight-vertex $R$ matrices (\ref{8VR3A}) and (\ref{8VR3B}) by fixing
the value of $\eta$ according to the expression $\omega\eta=iF\left(\sqrt{1-k^{2}}\right)$,
where $F(k)$ denotes the \emph{complete elliptic integral of first
kind} whose modulus is $k$. In this case, we get that $\text{sn}\left(\omega\eta|k\right)\rightarrow\infty$,
while $\text{cn}\left(\omega\eta|k\right)/\text{sn}\left(\omega\eta|k\right)\rightarrow-i$
and $\text{dn}\left(\omega\eta|k\right)/\text{sn}\left(\omega\eta|k\right)\rightarrow-ik$.
Noticing further that in this case we get $\beta_{1}\rightarrow0$
but $\beta_{1}\text{sn}\left(\omega\eta|k\right)\rightarrow\left(\epsilon/k\right)\sqrt{\delta_{1}\delta_{2}}$
for (\ref{8VR3A}) and $\beta_{1}\text{sn}\left(\omega\eta|k\right)\rightarrow\epsilon\sqrt{\delta_{1}\delta_{2}}$
for (\ref{8VR3B}), we can verify that this value for $\omega\eta$
leads, respectively, to the following unusual six-vertex $R$ matrices:
\begin{equation}
R(u)=\begin{pmatrix}\dfrac{\mathrm{cn}\left(\omega u|k\right)}{\mathrm{dn}\left(\omega u|k\right)}-i\epsilon\,\mathrm{sn}\left(\omega u|k\right) & 0 & 0 & \epsilon k\sqrt{\dfrac{\delta_{1}}{\delta_{2}}}\dfrac{\mathrm{sn}\left(\omega u|k\right)\mathrm{cn}\left(\omega u|k\right)}{\mathrm{dn}\left(\omega u|k\right)}\\
0 & 0 & 1 & 0\\
0 & 1 & 0 & 0\\
\epsilon k\sqrt{\dfrac{\delta_{2}}{\delta_{1}}}\dfrac{\mathrm{sn}\left(\omega u|k\right)\mathrm{cn}\left(\omega u|k\right)}{\mathrm{dn}\left(\omega u|k\right)} & 0 & 0 & \dfrac{\mathrm{cn}\left(\omega u|k\right)}{\mathrm{dn}\left(\omega u|k\right)}+i\epsilon\,\mathrm{sn}\left(\omega u|k\right)
\end{pmatrix},\qquad\omega=i\alpha_{1},\qquad k=-i\frac{\sqrt{\delta_{1}\delta_{2}}}{\alpha_{1}},\label{8V3R6UA}
\end{equation}
 and 
\begin{equation}
R(u)=\begin{pmatrix}\dfrac{\mathrm{dn}\left(\omega u|k\right)}{\mathrm{cn}\left(\omega u|k\right)}-i\epsilon k\,\mathrm{sn}\left(\omega u|k\right) & 0 & 0 & \epsilon\sqrt{\dfrac{\delta_{1}}{\delta_{2}}}\dfrac{\mathrm{sn}\left(\omega u|k\right)\mathrm{dn}\left(\omega u|k\right)}{\mathrm{cn}\left(\omega u|k\right)}\\
0 & 0 & 1 & 0\\
0 & 1 & 0 & 0\\
\epsilon\sqrt{\dfrac{\delta_{2}}{\delta_{1}}}\dfrac{\mathrm{sn}\left(\omega u|k\right)\mathrm{dn}\left(\omega u|k\right)}{\mathrm{cn}\left(\omega u|k\right)} & 0 & 0 & \dfrac{\mathrm{dn}\left(\omega u|k\right)}{\mathrm{cn}\left(\omega u|k\right)}+i\epsilon k\,\mathrm{sn}\left(\omega u|k\right)
\end{pmatrix},\qquad\omega=\sqrt{\delta_{1}\delta_{2}},\qquad k=i\frac{\alpha_{1}}{\sqrt{\delta_{1}\delta_{2}}}.\label{8V3R6UB}
\end{equation}
This limit was already discussed in \cite{KhachatryanSedrakyan2013}
and it is related to an unusual elliptic six-vertex $R$ matrix found
by the authors of that work. The $R$ matrices above can be compared
with the elliptic $R$ matrix (\ref{6VUR2}), which has the same shape
as them (see Section \ref{Sec6VU}). 

Other elliptic reduced solutions can be obtained by giving special
values for the elliptic modulus $k$ -- or, which is the same, by
considering solutions of the differential equation $\left(\mathrm{d}y/\mathrm{d}x\right)^{2}=Ay^{4}+By^{2}+C$
for particular values of the coefficients $A$, $B$ and $C$. Regarding
on these possibilities, we mention the work \cite{EbaidAly2012} in
which the authors had presented a table with fifty-two particular
solutions of this differential equation -- each one of them would
give place to a particular elliptic $R$ matrix.

The most interesting way of deriving reduced solutions from the general
ones is, however, to take special limits for the elliptic modulus
$k$, so that the elliptic functions degenerate into trigonometric
ones. Indeed, trigonometric $R$ matrices can be derived from the
elliptic ones by taking one of the following well-known limits of
the elliptic functions -- see \cite{NIST2010}: \begin{subequations}
\begin{align}
\lim_{k\rightarrow0}\text{sn}(x,k) & =\sin(x), & \lim_{k\rightarrow0}\text{cn}(x,k) & =\cos(x), & \lim_{k\rightarrow0}\text{dn}(x,k) & =1,\\
\lim_{k\rightarrow1}\text{sn}(x,k) & =\tanh(x), & \lim_{k\rightarrow1}\text{cn}(x,k) & =\text{sech}(x), & \lim_{k\rightarrow1}\text{dn}(x,k) & =\text{sech}(x),
\end{align}
 \end{subequations} (care should be taken in evaluating these these
limits, however, because the modulus $k$ of the elliptic functions
appearing in the $R$ matrices generally depends on complicated expressions
of the solutions parameters). In the same fashion, rational $R$ matrices
can be obtained from the trigonometric ones by taking some limits,
usually letting $\omega\rightarrow0$, which degenerate the trigonometric
functions into rational ones. These degenerated $R$ matrices will
be reported in the what follows (many of these degenerated solutions
were already reported in \cite{KhachatryanSedrakyan2013}).

\subsection{From elliptic $R$ matrices to trigonometric ones}

Let us begin our analysis with the elliptic $R$ matrix of the unusual
six-vertex model given by (\ref{6VUR2}). Here the limit $k\rightarrow0$
is achieved by setting either $\delta_{1}=0$ or $\delta_{2}=0$.
In any case, we get that $\eta\rightarrow\infty$ and $\omega\rightarrow2\epsilon i\alpha_{1}$
where $\epsilon=\text{sign}(\alpha_{1})$; whence we obtain the following
reduced solutions: 
\begin{equation}
R(u)=\left(\begin{array}{cccc}
\mathrm{e}^{\alpha_{1}u} & 0 & 0 & 0\\
0 & 0 & 1 & 0\\
0 & 1 & 0 & 0\\
\frac{\delta_{2}}{\alpha_{1}}\sinh\left(\alpha_{1}u\right)\cosh\left(\alpha_{1}u\right) & 0 & 0 & \mathrm{e}^{-\alpha_{1}u}
\end{array}\right),\qquad\text{and}\qquad R(u)=\left(\begin{array}{cccc}
\mathrm{e}^{\alpha_{1}u} & 0 & 0 & \frac{\delta_{1}}{\alpha_{1}}\sinh\left(\alpha_{1}u\right)\cosh\left(\alpha_{1}u\right)\\
0 & 0 & 1 & 0\\
0 & 1 & 0 & 0\\
0 & 0 & 0 & \mathrm{e}^{-\alpha_{1}u}
\end{array}\right).\label{Reduced6U0}
\end{equation}
The limit $k\rightarrow1$, on the other hand, can be achieved by
making either $\alpha_{1}=0$ or $\alpha_{1}=\epsilon i\sqrt{\delta_{1}\delta_{2}}$.
In the first case we get the $R$ matrix,
\begin{equation}
R(u)=\left(\begin{array}{cccc}
\sec\left(u\sqrt{\delta_{1}\delta_{2}}\right) & 0 & 0 & \sqrt{\frac{\delta_{1}}{\delta_{2}}}\tan\left(u\sqrt{\delta_{1}\delta_{2}}\right)\\
0 & 0 & 1 & 0\\
0 & 1 & 0 & 0\\
\sqrt{\frac{\delta_{2}}{\delta_{1}}}\tan\left(u\sqrt{\delta_{1}\delta_{2}}\right) & 0 & 0 & \sec\left(u\sqrt{\delta_{1}\delta_{2}}\right)
\end{array}\right),\label{Reduced6UA}
\end{equation}
 while, in the second case, we get, 
\begin{equation}
R(u)=\left(\begin{array}{cccc}
1+i\tanh\left(u\sqrt{\delta_{1}\delta_{2}}\right) & 0 & 0 & \sqrt{\frac{\delta_{1}}{\delta_{2}}}\tanh\left(u\sqrt{\delta_{1}\delta_{2}}\right)\\
0 & 0 & 1 & 0\\
0 & 1 & 0 & 0\\
\sqrt{\frac{\delta_{2}}{\delta_{1}}}\tanh\left(u\sqrt{\delta_{1}\delta_{2}}\right) & 0 & 0 & 1-i\tanh\left(u\sqrt{\delta_{1}\delta_{2}}\right)
\end{array}\right).\label{Reduced6U2B}
\end{equation}
We remark that the limits of (\ref{8V3R6UA}) and (\ref{8V3R6UB})
for $k\rightarrow0$ (which requires $\delta_{1}=0$ or $\delta_{2}=0$
in the first case and $\alpha_{1}=0$ in the second) correspond to
the same $R$ matrices as given by (\ref{Reduced6U0}) and (\ref{Reduced6UA}),
respectively, while their limits for $k\rightarrow1$ (which in any
case is achieved as $i\alpha_{1}=\sqrt{\delta_{1}\delta_{2}}$) are
both the same and they are given by (\ref{Reduced6U2B}) with $a_{1}$
and $a_{2}$ interchanged (this swapping between $a_{1}$ and $a_{2}$
can be explained by the symmetry of the solutions (\ref{8V3R6UA})
and (\ref{8V3R6UB}) regarding the inversion of the elliptic modulus
$k$ -- see also Footnote \ref{FootnoteInversion}). We can conclude,
therefore, that all the three elliptic unusual six-vertex $R$ matrices
(\ref{6VUR2}), (\ref{8V3R6UA}) and (\ref{8V3R6UB}) have essentially
the same trigonometric limits. 

Now, let us analyze the trigonometric reductions of the eight-vertex
elliptic $R$ matrices. For the $R$ matrix (\ref{8VR1}) in which
$a_{2}=a_{1}$, the limit $k\rightarrow0$ can be achieved by letting
either $\delta_{1}\rightarrow0$ or $\delta_{2}\rightarrow0$. However,
we can verify that these limits lead to the seven-vertex $R$ matrices
(\ref{7V1}) and (\ref{7V2}), respectively (with $i\omega$ in place
of $\omega$), so that anything new is obtained here. The limit $k\rightarrow1$,
on the other hand, can be achieved as $\alpha_{1}\rightarrow\epsilon\left(\beta_{1}-\sqrt{\delta_{1}\delta_{2}}\right)$
and it provides the following trigonometric eight-vertex $R$ matrix:
\begin{equation}
R(u)=\left(\begin{array}{cccc}
\dfrac{\tanh\left[\omega\left(\epsilon u+\eta\right)\right]}{\tanh\left(\omega\eta\right)} & 0 & 0 & \sqrt{\frac{\delta_{1}}{\delta_{2}}}\tanh\left(\omega u\right)\tanh\left[\omega\left(\epsilon u+\eta\right)\right]\\
0 & \dfrac{\tanh\left(\omega u\right)}{\tanh\left(\omega\eta\right)} & 1 & 0\\
0 & 1 & \dfrac{\tanh\left(\omega u\right)}{\tanh\left(\omega\eta\right)} & 0\\
\sqrt{\frac{\delta_{2}}{\delta_{1}}}\tanh\left(\omega u\right)\tanh\left[\omega\left(\epsilon u+\eta\right)\right] & 0 & 0 & \dfrac{\tanh\left[\omega\left(\epsilon u+\eta\right)\right]}{\tanh\left(\omega\eta\right)}
\end{array}\right),\label{ReducedT8V1}
\end{equation}
where $\omega=\sqrt{\beta_{1}\left(\beta_{1}-\epsilon\alpha_{1}\right)}=\sqrt{\beta_{1}\sqrt{\delta_{1}\delta_{2}}}$
and $\tanh\left(\omega\eta\right)=\epsilon\omega/\beta_{1}$ so that
$\text{sech}\left(\omega\eta\right)=\epsilon\alpha_{1}/\beta_{1}$.

Besides, for the $R$ matrices given by (\ref{8VR3A}) and (\ref{8VR3B}),
in which $a_{2}\neq a_{1}$, we have the following: for $k\rightarrow0$,
the limit of (\ref{8VR3A}) if found as we make either $\delta_{1}=0$
or $\delta_{2}=0$; however, we can verify that the two seven-vertex
$R$ matrices obtained in this way are equivalent to the $R$ matrices
(\ref{7V5}) and (\ref{7V6}), after we replace $\omega$ by $i\omega$
and $u$ by $\epsilon u$, so nothing new is obtained here again.
The limit of (\ref{8VR3B}) for $k\rightarrow0$, on the other hand,
is achieved when $\beta_{1}=\epsilon\alpha_{1}$ and it leads to the
following trigonometric eight-vertex $R$ matrix:
\begin{equation}
R(u)=\left(\begin{array}{cccc}
\sec\left(\omega u\right)+\epsilon\dfrac{\sin\left(\omega u\right)}{\sin\left(\omega\eta\right)} & 0 & 0 & \epsilon\dfrac{\delta_{1}}{\beta_{1}}\dfrac{\tan\left(\omega u\right)}{\sin\left(\omega\eta\right)}\\
0 & \dfrac{\sin\left(\omega u\right)}{\sin\left(\omega\eta\right)} & 1 & 0\\
0 & 1 & \dfrac{\sin\left(\omega u\right)}{\sin\left(\omega\eta\right)} & 0\\
\epsilon\dfrac{\delta_{2}}{\beta_{1}}\dfrac{\tan\left(\omega u\right)}{\sin\left(\omega\eta\right)} & 0 & 0 & \sec\left(\omega u\right)-\epsilon\dfrac{\sin\left(\omega u\right)}{\sin\left(\omega\eta\right)}
\end{array}\right),\label{ReducedT8V3}
\end{equation}
where $\omega=\sqrt{\delta_{1}\delta_{2}}$ and $\sin\left(\omega\eta\right)=\omega/\beta_{1}$.
Finally, it remains to consider the limit $k\rightarrow1$ of the
$R$ matrices (\ref{8VR3A}) and (\ref{8VR3B}). Here, in both cases
the limit $k\rightarrow1$ is achieved by imposing the additional
constraint $\delta_{1}\delta_{2}=\beta_{1}^{2}-\alpha_{1}^{2}$ and
it leads to the same trigonometric $R$ matrix, namely,

\begin{equation}
R(u)=\left(\begin{array}{cccc}
1+\epsilon\dfrac{\tanh\left(\omega u\right)}{\sinh\left(\omega\eta\right)} & 0 & 0 & \dfrac{\delta_{1}}{\beta_{1}}\dfrac{\tanh\left(\omega u\right)}{\tanh\left(\omega\eta\right)}\\
0 & \dfrac{\tanh\left(\omega u\right)}{\tanh\left(\omega\eta\right)} & 1 & 0\\
0 & 1 & \dfrac{\tanh\left(\omega u\right)}{\tanh\left(\omega\eta\right)} & 0\\
\dfrac{\delta_{2}}{\beta_{1}}\dfrac{\tanh\left(\omega u\right)}{\tanh\left(\omega\eta\right)} & 0 & 0 & 1-\epsilon\dfrac{\tanh\left(\omega u\right)}{\sinh\left(\omega\eta\right)}
\end{array}\right),
\end{equation}
where $\epsilon=\text{sign}(\alpha_{1})$, $\omega=\sqrt{\delta_{1}\delta_{2}}=\sqrt{\beta_{1}^{2}-\alpha_{1}^{2}}$
and $\tanh\left(\omega\eta\right)=\omega/\beta_{1}$. 

\subsection{From trigonometric $R$ matrices to rational ones}

Rational $R$ matrices can be derived from the trigonometric ones
from special limits of its free-parameters, usually through the limit
$\omega\rightarrow0$. As we shall see, all trigonometric $R$ matrices
have a non-trivial rational limit with the only exception being the
four-vertex $R$ matrix (\ref{4V}), whose rational limit is $R(u)=P$.

Let us begin our analysis with the usual six-vertex $R$ matrices
(\ref{6VR1}) and (\ref{6VR2}). Setting $\gamma_{2}=\gamma_{1}=0$
to eliminate the exponentials and then taking limit $\omega\rightarrow0$,
which implies $\alpha_{1}=\sqrt{\beta_{1}\beta_{2}}$, we shall get
respectively the following rational $R$ matrices: 
\begin{equation}
R(u)=\left(\begin{array}{cccc}
1+u\sqrt{\beta_{1}\beta_{2}} & 0 & 0 & 0\\
0 & \beta_{1}u & 1 & 0\\
0 & 1 & \beta_{2}u & 0\\
0 & 0 & 0 & 1+u\sqrt{\beta_{1}\beta_{2}}
\end{array}\right),\qquad\text{and}\qquad R(u)=\left(\begin{array}{cccc}
1+u\sqrt{\beta_{1}\beta_{2}} & 0 & 0 & 0\\
0 & \beta_{1}u & 1 & 0\\
0 & 1 & \beta_{2}u & 0\\
0 & 0 & 0 & 1-u\sqrt{\beta_{1}\beta_{2}}
\end{array}\right).
\end{equation}

Now, let us take the unusual six-vertex $R$ matrix (\ref{6UR1}).
Here the limit $\omega\rightarrow0$ is achieved by imposing the additional
constraint $\alpha_{1}=\epsilon\sqrt{\delta_{1}\delta_{2}}$. This
provides the following $R$ rational matrix: 
\begin{equation}
R(u)=\left(\begin{array}{cccc}
\dfrac{1}{1-\epsilon\sqrt{\delta_{1}\delta_{2}}u} & 0 & 0 & \dfrac{\delta_{1}u}{1-\epsilon\sqrt{\delta_{1}\delta_{2}}u}\\
0 & 0 & 1 & 0\\
0 & 1 & 0 & 0\\
\dfrac{\delta_{2}u}{1-\epsilon\sqrt{\delta_{1}\delta_{2}}u} & 0 & 0 & \dfrac{1}{1-\epsilon\sqrt{\delta_{1}\delta_{2}}u}
\end{array}\right).\label{R6VURational}
\end{equation}
This solution also corresponds to the case $\xi_{1}=\xi_{2}$ in the
differential equation (\ref{6VUedoY}). For the remaining unusual
six-vertex $R$ matrices given by (\ref{6VUR2A}), (\ref{R6U34})
and (\ref{R6U56}), we have the following rational limits: 
\begin{equation}
R(u)=\left(\begin{array}{cccc}
1 & 0 & 0 & 0\\
0 & 0 & 1 & 0\\
0 & 1 & 0 & 0\\
\delta_{2}u & 0 & 0 & 1
\end{array}\right),\qquad\text{and}\qquad R(u)=\left(\begin{array}{cccc}
1 & 0 & 0 & \delta_{1}u\\
0 & 0 & 1 & 0\\
0 & 1 & 0 & 0\\
0 & 0 & 0 & 1
\end{array}\right).\label{5VRationalDelta}
\end{equation}

For the eight-vertex $R$ matrices, we have the following: the rational
limit of (\ref{8VR1}) is found as we take the limit $\omega\rightarrow0$,
which implies that $\beta_{1}=\epsilon\alpha_{1}$ and either $\delta_{1}=0$
or $\delta_{2}=0$. This gives us two rational $R$ matrices with
seven non-null entries: 
\begin{equation}
R(u)=\left(\begin{array}{cccc}
1+\alpha_{1}u & 0 & 0 & 0\\
0 & \epsilon\alpha_{1}u & 1 & 0\\
0 & 1 & \epsilon\alpha_{1}u & 0\\
\delta_{2}\left(1+\alpha_{1}u\right)u & 0 & 0 & 1+\epsilon\frac{u}{\eta}
\end{array}\right),\qquad\text{and}\qquad R(u)=\left(\begin{array}{cccc}
1+\alpha_{1}u & 0 & 0 & \delta_{1}\left(1+\alpha_{1}u\right)u\\
0 & \epsilon\alpha_{1}u & 1 & 0\\
0 & 1 & \epsilon\alpha_{1}u & 0\\
0 & 0 & 0 & 1+\epsilon\frac{u}{\eta}
\end{array}\right).\label{Rational8V1}
\end{equation}
Besides, the rational limit of the $R$ matrix (\ref{8VR1}) is achieved
as $\beta_{1}\rightarrow0$ and $\delta_{1}\rightarrow0$ or as $\beta_{1}\rightarrow0$
and $\delta_{1}\rightarrow0$; these two cases lead respectively to
the same $R$ matrices given at (\ref{5VRationalDelta}). Finally,
for the $R$ matrices (\ref{8VR3A}) and (\ref{8VR3B}) the rational
limit is obtained as $\omega\rightarrow0$, which implies $\beta_{1}=\epsilon\alpha_{1}$
and, either, $\delta_{1}=0$ or $\delta_{2}=0$. This provides us
with two other rational $R$ matrices with seven non-null entries:
\begin{equation}
R(u)=\left(\begin{array}{cccc}
1+\alpha_{1}u & 0 & 0 & 0\\
0 & \epsilon\alpha_{1}u & 1 & 0\\
0 & 1 & \epsilon\alpha_{1}u & 0\\
\delta_{2}u & 0 & 0 & 1-\epsilon\alpha_{1}u
\end{array}\right),\qquad\text{and}\qquad R(u)=\left(\begin{array}{cccc}
1+\alpha_{1}u & 0 & 0 & \delta_{1}u\\
0 & \epsilon\alpha_{1}u & 1 & 0\\
0 & 1 & \epsilon\alpha_{1}u & 0\\
0 & 0 & 0 & 1-\epsilon\alpha_{1}u
\end{array}\right).\label{Rational8V3}
\end{equation}

For the five-vertex $R$ matrices the rational limit is found by taking
simultaneously the limits $\alpha_{2}\rightarrow\alpha_{1}\rightarrow0$
and $\gamma_{2}\rightarrow\gamma_{1}\rightarrow0$. In this way, the
$R$ matrices given by (\ref{5V12}) and (\ref{5V34}) reduce to the
following rational $R$ matrices: 
\begin{equation}
R(u)=\left(\begin{array}{cccc}
1 & 0 & 0 & 0\\
0 & 0 & 1 & 0\\
0 & 1 & \beta_{2}u & 0\\
0 & 0 & 0 & 1
\end{array}\right),\qquad\text{and}\qquad R(u)=\left(\begin{array}{cccc}
1 & 0 & 0 & 0\\
0 & \beta_{1}u & 1 & 0\\
0 & 1 & 0 & 0\\
0 & 0 & 0 & 1
\end{array}\right),
\end{equation}
while the $R$ matrices presented in (\ref{5U12}), (\ref{5U3}),
(\ref{5U4}) and in (\ref{5U56}) reduce respectively to the same
$R$ matrices given by (\ref{5VRationalDelta}).

Finally, it can be verified that the seven-vertex $R$ matrices (\ref{7V1})
and (\ref{7V2}) reduce to same rational $R$ matrices given at (\ref{Rational8V1}),
while the seven-vertex $R$ matrices (\ref{7V3}) and (\ref{7V4})
reduce to the same rational $R$ matrix given at (\ref{5VRationalDelta}).
The seven-vertex $R$ matrices (\ref{7V5}), (\ref{7V6}), on the
other hand, reduce respectively to the $R$ matrices given at (\ref{Rational8V3}),
while the $R$ matrices (\ref{7V7}) and (\ref{7V8}) reduce to the
following $R$ matrices: 
\begin{equation}
R(u)=\left(\begin{array}{cccc}
1+\alpha_{1}u & 0 & 0 & 0\\
0 & -\epsilon\alpha_{1}u & 1 & 0\\
0 & 1 & -\epsilon\alpha_{1}u & 0\\
\delta_{2}u & 0 & 0 & 1-\alpha_{1}u
\end{array}\right),\qquad\text{and}\qquad R(u)=\left(\begin{array}{cccc}
1+\alpha_{1}u & 0 & 0 & \delta_{1}u\\
0 & -\epsilon\alpha_{1}u & 1 & 0\\
0 & 1 & -\epsilon\alpha_{1}u & 0\\
0 & 0 & 0 & 1-\alpha_{1}u
\end{array}\right).
\end{equation}

\section{Symmetries and properties of the solutions \label{AppendixS}}

Regular solutions of the \noun{ybe} can satisfy several symmetries.
These symmetries can be discovered either by an algebraic approach
-- analyzing, for instance, the shape of the $R$ matrix and the
ratios between its elements -- or through a geometric analysis --
studying, in this case, the properties of the manifolds associated
with the solutions. 

The unitarity, permutation, transposition, permutation-transposition
and crossing symmetries that we commented in Section \ref{SecYBE}
depend mainly on the ratios of the $R$ matrix elements and they can
be said to be of the algebraic type. The solutions we found in this
work, however, generally do not enjoy the majority of these symmetries.
This, of course, is due to the fact that we assumed a priori none
of these symmetries, which led us to quite asymmetric $R$ matrices.
Nonetheless, the $R$ matrices can in general satisfy a given required
symmetry if we impose the necessary restrictions on the free-parameters
of the solutions. In fact, from the definitions given to these symmetries
in Section \ref{SecYBE}, we can verify that the permutation symmetry
requires $b_{2}=b_{1}$ and $c_{2}=c_{1}$, while the transposition
symmetry requires $c_{2}=c_{1}$ and $d_{2}=d_{1}$. Besides, to the
solution satisfy the transposition-permutation symmetry, it is necessary
that $b_{2}=b_{1}$ and $d_{2}=d_{1}$ and, finally, other more complicated
relations are necessary for the solution to have the unitary and crossing
symmetries. Notwithstanding, we highlight that all the solutions we
found in this work remarkably satisfy the unitarity condition regardless
of any additional constraint (i.e., in their most general form). In
Table \ref{TableR} we have classified the solutions into thirty-one
families, according to the different ratios of the $R$ matrix elements.
The algebraic symmetries satisfied by these $R$ matrices (within
the constraints necessary for the solutions to have a given symmetry,
if necessary) are presented in Table \ref{TableS}.

The geometric symmetries of the solutions, on the other hand, can
be achieved through the analysis of their invariants, that is, the
relations among the $R$ matrix elements that are independent on the
spectral parameters $u$ and $v$. These invariants fix the manifolds
associated with the solutions and they are a consequence of the \noun{ybe}.
Here again, the differential forms of the \noun{ybe} given by equations
(\ref{DYB1}) and (\ref{DYB2}) provide a more direct way of deriving
these invariants. To see why, we can proceed as follows: first, we
can eliminate all the derivatives from the systems $E$ and $F$,
after which they reduce to two systems of polynomial equations only.
Then, methods of algebraic geometry and commutative algebra -- e.g.,
Gröbner basis, Hilbert series, multivariate resultants, and so on
-- can be used to study the ideals generated by these polynomial
equations \cite{CoxLittleOshea2015}. In fact, we can show in this
way that the Hilbert dimension corresponding to each affine variety
is usually positive, which means that the system of equations are
satisfied before all the unknowns are fixed. This explains why we
need to solve one or more differential equations at the end of the
calculations to find the remaining elements of the $R$ matrix.

As we shall see below, the following quantity plays a key role in
the analysis of the eight-vertex model invariants and its descendants
\cite{FanWu1969,GalleasMartins2002}:
\begin{equation}
\varPhi=a_{1}a_{2}+b_{1}b_{2}-c_{1}c_{2}-d_{1}d_{2}.\label{FreeF}
\end{equation}
In fact, the eight-vertex models are usually classified into two main
classes according to whether $\varPhi$ is zero or not: the case $\varPhi=0$
corresponds to the so-called \emph{free-fermion} models (or \emph{Felderhof-type}
models, after \cite{Felderhof1973}), while the case $\varPhi\neq0$
corresponds to the \emph{non-free-fermion} models (or\emph{ Baxter-type}
models, after \cite{Baxter1985}). If we take the derivative of (\ref{FreeF})
with respect to $u$ and evaluate the result at $u=0$ we shall get
the quantity 
\begin{equation}
\varphi=\alpha_{1}+\alpha_{2}-\gamma_{1}-\gamma_{2}.
\end{equation}
Thus, $\varphi=0$ provides a necessary condition (often sufficient)
for a model to be of the free-fermion type. Due to its importance,
we indicate if a given vertex model is or not of the free-fermion
type in Table \ref{TableR}. 

In the sequel, we shall analyze in more details the invariants and
manifolds associated with the six and eight-vertex models. A similar
analysis can be done for the four, five and seven-vertex models but
it will be concealed.

\begin{table}[H]
\begin{center}%
\begin{tabular}{lllllll}
\hline 
Model & $U$ & $P$ & $T$ & $PT$ & $C$ & Crossing parameter $\zeta$\tabularnewline
\hline 
\noun{4V} & \ding{51} & $\gamma_{2}=\gamma_{1}$ & $\gamma_{2}=\gamma_{1}$ & \ding{51} & \ding{55} & \ding{55}\tabularnewline
5V1 & \ding{51} & \ding{55} & $\gamma_{2}=\gamma_{1}$ & \ding{55} & \ding{55} & \ding{55}\tabularnewline
5V2 & \ding{51} & \ding{55} & $\gamma_{2}=\gamma_{1}$ & \ding{55} & \ding{55} & \ding{55}\tabularnewline
5V3 & \ding{51} & \ding{55} & $\gamma_{1}=\left(\alpha_{1}+\alpha_{2}\right)/2$ & \ding{55} & \ding{55} & \ding{55}\tabularnewline
5V4 & \ding{51} & \ding{55} & $\gamma_{1}=\left(\alpha_{1}+\alpha_{2}\right)/2$ & \ding{55} & \ding{55} & \ding{55}\tabularnewline
5U1 & \ding{51} & $\gamma_{2}=\gamma_{1}$ & \ding{55} & \ding{55} & \ding{55} & \ding{55}\tabularnewline
5U2 & \ding{51} & $\gamma_{2}=\gamma_{1}$ & \ding{55} & \ding{55} & \ding{55} & \ding{55}\tabularnewline
5U3 & \ding{51} & \ding{51} & \ding{55} & \ding{55} & \ding{55} & \ding{55}\tabularnewline
5U4 & \ding{51} & \ding{51} & \ding{55} & \ding{55} & \ding{55} & \ding{55}\tabularnewline
5U5 & \ding{51} & \ding{51} & \ding{55} & \ding{55} & \ding{55} & \ding{55}\tabularnewline
5U6 & \ding{51} & \ding{51} & \ding{55} & \ding{55} & \ding{55} & \ding{55}\tabularnewline
6V1 & \ding{51} & $\beta_{2}=\beta_{1}$, $\gamma_{2}=\gamma_{1}$ & $\gamma_{2}=\gamma_{1}$ & $\beta_{2}=\beta_{1}$ & \multirow{1}{*}{$\beta_{2}=\beta_{1}$, $\gamma_{2}=\gamma_{1}$} & $\coth(\omega\zeta)=$$(\alpha_{1}-\gamma_{1})/(2\omega)$\tabularnewline
6V2 & \ding{51} & $\beta_{2}=\beta_{1}$, $\gamma_{2}=\gamma_{1}$ & $\gamma_{2}=\gamma_{1}$ & $\beta_{2}=\beta_{1}$ & $\beta_{2}=\beta_{1}$, $\gamma_{2}=\gamma_{1}$ & $\coth(\omega\zeta)=0$\tabularnewline
6U1 & \ding{51} & \ding{51} & $\delta_{2}=\delta_{1}$ & $\delta_{2}=\delta_{1}$ & \ding{51} & $\coth(\omega\zeta)=\alpha_{1}/\omega$\tabularnewline
6U2 & \ding{51} & \ding{51} & $\delta_{2}=\delta_{1}$ & $\delta_{2}=\delta_{1}$ & \ding{55} & \ding{55}\tabularnewline
6U3 & \ding{51} & \ding{55} & \ding{55} & \ding{55} & \ding{55} & \ding{55}\tabularnewline
6U4 & \ding{51} & \ding{55} & \ding{55} & \ding{55} & \ding{55} & \ding{55}\tabularnewline
6U5 & \ding{51} & \ding{55} & \ding{55} & \ding{55} & \ding{55} & \ding{55}\tabularnewline
6U6 & \ding{51} & \ding{55} & \ding{55} & \ding{55} & \ding{55} & \ding{55}\tabularnewline
7V1 & \ding{51} & \ding{51} & \ding{55} & \ding{55} & \ding{51} & $\coth(\omega\zeta)=\alpha_{1}/\omega$\tabularnewline
7V2 & \ding{51} & \ding{51} & \ding{55} & \ding{55} & \ding{51} & $\coth(\omega\zeta)=\alpha_{1}/\omega$\tabularnewline
7V3 & \ding{51} & \ding{55} & \ding{55} & \ding{55} & \ding{55} & \ding{55}\tabularnewline
7V4 & \ding{51} & \ding{55} & \ding{55} & \ding{55} & \ding{55} & \ding{55}\tabularnewline
7V5 & \ding{51} & \ding{51} & \ding{55} & \ding{55} & \ding{51} & $\coth(\omega\zeta)=0$\tabularnewline
7V6 & \ding{51} & \ding{51} & \ding{55} & \ding{55} & \ding{51} & $\coth(\omega\zeta)=0$\tabularnewline
7V7 & \ding{51} & \ding{55} & \ding{55} & \ding{55} & \ding{55} & \ding{55}\tabularnewline
7V8 & \ding{51} & \ding{55} & \ding{55} & \ding{55} & \ding{55} & \ding{55}\tabularnewline
8V1 & \ding{51} & \ding{51} & $\delta_{2}=\delta_{1}$ & $\delta_{2}=\delta_{1}$ & \ding{51} & $\text{cn}\left(\omega\zeta|k\right)\text{dn}\left(\omega\zeta|k\right)=\epsilon\alpha_{1}/\beta_{1}$\tabularnewline
8V2 & \ding{51} & \ding{55} & $\delta_{2}=\delta_{1}$ & \ding{55} & \ding{55} & \ding{55}\tabularnewline
8V3 & \ding{51} & \ding{51} & $\delta_{2}=\delta_{1}$ & $\delta_{2}=\delta_{1}$ & \ding{51} & $\text{sn}\left(\omega\zeta|k\right)=1$\tabularnewline
8V4 & \ding{51} & \ding{51} & $\delta_{2}=\delta_{1}$ & $\delta_{2}=\delta_{1}$ & \ding{51} & $\text{sn}\left(\omega\zeta|k\right)=1$\tabularnewline
\hline 
\end{tabular}\end{center}

\caption{The unitarity $(U)$, permutation $(P)$, transposition $(T)$, permutation-transposition
$(PT)$ and crossing $(C)$ symmetries of the solutions, among with
the relation defining the crossing parameter $(\zeta)$, when available.
The symbol \ding{51} indicates that the solution has the corresponding
symmetry without any constraint. If some conditions are necessary
for the solution to have a given symmetry (in such a way that the
shape of the $R$ matrix remains the same), we write them down. Finally,
the symbol \ding{55} means that the solution cannot have the given
symmetry at all. }

\label{TableS}
\end{table}

\subsection{Invariants of the usual six-vertex model}

Let us begin with the usual six-vertex model, in which case we must
take $d_{1}=d_{2}=0$. After the derivatives are eliminated from the
systems (\ref{DYB1}) and (\ref{DYB2}), only a few equations do not
vanish. Some of them fix the ratios of the $R$ matrix elements, providing
the already derived relations (see Section \ref{Sec6V}):
\begin{equation}
\frac{a_{2}}{a_{1}}=1+\left(\frac{\alpha_{1}-\alpha_{2}}{\beta_{1}}\right)b_{1},\qquad\frac{b_{2}}{b_{1}}=\frac{\beta_{2}}{\beta_{1}},\qquad\frac{c_{2}}{c_{1}}=\mathrm{e}^{(\gamma_{2}-\gamma_{1})u}.\label{6VRatios}
\end{equation}
Among the remaining equations, $E_{2,3}$ and $E_{6,7}$ are of particular
importance. After simplification, they become:\begin{subequations}
\begin{align}
\beta_{1}\left(a_{1}a_{2}+b_{1}b_{2}-c_{1}c_{2}\right) & =\left(\alpha_{1}+\alpha_{2}-\gamma_{1}-\gamma_{2}\right)a_{1}b_{1},\\
\beta_{2}\left(a_{1}a_{2}+b_{1}b_{2}-c_{1}c_{2}\right) & =\left(\alpha_{1}+\alpha_{2}-\gamma_{1}-\gamma_{2}\right)a_{2}b_{2}.
\end{align}
 \end{subequations} From these two equations, the following invariants
directly follow: 
\begin{equation}
\frac{a_{1}a_{2}+b_{1}b_{2}-c_{1}c_{2}}{a_{1}b_{1}+a_{2}b_{2}}=\frac{\alpha_{1}+\alpha_{2}-\gamma_{1}-\gamma_{2}}{\beta_{1}+\beta_{2}},\qquad\frac{a_{1}a_{2}+b_{1}b_{2}-c_{1}c_{2}}{a_{1}b_{1}-a_{2}b_{2}}=\frac{\alpha_{1}+\alpha_{2}-\gamma_{1}-\gamma_{2}}{\beta_{1}-\beta_{2}},
\end{equation}
 from which we can deduce as well that 
\begin{equation}
\frac{a_{1}b_{1}+a_{2}b_{2}}{a_{1}b_{1}-a_{2}b_{2}}=\frac{\beta_{1}+\beta_{2}}{\beta_{1}-\beta_{2}}.
\end{equation}

The possible manifolds associated with the six-vertex models are found
after we multiply $E_{2,3}$ by $\beta_{2}$ and $E_{6,7}$ by $\beta_{1}$
and take the sum or the difference of them. In fact, this provide
us with the expressions: \begin{subequations}
\begin{align}
\left(\alpha_{1}+\alpha_{2}-\gamma_{1}-\gamma_{2}\right)\left(a_{1}b_{1}\beta_{2}-a_{2}b_{2}\beta_{1}\right) & =0,\\
\left(\alpha_{1}+\alpha_{2}-\gamma_{1}-\gamma_{2}\right)\left(a_{2}b_{2}\beta_{1}+a_{1}b_{1}\beta_{2}\right) & =2\left(a_{1}a_{2}+b_{1}b_{2}-c_{1}c_{2}\right)\beta_{1}\beta_{2}.
\end{align}
\end{subequations} Therefore, we can see that there are two possibilities,
namely, either $\varphi=\alpha_{1}+\alpha_{2}-\gamma_{1}-\gamma_{2}=0$
or $a_{1}b_{1}\beta_{2}-a_{2}b_{2}\beta_{1}=0$. The first case corresponds
to a free-fermion solution which was named 6V2 in Table \ref{TableR}.
In the second case, we get a manifold characterized by 
\begin{equation}
\frac{a_{2}b_{2}}{a_{1}b_{1}}=\frac{\beta_{2}}{\beta_{1}},
\end{equation}
which actually means that $a_{2}=a_{1}$ after we use (\ref{6VRatios}).
Assuming $\varphi\neq0$ we get a manifold belonging to the non-free-fermion
solution that corresponds to the $R$ matrix named 6V1 in Table \ref{TableR}
(notice that the factor $a_{2}b_{2}\beta_{1}+a_{1}b_{1}\beta_{2}$
cannot be zero because this would imply $a_{2}=-a_{1}$ in view of
(\ref{6VRatios}), which incompatible with the regularity condition).

\subsection{Invariants of the unusual six-vertex model where $b_{1}=0$ and $b_{2}=0$}

In the case of the elliptic unusual six-vertex model presented in
 Section \ref{Sec6VU}, we have $b_{1}=b_{2}=0$ but $d_{1}\neq0$
and $d_{2}\neq0$. After the derivatives are eliminated from the systems
$E$ and $F$, some equations can be used to fix the ratios between
the $R$ matrix elements, which are:
\begin{equation}
\frac{a_{2}}{a_{1}}=\sqrt{1+2\left(\dfrac{\alpha_{2}-\alpha_{1}}{\delta_{1}}\right)\dfrac{c_{1}d_{1}}{a_{1}^{2}}},\qquad\frac{c_{2}}{c_{1}}=1,\qquad\frac{d_{2}}{d_{1}}=\frac{\delta_{2}}{\delta_{1}},\label{6URatios}
\end{equation}
In fact, we get from $E_{1,1}$ and $F_{1,1}$ for instance, the invariant,
\begin{equation}
\frac{c_{1}d_{2}}{c_{2}d_{1}}=\frac{\delta_{2}}{\delta_{1}},\label{InvC1D1-1}
\end{equation}
which actually means $c_{2}=c_{1}$ and $d_{2}=\left(\delta_{2}/\delta_{1}\right)d_{2}$,
as stated above. Moreover, from $E_{1,4}$, $E_{4,1}$, $E_{5,8}$
and $E_{8,5}$ we get as well the invariants:
\begin{equation}
\frac{a_{1}^{2}-a_{2}^{2}}{2c_{1}d_{2}}=\frac{\alpha_{1}-\alpha_{2}}{\delta_{2}},\qquad\frac{a_{1}^{2}-a_{2}^{2}}{2c_{2}d_{1}}=\frac{\alpha_{1}-\alpha_{2}}{\delta_{1}},\label{Inv1458-1}
\end{equation}
 from which we obtain the ratio between $a_{2}$ and $a_{1}$ given
by the first equation in (\ref{6URatios}). 

The possible manifolds associated with this unusual six-vertex model
can be found from the analysis of the equations $E_{2,8}$ and $F_{1,7}$.
In fact, taking the difference between $E_{2,8}$ and $E_{1,7}$,
we shall get, after we use (\ref{6URatios}), (\ref{InvC1D1}) and
(\ref{Inv6U}), the following relation: 
\begin{equation}
\left(\alpha_{2}-\alpha_{1}\right)\left(a_{1}a_{2}-c_{1}c_{2}-d_{1}d_{2}\right)=0.
\end{equation}
Therefore, we have two possibilities: if $\alpha_{2}=\alpha_{1}$
we get the solution given by equation (\ref{6UR1}), which is a non-free-fermion
solution (solution 6U1 in Table \ref{TableR}). In this case, the
following invariants hold:
\begin{equation}
\frac{a_{1}^{2}-c_{1}^{2}-d_{1}d_{2}}{2d_{1}}=\frac{\alpha_{1}-\gamma_{1}}{\delta_{1}},\qquad\frac{a_{1}^{2}-c_{1}^{2}-d_{1}d_{2}}{2d_{2}}=\frac{\alpha_{1}-\gamma_{1}}{\delta_{2}}.
\end{equation}
The other possibility, on the other hand, means that $\varPhi=0$,
that is, it belongs to a free-fermion solution. This case corresponds
to the solution 6U2 of Table \ref{TableR}. 

We shall not discuss the invariants of the unusual six-vertex $R$
matrices given by (\ref{R6U34}) and (\ref{R6U56}) because the analysis
is similar to the previous cases.

\subsection{Invariants of the eight-vertex model}

In the case of the eight-vertex model, several other equations survive
after the derivatives are eliminated from the systems $E$ and $F$.
Some of them provide the ratios derived in  Section \ref{Sec8V},
namely, 
\begin{equation}
\frac{a_{2}}{a_{1}}=1+\left(\frac{\alpha_{1}-\alpha_{2}}{\beta_{1}}\right)b_{1},\qquad\frac{b_{2}}{b_{1}}=\frac{\beta_{2}}{\beta_{1}}=\pm1,\qquad\frac{c_{2}}{c_{1}}=1,\qquad\frac{d_{2}}{d_{1}}=\frac{\delta_{2}}{\delta_{1}}.\label{8VRatios}
\end{equation}
As in the previous case, from $E_{1,1}$ and $F_{1,1}$ we get the
invariant,
\begin{equation}
\frac{c_{1}d_{2}}{c_{2}d_{1}}=\frac{\delta_{2}}{\delta_{1}},\label{InvC1D1}
\end{equation}
and, from $E_{1,4}$, $E_{4,1}$, $E_{5,8}$ and $E_{8,5}$ we also
get that,
\begin{equation}
\frac{a_{1}^{2}-a_{2}^{2}}{2c_{1}d_{2}}=\frac{\alpha_{1}-\alpha_{2}}{\delta_{2}},\qquad\frac{a_{1}^{2}-a_{2}^{2}}{2c_{2}d_{1}}=\frac{\alpha_{1}-\alpha_{2}}{\delta_{1}},\label{Inv6U}
\end{equation}
where we used the relation $b_{2}^{2}=b_{1}^{2}$, which always holds
for regular solutions of the eight-vertex models. 

Here, the equations $E_{2,3}$ and $E_{6,7}$ are again of particular
interest. In this case, they are: \begin{subequations}
\begin{align}
\left(\alpha_{1}+\alpha_{2}-\gamma_{1}-\gamma_{2}\right)a_{1}b_{1} & =\beta_{1}\left(a_{1}a_{2}+b_{1}b_{2}-c_{1}c_{2}\right)+\beta_{2}d_{1}d_{2}-\delta_{2}\left(a_{2}b_{1}+a_{1}b_{2}\right)\left(d_{1}/c_{1}\right),\\
\left(\alpha_{1}+\alpha_{2}-\gamma_{1}-\gamma_{2}\right)a_{2}b_{2} & =\beta_{2}\left(a_{1}a_{2}+b_{1}b_{2}-c_{1}c_{2}\right)+\beta_{1}d_{1}d_{2}-\delta_{2}\left(a_{2}b_{1}+a_{1}b_{2}\right)\left(d_{1}/c_{1}\right).
\end{align}
\end{subequations} Taking the sum and the difference of them, we
get, \begin{subequations}\label{8VFE}
\begin{align}
\left(\alpha_{1}+\alpha_{2}-\gamma_{1}-\gamma_{2}\right)\left(a_{1}b_{1}+a_{2}b_{2}\right) & =\left(\beta_{1}+\beta_{2}\right)\left(a_{1}a_{2}+b_{1}b_{2}-c_{1}c_{2}+d_{1}d_{2}\right)-2\delta_{2}\left(a_{1}b_{2}+a_{2}b_{1}\right)\left(d_{1}/c_{1}\right),\\
\left(\alpha_{1}+\alpha_{2}-\gamma_{1}-\gamma_{2}\right)\left(a_{1}b_{1}-a_{2}b_{2}\right) & =\left(\beta_{1}-\beta_{2}\right)\left(a_{1}a_{2}+b_{1}b_{2}-c_{1}c_{2}-d_{1}d_{2}\right).
\end{align}
 \end{subequations}

From this, we can analyze the possible manifolds associated with the
eight-vertex model. Considering first the free-fermion case, where
$\varPhi=0$, we can see from (\ref{8VFE}) that there is only two
possibilities here -- in the same fashion as in the usual six-vertex
model --, namely, we should have either $\varphi=0$ or $a_{1}b_{1}-a_{2}b_{2}=0$.
For the first case, it is more convenient to consider the subcases
$\beta_{2}=\beta_{1}$ and $\beta_{2}=-\beta_{1}$ separately. For
$\beta_{2}=-\beta_{1}$ we obtain the invariant, 
\begin{equation}
\frac{a_{2}b_{1}}{a_{1}b_{2}}=-1,
\end{equation}
which means that $a_{2}=a_{1}$ and $b_{2}=-b_{1}$; therefore, this
manifold corresponds to the solution named 8V2 in Table \ref{TableR}.
For the other possibility, $\beta_{2}=\beta_{1}$, it follows that
\begin{equation}
\frac{\left(a_{1}+a_{2}\right)b_{1}}{2c_{1}d_{2}}=\frac{\beta_{1}}{\delta_{2}},
\end{equation}
so that this manifold is associated with to the solutions 8V3 and
8V4 of Table \ref{TableR}. The last possibility $a_{1}b_{1}-a_{2}b_{2}=0$
leads us to the invariant 
\begin{equation}
\frac{a_{2}b_{1}}{a_{1}b_{2}}=1,
\end{equation}
which actually means that $b_{2}=b_{1}$ and $a_{2}=a_{1}$. In this
case, however, other equations imply that $\varphi=0$ as well, so
that we lie into a particular case of the solutions above. Notice
that for the free-fermion case, the following invariants always hold:
\begin{equation}
\frac{a_{1}b_{2}+a_{2}b_{1}}{c_{1}d_{2}}=\frac{\beta_{1}+\beta_{1}}{\delta_{2}},\qquad\text{and}\qquad\frac{a_{1}^{2}-a_{2}^{2}}{a_{1}b_{2}+a_{2}b_{1}}=\frac{\alpha_{1}-\alpha_{2}}{\beta_{1}+\beta_{1}}.
\end{equation}

Now, let us consider the non-free-fermion case in which $\varPhi\neq0$.
Here we should assume that $\varphi\neq0$. Besides, the possibility
$\beta_{2}=-\beta_{1}$ leads to $d_{1}=0$ if the solution is regular,
so that there is only one possibility here, namely, $\beta_{2}=\beta_{1}$.
By its turn, this lead us to the invariant, 
\begin{equation}
\frac{a_{2}b_{2}}{a_{1}b_{1}}=1,
\end{equation}
which implies that $a_{2}=a_{1}$ and $b_{2}=b_{1}$. Therefore, this
manifold is associated with the solution 8V1 of Table \ref{TableR}.
Finally, notice that the following invariants always hold for the
non-free-fermion solutions:
\begin{equation}
\frac{a_{1}b_{1}}{c_{1}d_{1}}=\frac{\beta_{1}}{\delta_{1}},\qquad\frac{a_{1}^{2}+b_{1}^{2}-c_{1}^{2}-d_{1}d_{2}}{2c_{1}d_{1}}=\frac{\alpha_{1}-\gamma_{1}}{\delta_{1}},\qquad\frac{a_{1}^{2}+b_{1}^{2}-c_{1}^{2}-d_{1}d_{2}}{2a_{1}b_{1}}=\frac{\alpha_{1}-\gamma_{1}}{\beta_{1}},
\end{equation}
as we can see from the equations $E_{1,6}$ and $E_{1,7}$, for instance,
after we make use of the relations (\ref{8VRatios}), (\ref{InvC1D1})
and (\ref{Inv6U}). 

Several other invariants can be found for the eight-vertex model as
well. We shall not go further on this subject, however, because these
relations do not provide much more than we already have. 

\section{Hamiltonians\label{AppendixH}}

Another advantage of the differential method is that it provides in
a straightforward way the Hamiltonian of the system. In fact, for
periodic boundary conditions in a lattice of length $L$, the Hamiltonian
is given by \cite{Kulish1996,KulishSklyanin1982,Baxter1985}:
\begin{equation}
\mathcal{H}=\sum_{n=1}^{L-1}\mathcal{H}_{n,n+1}+\mathcal{H}_{L,1},\qquad\text{where,}\qquad\mathcal{H}_{n,n+1}=\left.\frac{\mathrm{d}R_{n,n+1}(u)}{\mathrm{d}u}\right|_{u=0}P_{n,n+1}.
\end{equation}
Notice that 
\begin{equation}
\mathcal{H}_{n,n+1}=H_{n,n+1}P_{n,n+1}=\begin{pmatrix}\alpha_{1} & 0 & 0 & \delta_{1}\\
0 & \gamma_{1} & \beta_{1} & 0\\
0 & \beta_{2} & \gamma_{2} & 0\\
\delta_{2} & 0 & 0 & \alpha_{2}
\end{pmatrix},\label{Hamilton1}
\end{equation}
where $H_{n,n+1}$ is the same matrix given in (\ref{H}). 

It is perhaps more convenient to rewrite the local Hamiltonian in
terms of the Pauli matrices -- as this is the standard way in which
the Hamiltonian is usually presented. In this new basis, the most
general Hamiltonian of the eight-vertex model becomes: 
\begin{multline}
\mathcal{H}_{n,n+1}=\tfrac{1}{4}\left\{ J_{x}\sigma_{n}^{x}\sigma_{n+1}^{x}+J_{y}\sigma_{n}^{y}\sigma_{n+1}^{y}+J_{z}\sigma_{n}^{z}\sigma_{n+1}^{z}+J_{t}\sigma_{n}^{t}\sigma_{n+1}^{t}\right.\\
\left.+\left(J_{tz}+J_{zt}\right)\sigma_{n}^{z}\sigma_{n+1}^{t}+\left(J_{zt}-J_{tz}\right)\sigma_{n}^{t}\sigma_{n+1}^{z}+i\left(J_{yx}-J_{xy}\right)\sigma_{n}^{x}\sigma_{n+1}^{y}+i\left(J_{xy}+J_{yx}\right)\sigma_{n}^{y}\sigma_{n+1}^{x}\right\} ,\label{HPauli}
\end{multline}
where $\sigma_{k}^{x}$, $\sigma_{k}^{x}$ and $\sigma_{k}^{x}$ are
the Pauli matrices with values in $\mathrm{End}(V_{k})$, with $\sigma_{k}^{t}$
denoting the identity matrix and where $\sigma_{n}^{i}\sigma_{n+1}^{j}$
means $\sigma_{n}^{i}\otimes\sigma_{n+1}^{j}$. In a matrix form,
(\ref{HPauli}) becomes:
\begin{equation}
\mathcal{H}_{n,n+1}=\frac{1}{4}\left(\begin{array}{cccc}
J_{t}+2J_{zt}+J_{z} & 0 & 0 & J_{x}+2J_{yx}-J_{y}\\
0 & J_{t}+2J_{tz}-J_{z} & J_{x}+2J_{xy}+J_{y} & 0\\
0 & J_{x}-2J_{xy}+J_{y} & J_{t}-2J_{tz}-J_{z} & 0\\
J_{x}-2J_{yx}-J_{y} & 0 & 0 & J_{t}-2J_{zt}+J_{z}
\end{array}\right).\label{Hamilton2}
\end{equation}

From this we can easily find how the two sets of parameters in (\ref{Hamilton1})
and (\ref{Hamilton2}) are related with each other. In fact, we have
that, \begin{subequations}
\begin{align}
\alpha_{1} & =J_{t}+2J_{tz}+J_{z}, & \beta_{1} & =J_{x}+2J_{xy}+J_{y}, & \gamma_{1} & =J_{t}+2J_{tz}-J_{z}, & \delta_{1} & =J_{x}+2J_{yx}-J_{y},\\
\alpha_{2} & =J_{t}-2J_{zt}+J_{z}, & \beta_{2} & =J_{x}-2J_{xy}+J_{y}, & \gamma_{2} & =J_{t}-2J_{tz}-J_{z}, & \delta_{2} & =J_{x}-2J_{yx}-J_{y},
\end{align}
 \end{subequations} or, conversely, \begin{subequations}
\begin{align}
J_{x} & =\beta_{1}+\beta_{2}+\delta_{1}+\delta_{2}, & J_{z} & =\alpha_{1}+\alpha_{2}-\gamma_{1}-\gamma_{2}, & J_{zt} & =\alpha_{1}-\alpha_{2}, & J_{xy}=\beta_{1}-\beta_{2},\label{JH}\\
J_{y} & =\beta_{1}+\beta_{2}-\delta_{1}-\delta_{2}, & J_{t} & =\alpha_{1}+\alpha_{2}+\gamma_{1}+\gamma_{2}, & J_{tz} & =\gamma_{1}-\gamma_{2}, & J_{yx}=\delta_{1}-\delta_{2}.
\end{align}
 \end{subequations}

In Table \ref{TableH} we present the values of the $J$'s coefficients
for each vertex model listed in Table \ref{TableR}. We can see from
the Table \ref{TableH} that the four-vertex $R$ matrix corresponds
to the Ising spin chain, in which the interaction occurs only in the
\noun{z} direction. Besides, all the free-fermion solutions are associated
with either the \noun{xx} or \noun{xy} Heisenberg spin chains. Finally,
the non-free-fermion solutions are related to either the \noun{xxz}
or \noun{xyz} Heisenberg spin chains (the\noun{ xxx} spin chain is
related to some of the reduced solutions). 

\begin{table}
\begin{center}%
\resizebox{\textwidth}{!}{
\begin{tabular}{llllllllll}
\hline 
Model & $J_{x}$ & $J_{y}$ & $J_{z}$ & $J_{t}$ & $J_{zt}$ & $J_{tz}$ & $J_{xy}$ & $J_{yx}$ & Spin chain\tabularnewline
\hline 
4V & $0$ & $0$ & $\alpha_{1}+\alpha_{2}-\gamma_{1}-\gamma_{2}$ & $\alpha_{1}+\alpha_{2}+\gamma_{1}+\gamma_{2}$ & $\alpha_{1}-\alpha_{2}$ & $\gamma_{1}-\gamma_{2}$ & $0$ & $0$ & \noun{z}\tabularnewline
5V1 & $\beta_{2}$ & $\beta_{2}$ & $2\alpha_{1}-\gamma_{1}-\gamma_{2}$ & $2\alpha_{1}+\gamma_{1}+\gamma_{2}$ & $0$ & $\gamma_{1}-\gamma_{2}$ & $-\beta_{2}$ & $0$ & \noun{xxz}\tabularnewline
5V2 & $\beta_{1}$ & $\beta_{1}$ & $2\alpha_{1}-\gamma_{1}-\gamma_{2}$ & $2\alpha_{1}+\gamma_{1}+\gamma_{2}$ & $0$ & $\gamma_{1}-\gamma_{2}$ & $\beta_{1}$ & $0$ & \noun{xxz}\tabularnewline
5V3 & $\beta_{2}$ & $\beta_{2}$ & $0$ & $2\alpha_{1}+2\alpha_{2}$ & $\alpha_{1}-\alpha_{2}$ & $2\gamma_{1}-\alpha_{1}-\alpha_{2}$ & $-\beta_{2}$ & $0$ & \noun{xx}\tabularnewline
5V4 & $\beta_{1}$ & $\beta_{1}$ & $0$ & $2\alpha_{1}+2\alpha_{2}$ & $\alpha_{1}-\alpha_{2}$ & $2\gamma_{1}-\alpha_{1}-\alpha_{2}$ & $\beta_{1}$ & $0$ & \noun{xx}\tabularnewline
5U1 & $\delta_{2}$ & $-\delta_{2}$ & $0$ & $2\gamma_{1}+2\gamma_{2}$ & $0$ & $\gamma_{1}-\gamma_{2}$ & $0$ & $-\delta_{2}$ & \noun{xy}\tabularnewline
5U2 & $\delta_{1}$ & $-\delta_{1}$ & $0$ & $2\gamma_{1}+2\gamma_{2}$ & $0$ & $\gamma_{1}-\gamma_{2}$ & $0$ & $\delta_{1}$ & \noun{xy}\tabularnewline
5U3 & $\delta_{2}$ & $-\delta_{2}$ & $0$ & $2\alpha_{1}+2\alpha_{2}$ & $\alpha_{1}-\alpha_{2}$ & $0$ & $0$ & $-\delta_{2}$ & \noun{xy}\tabularnewline
5U4 & $\delta_{1}$ & $-\delta_{1}$ & $0$ & $2\alpha_{1}+2\alpha_{2}$ & $\alpha_{1}-\alpha_{2}$ & $0$ & $0$ & $\delta_{1}$ & \noun{xy}\tabularnewline
5U5 & $\delta_{2}$ & $-\delta_{2}$ & $2\alpha_{1}-2\gamma_{1}$ & $2\alpha_{1}+2\gamma_{1}$ & $0$ & $0$ & $0$ & $-\delta_{2}$ & \noun{xyz}\tabularnewline
5U6 & $\delta_{1}$ & $-\delta_{1}$ & $2\alpha_{1}-2\gamma_{1}$ & $2\alpha_{1}+2\gamma_{1}$ & $0$ & $0$ & $0$ & $\delta_{1}$ & \noun{xyz}\tabularnewline
6V1 & $\beta_{1}+\beta_{2}$ & $\beta_{1}+\beta_{2}$ & $2\alpha_{1}-\gamma_{1}-\gamma_{2}$ & $2\alpha_{1}+\gamma_{1}+\gamma_{2}$ & $0$ & $\gamma_{1}-\gamma_{2}$ & $\beta_{1}-\beta_{2}$ & $0$ & \noun{xxz}\tabularnewline
6V2 & $\beta_{1}+\beta_{2}$ & $\beta_{1}+\beta_{2}$ & $0$ & $2\alpha_{1}+2\alpha_{2}$ & $\alpha_{1}-\alpha_{2}$ & $2\gamma_{1}-\alpha_{1}-\alpha_{2}$ & $\beta_{1}-\beta_{2}$ & $0$ & \noun{xx}\tabularnewline
6U1 & $\delta_{1}+\delta_{2}$ & $-\delta_{1}-\delta_{2}$ & $2\alpha_{1}$ & $2\alpha_{1}$ & $0$ & $0$ & $0$ & $\delta_{1}-\delta_{2}$ & \noun{xyz}\tabularnewline
6U2 & $\delta_{1}+\delta_{2}$ & $-\delta_{1}-\delta_{2}$ & $0$ & $0$ & $2\alpha_{1}$ & $0$ & $0$ & $\delta_{1}-\delta_{2}$ & \noun{xy}\tabularnewline
6U3 & $\epsilon(\alpha_{2}-\alpha_{1})+\delta_{2}$ & $\epsilon(\alpha_{2}-\alpha_{1})-\delta_{2}$ & $0$ & $2\alpha_{1}+2\alpha_{2}$ & $\alpha_{1}-\alpha_{2}$ & $\alpha_{1}-\alpha_{2}$ & $\epsilon(\alpha_{1}-\alpha_{2})$ & $-\delta_{2}$ & \noun{xy}\tabularnewline
6U4 & $\epsilon(\alpha_{2}-\alpha_{1})+\delta_{1}$ & $\epsilon(\alpha_{2}-\alpha_{1})-\delta_{1}$ & $0$ & $2\alpha_{1}+2\alpha_{2}$ & $\alpha_{1}-\alpha_{2}$ & $\alpha_{2}-\alpha_{1}$ & $\epsilon(\alpha_{1}-\alpha_{2})$ & $\delta_{1}$ & \noun{xy}\tabularnewline
6U5 & $\epsilon(\alpha_{2}-\alpha_{1})+\delta_{2}$ & $\epsilon(\alpha_{2}-\alpha_{1})-\delta_{2}$ & $0$ & $2\alpha_{1}+2\alpha_{2}$ & $\alpha_{1}-\alpha_{2}$ & $\alpha_{2}-\alpha_{1}$ & $\epsilon(\alpha_{2}-\alpha_{1})$ & $-\delta_{2}$ & \noun{xy}\tabularnewline
6U6 & $\epsilon(\alpha_{2}-\alpha_{1})-\delta_{1}$ & $\epsilon(\alpha_{2}-\alpha_{1})-\delta_{1}$ & $0$ & $2\alpha_{1}+2\alpha_{2}$ & $\alpha_{1}-\alpha_{2}$ & $\alpha_{1}-\alpha_{2}$ & $\epsilon(\alpha_{2}-\alpha_{1})$ & $\delta_{1}$ & \noun{xy}\tabularnewline
7V1 & $2\beta_{1}+\delta_{2}$ & $2\beta_{1}-\delta_{2}$ & $2\alpha_{1}$ & $2\alpha_{1}$ & $0$ & $0$ & $0$ & $-\delta_{2}$ & \noun{xyz}\tabularnewline
7V2 & $2\beta_{1}+\delta_{1}$ & $2\beta_{1}-\delta_{1}$ & $2\alpha_{1}$ & $2\alpha_{1}$ & $0$ & $0$ & $0$ & $\delta_{1}$ & \noun{xyz}\tabularnewline
7V3 & $\delta_{2}$ & $-\delta_{2}$ & $0$ & $2\gamma_{1}+2\gamma_{2}$ & $0$ & $\gamma_{1}-\gamma_{2}$ & $2\beta_{1}$ & $-\delta_{2}$ & \noun{xy}\tabularnewline
7V4 & $\delta_{1}$ & $-\delta_{1}$ & $0$ & $2\gamma_{1}+2\gamma_{2}$ & $0$ & $\gamma_{1}-\gamma_{2}$ & $2\beta_{1}$ & $\delta_{1}$ & \noun{xy}\tabularnewline
7V5 & $2\beta_{1}+\delta_{2}$ & $2\beta_{1}-\delta_{2}$ & $0$ & $0$ & $2\alpha_{1}$ & $0$ & $0$ & $-\delta_{2}$ & \noun{xy}\tabularnewline
7V6 & $2\beta_{1}+\delta_{1}$ & $2\beta_{1}-\delta_{1}$ & $0$ & $0$ & $2\alpha_{1}$ & $0$ & $0$ & $\delta_{1}$ & \noun{xy}\tabularnewline
7V7 & $-2\epsilon\alpha_{1}+\delta_{2}$ & $-2\epsilon\alpha_{1}-\delta_{2}$ & $0$ & $0$ & $2\alpha_{1}$ & $2\alpha_{1}+2\epsilon\beta_{1}$ & $2\epsilon\alpha_{1}+2\beta_{1}$ & $-\delta_{2}$ & \noun{xy}\tabularnewline
7V8 & $-2\epsilon\alpha_{1}+\delta_{1}$ & $-2\epsilon\alpha_{1}-\delta_{1}$ & $0$ & $0$ & $2\alpha_{1}$ & $-2\alpha_{1}-2\epsilon\beta_{1}$ & $2\epsilon\alpha_{1}+2\beta_{1}$ & $\delta_{1}$ & \noun{xy}\tabularnewline
8V1 & $2\beta_{1}+\delta_{1}+\delta_{2}$ & $2\beta_{1}-\delta_{1}-\delta_{2}$ & $2\alpha_{1}$ & $2\alpha_{1}$ & $0$ & $0$ & $0$ & $\delta_{1}-\delta_{2}$ & \noun{xyz}\tabularnewline
8V2 & $\delta_{1}+\delta_{2}$ & $-\delta_{1}-\delta_{2}$ & $0$ & $0$ & $0$ & $0$ & $2\beta_{1}$ & $\delta_{1}-\delta_{2}$ & \noun{xy}\tabularnewline
8V3 & $2\beta_{1}+\delta_{1}+\delta_{2}$ & $2\beta_{1}-\delta_{1}-\delta_{2}$ & $0$ & $0$ & $2\alpha_{1}$ & $0$ & $0$ & $\delta_{1}-\delta_{2}$ & \noun{xy}\tabularnewline
8V4 & $2\beta_{1}+\delta_{1}+\delta_{2}$ & $2\beta_{1}-\delta_{1}-\delta_{2}$ & $0$ & $0$ & $2\alpha_{1}$ & $0$ & $0$ & $\delta_{1}-\delta_{2}$ & \noun{xy}\tabularnewline
\hline 
\end{tabular}}
\end{center}

\caption{Hamiltonian coefficients for each vertex model and the corresponding
spin chains. Notice that the free-fermion models are characterized
by $J_{z}=0$.}

\label{TableH}
\end{table}

\section{Classical limits\label{AppendixClassical}}

We finish this paper by presenting the classical limits of the $R$
matrices classified in Table \ref{TableR}. As commented in Section
\ref{SecYBE}, the \noun{ybe} plays a fundamental role in theory of
quantum integrable systems. For classical integrable systems, an analogous
role is played by the so-called \emph{classical Yang-Baxter equation}:
a first-order approximation of the (quantum) \noun{ybe} (\ref{ybeA})
that reads: 
\begin{equation}
\left[r_{12}(u),r_{13}(u+v)\right]+\left[r_{12}(u),r_{23}(v)\right]+\left[r_{13}(u+v),r_{23}(v)\right]=0.\label{CYBE}
\end{equation}
The solutions of the classical \noun{ybe} are called \emph{classical
$r$ matrice}s. Given a solution $R(u,h)$ of the (quantum) \noun{ybe}
(\ref{ybeA}) that also depends smoothly on a certain parameter $h$,
besides the spectral parameter $u$, we say that this $R$ matrix
has a classical limit if the following condition holds: 
\begin{equation}
\lim_{h\rightarrow0}R\left(u,h\right)=R(h,0)=f(u)I,\label{FI}
\end{equation}
where $I$ is the identity matrix and $f(u)$ is any differentiable
complex function. In this case, the classical $r$ matrix is defined
by the formula:
\begin{equation}
r(u)=\left.\frac{\mathrm{d}}{\mathrm{d}h}R\left(u,h\right)\right|_{h=0},
\end{equation}
and we can verify that it satisfies the classical \noun{ybe} (\ref{CYBE}).
In fact, (\ref{CYBE}) follows after we differentiate twice the quantum\noun{
ybe} (\ref{ybeA}) with respect to $h$, evaluate the result at $h=0$
and use the property (\ref{FI}). 

The classical \noun{ybe} was introduced by Sklyanin in \cite{Sklyanin1979}
and Belavin \& Drinfel'd in \cite{BelavinDrinfeld1982} have classified
the non-degenerated solutions of the classical \noun{ybe} for finite
dimensional simple Lie algebras (a non degenerated solution of (\ref{CYBE})
is such that $\det r(u)\neq0$). After that, Jimbo \cite{Jimbo1985}
and Drinfel'd \cite{Drinfeld1985,Drinfeld1988B} independently introduced
the concept of \emph{quantum groups} as deformations of Lie algebras
that allowed the construction of $R$ matrices from the classical
ones \cite{Jimbo1986A}. Since then, the classical \noun{ybe} has
appeared in connection with many important topics of theoretical physics
and mathematics -- see, for instance, \cite{Jimbo1990} and references
therein. 

In this section we shall present the classical limit of the $R$ matrices
associated with two-state systems. For the cases where this limit
exists, we have identified the parameter $h$ with $\eta$ and normalized\footnote{Different normalizations lead to different $r$ matrices that differ
from each other only by the addition of a term proportional to the
identity. These $r$ matrices, however, can be regarded as equivalent
thanks to the following additive property of the classical \noun{ybe}:
if $r(u)$ is a solution of (\ref{CYBE}), then the matrix $\bar{r}(u)=r(u)+f(u)I$
is also a solution of it. } the $R$ matrices so that we have $f(u)=1$. We remark, however,
that the majority of the $R$ matrices reported in this work does
not have a classical limit as well. This is because the condition
(\ref{FI}) is too restrictive\footnote{\label{FootnoteClassical}Some solutions can have a classical limit
if the condition (\ref{FI}) is weakened. For example, it can be verified
that the classical limit of the asymmetric six-vertex $R$ matrix
(\ref{6VR2}) is given by the same $r$ matrix (\ref{6VCR}), although
the condition (\ref{FI}) is not satisfied in this case. Also, for
the six-vertex $R$ matrices (\ref{6VR1}) and (\ref{6VR2}), the
classical \noun{ybe} is still satisfied without the requirement $\beta_{2}=\beta_{1}$.
These cases, however, fall outside the Belavin-Drinfel'd classification
so that we shall not analyze these possibilities further. }. In fact, any $R$ matrix whose elements $b_{1}$ or $b_{2}$ is
zero cannot have a classical limit because (\ref{FI}) can never be
satisfied. Thus, the four-vertex $R$ matrix, all the five-vertex
$R$ matrices and the unusual six-vertex $R$ matrices cannot have
such a classical limit. Besides, the condition (\ref{FI}) is satisfied
only if $a_{2}(u,0)=a_{1}(u,0)$, $b_{2}(u,0)=b_{1}(u,0)$ etc., which
implies that the asymmetric six-vertex $R$ matrix (\ref{6VR2}),
all the seven-vertex $R$ matrices except (\ref{7V1}) and (\ref{7V2}),
and, finally, the eight-vertex $R$ matrices (\ref{8VR2}), (\ref{8VR3A})
and (\ref{8VR3B}) cannot have such a classical limit. In short, the
only $R$ matrices that admit a classical limit are the six-vertex
$R$ matrix (\ref{6VR1}), the seven-vertex $R$ matrices (\ref{7V1})
and (\ref{7V2}) and the eight-vertex $R$ matrix (\ref{8VR1}). The
classical limits of these $R$ matrices will be presented below.

For the six-vertex $R$ matrix (\ref{6VR1}), we have the following
classical $r$ matrix: 
\begin{equation}
r(u)=\left(\begin{array}{cccc}
\omega\coth\left(\omega u\right) & 0 & 0 & 0\\
0 & 0 & \mathrm{e}^{\frac{1}{2}\left(\gamma_{1}-\gamma_{2}\right)}\omega\,\text{cosech}\left(\omega u\right) & 0\\
0 & \mathrm{e}^{\frac{1}{2}\left(\gamma_{1}-\gamma_{2}\right)}\omega\,\text{cosech}\left(\omega u\right) & 0 & 0\\
0 & 0 & 0 & \omega\coth\left(\omega u\right)
\end{array}\right),\label{6VCR}
\end{equation}
where we have made $\beta_{2}=\beta_{1}$ so that (\ref{FI}) is satisfied
(see Footnote \ref{FootnoteClassical}). 

For the seven-vertex $R$ matrices (\ref{7V1}) and (\ref{7V2}),
we have respectively the following classical $r$ matrices:
\begin{equation}
r(u)=\left(\begin{array}{cccc}
\omega\coth\left(\omega u\right) & 0 & 0 & 0\\
0 & 0 & \omega\,\text{cosech}\left(\omega u\right) & 0\\
0 & \omega\,\text{cosech}\left(\omega u\right) & 0 & 0\\
\dfrac{\delta_{2}\beta_{1}}{\omega}\sinh\left(\omega u\right) & 0 & 0 & \omega\coth\left(\omega u\right)
\end{array}\right),\label{7VCR1}
\end{equation}
 and 
\begin{equation}
r(u)=\left(\begin{array}{cccc}
\omega\coth\left(\omega u\right) & 0 & 0 & \dfrac{\delta_{1}\beta_{1}}{\omega}\sinh\left(\omega u\right)\\
0 & 0 & \omega\,\text{cosech}\left(\omega u\right) & 0\\
0 & \omega\,\text{cosech}\left(\omega u\right) & 0 & 0\\
0 & 0 & 0 & \omega\coth\left(\omega u\right)
\end{array}\right).\label{7VCR2}
\end{equation}

For the eight-vertex $R$ matrix (\ref{8VR1}) we have the following
classical $r$ matrix:
\begin{equation}
r(u)=\left(\begin{array}{cccc}
\omega\dfrac{\text{cn}\left(\omega u|k\right)\text{dn}\left(\omega u|k\right)}{\text{sn}\left(\omega u|k\right)} & 0 & 0 & \omega k\sqrt{\dfrac{\delta_{1}}{\delta_{2}}}\,\text{sn}\left(\omega u|k\right)\\
0 & 0 & \dfrac{\omega}{\text{sn}\left(\omega u|k\right)} & 0\\
0 & \dfrac{\omega}{\text{sn}\left(\omega u|k\right)} & 0 & 0\\
\omega k\sqrt{\dfrac{\delta_{2}}{\delta_{1}}}\text{sn}\left(\omega u|k\right) & 0 & 0 & \omega\dfrac{\text{cn}\left(\omega u|k\right)\text{dn}\left(\omega u|k\right)}{\text{sn}\left(\omega u|k\right)}
\end{array}\right).\label{8VCR}
\end{equation}

Finally, we can compare our results with the Belavin-Drinfel'd classification
\cite{BelavinDrinfeld1982}. For two-states systems, the $r$ matrices
correspond to the $sl(2)$ Lie algebra and for this case Belavin \&
Drinfel'd have shown that, up to certain equivalences, there is only
two trigonometric solutions and only one elliptic solution, besides
their reduced rational solutions. The two trigonometric solutions
are associated, respectively, with the six and seven-vertex models
and the corresponding $r$ matrices are indeed equivalent to the $r$
matrices we found here -- namely, that ones given by (\ref{6VCR}),
(\ref{7VCR1}) and (\ref{7VCR2}), respectively. The elliptic solution,
by its turn, is associated with Baxter's eight-vertex model and the
corresponding classical $r$ matrix is indeed equivalent to the $r$
matrix (\ref{8VCR}). We conclude therefore that our classification
completely agrees with the Belavin-Drinfel'd analysis. 

\section{The differential Yang-Baxter equations\label{ApendixYB}}

We list below the non-null functional equations corresponding to the
the \noun{ybe} (\ref{ybeA}) for the general eight-vertex model. The
functional equations for the four, five, six and seven-vertex models
are contained in this system and can be obtained by zeroing the corresponding
elements. 
\begin{alignat*}{2}
Y_{1,1}\text{:} & \quad & a_{1}(u)a_{1}(u+v)a_{1}(v)+d_{1}(u)c_{2}(u+v)d_{2}(v) & =a_{1}(u)a_{1}(u+v)a_{1}(v)+d_{2}(u)c_{1}(u+v)d_{1}(v),\\
Y_{1,4}\text{:} &  & d_{1}(u)c_{2}(u+v)a_{2}(v)+a_{1}(u)a_{1}(u+v)d_{1}(v) & =c_{2}(u)d_{1}(u+v)a_{1}(v)+b_{1}(u)b_{1}(u+v)d_{1}(v),\\
Y_{1,6}\text{:} &  & a_{1}(u)d_{1}(u+v)b_{1}(v)+d_{1}(u)b_{2}(u+v)c_{2}(v) & =b_{2}(u)d_{1}(u+v)a_{1}(v)+c_{1}(u)b_{1}(u+v)d_{1}(v),\\
Y_{1,7}\text{:} &  & a_{1}(u)d_{1}(u+v)c_{1}(v)+d_{1}(u)b_{2}(u+v)b_{2}(v) & =a_{2}(u)c_{1}(u+v)d_{1}(v)+d_{1}(u)a_{1}(u+v)a_{1}(v),\\
Y_{2,2}\text{:} &  & a_{1}(u)b_{1}(u+v)b_{1}(v)+d_{1}(u)d_{2}(u+v)c_{2}(v) & =a_{1}(u)b_{1}(u+v)b_{1}(v)+d_{2}(u)d_{1}(u+v)c_{1}(v),\\
Y_{2,3}\text{:} &  & a_{1}(u)b_{1}(u+v)c_{1}(v)+d_{1}(u)d_{2}(u+v)b_{2}(v) & =b_{1}(u)a_{1}(u+v)c_{1}(v)+c_{2}(u)c_{1}(u+v)b_{1}(v),\\
Y_{2,5}\text{:} &  & a_{1}(u)c_{1}(u+v)a_{1}(v)+d_{1}(u)a_{2}(u+v)d_{2}(v) & =c_{1}(u)a_{1}(u+v)c_{1}(v)+b_{2}(u)c_{1}(u+v)b_{1}(v),\\
Y_{2,8}\text{:} &  & a_{1}(u)c_{1}(u+v)d_{1}(v)+d_{1}(u)a_{2}(u+v)a_{2}(v) & =a_{2}(u)d_{1}(u+v)c_{1}(v)+d_{1}(u)b_{1}(u+v)b_{1}(v),\\
Y_{3,2}\text{:} &  & b_{1}(u)a_{1}(u+v)c_{2}(v)+c_{1}(u)c_{2}(u+v)b_{1}(v) & =a_{1}(u)b_{1}(u+v)c_{2}(v)+d_{2}(u)d_{1}(u+v)b_{2}(v)\\
Y_{3,3}\text{:} &  & b_{1}(u)a_{1}(u+v)b_{2}(v)+c_{1}(u)c_{2}(u+v)c_{1}(v) & =b_{1}(u)a_{1}(u+v)b_{2}(v)+c_{2}(u)c_{1}(u+v)c_{2}(v),\\
Y_{3,5}\text{:} &  & c_{1}(u)b_{2}(u+v)a_{1}(v)+b_{1}(u)d_{1}(u+v)d_{2}(v) & =c_{1}(u)a_{1}(u+v)b_{2}(v)+b_{2}(u)c_{1}(u+v)c_{2}(v),\\
Y_{3,8}\text{:} &  & b_{1}(u)d_{1}(u+v)a_{2}(v)+c_{1}(u)b_{2}(u+v)d_{1}(v) & =a_{2}(u)d_{1}(u+v)b_{2}(v)+d_{1}(u)b_{1}(u+v)c_{2}(v),\\
Y_{4,1}\text{:} &  & c_{1}(u)d_{2}(u+v)a_{1}(v)+b_{1}(u)b_{1}(u+v)d_{2}(v) & =d_{2}(u)c_{1}(u+v)a_{2}(v)+a_{1}(u)a_{1}(u+v)d_{2}(v),\\
Y_{4,4}\text{:} &  & b_{1}(u)b_{1}(u+v)a_{2}(v)+c_{1}(u)d_{2}(u+v)d_{1}(v) & =b_{1}(u)b_{1}(u+v)a_{2}(v)+c_{2}(u)d_{1}(u+v)d_{2}(v),\\
Y_{4,6}\text{:} &  & c_{1}(u)a_{2}(u+v)b_{1}(v)+b_{1}(u)c_{1}(u+v)c_{2}(v) & =c_{1}(u)b_{1}(u+v)a_{2}(v)+b_{2}(u)d_{1}(u+v)d_{2}(v),\\
Y_{4,7}\text{:} &  & c_{1}(u)a_{2}(u+v)c_{1}(v)+b_{1}(u)c_{1}(u+v)b_{2}(v) & =a_{2}(u)c_{1}(u+v)a_{2}(v)+d_{1}(u)a_{1}(u+v)d_{2}(v),\\
Y_{5,2}\text{:} &  & c_{2}(u)a_{1}(u+v)c_{2}(v)+b_{2}(u)c_{2}(u+v)b_{1}(v) & =a_{1}(u)c_{2}(u+v)a_{1}(v)+d_{2}(u)a_{2}(u+v)d_{1}(v),\\
Y_{5,3}\text{:} &  & c_{2}(u)a_{1}(u+v)b_{2}(v)+b_{2}(u)c_{2}(u+v)c_{1}(v) & =c_{2}(u)b_{2}(u+v)a_{1}(v)+b_{1}(u)d_{2}(u+v)d_{1}(v),\\
Y_{5,5}\text{:} &  & b_{2}(u)b_{2}(u+v)a_{1}(v)+c_{2}(u)d_{1}(u+v)d_{2}(v) & =b_{2}(u)b_{2}(u+v)a_{1}(v)+c_{1}(u)d_{2}(u+v)d_{1}(v),\\
Y_{5,8}\text{:} &  & c_{2}(u)d_{1}(u+v)a_{2}(v)+b_{2}(u)b_{2}(u+v)d_{1}(v) & =d_{1}(u)c_{2}(u+v)a_{1}(v)+a_{2}(u)a_{2}(u+v)d_{1}(v),\\
Y_{6,1}\text{:} &  & b_{2}(u)d_{2}(u+v)a_{1}(v)+c_{2}(u)b_{1}(u+v)d_{2}(v) & =a_{1}(u)d_{2}(u+v)b_{1}(v)+d_{2}(u)b_{2}(u+v)c_{1}(v),\\
Y_{6,4}\text{:} &  & c_{2}(u)b_{1}(u+v)a_{2}(v)+b_{2}(u)d_{2}(u+v)d_{1}(v) & =c_{2}(u)a_{2}(u+v)b_{1}(v)+b_{1}(u)c_{2}(u+v)c_{1}(v),\\
Y_{6,6}\text{:} &  & b_{2}(u)a_{2}(u+v)b_{1}(v)+c_{2}(u)c_{1}(u+v)c_{2}(v) & =b_{2}(u)a_{2}(u+v)b_{1}(v)+c_{1}(u)c_{2}(u+v)c_{1}(v),\\
Y_{6,7}\text{:} &  & b_{2}(u)a_{2}(u+v)c_{1}(v)+c_{2}(u)c_{1}(u+v)b_{2}(v) & =a_{2}(u)b_{2}(u+v)c_{1}(v)+d_{1}(u)d_{2}(u+v)b_{1}(v),\\
Y_{7,1}\text{:} &  & a_{2}(u)c_{2}(u+v)d_{2}(v)+d_{2}(u)a_{1}(u+v)a_{1}(v) & =a_{1}(u)d_{2}(u+v)c_{2}(v)+d_{2}(u)b_{2}(u+v)b_{2}(v),\\
Y_{7,4}\text{:} &  & a_{2}(u)c_{2}(u+v)a_{2}(v)+d_{2}(u)a_{1}(u+v)d_{1}(v) & =c_{2}(u)a_{2}(u+v)c_{2}(v)+b_{1}(u)c_{2}(u+v)b_{2}(v),\\
Y_{7,6}\text{:} &  & a_{2}(u)b_{2}(u+v)c_{2}(v)+d_{2}(u)d_{1}(u+v)b_{1}(v) & =b_{2}(u)a_{2}(u+v)c_{2}(v)+c_{1}(u)c_{2}(u+v)b_{2}(v),\\
Y_{7,7}\text{:} &  & a_{2}(u)b_{2}(u+v)b_{2}(v)+d_{2}(u)d_{1}(u+v)c_{1}(v) & =a_{2}(u)b_{2}(u+v)b_{2}(v)+d_{1}(u)d_{2}(u+v)c_{2}(v),\\
Y_{8,2}\text{:} &  & a_{2}(u)d_{2}(u+v)c_{2}(v)+d_{2}(u)b_{1}(u+v)b_{1}(v) & =a_{1}(u)c_{2}(u+v)d_{2}(v)+d_{2}(u)a_{2}(u+v)a_{2}(v),\\
Y_{8,3}\text{:} &  & a_{2}(u)d_{2}(u+v)b_{2}(v)+d_{2}(u)b_{1}(u+v)c_{1}(v) & =b_{1}(u)d_{2}(u+v)a_{2}(v)+c_{2}(u)b_{2}(u+v)d_{2}(v),\\
Y_{8,5}\text{:} &  & d_{2}(u)c_{1}(u+v)a_{1}(v)+a_{2}(u)a_{2}(u+v)d_{2}(v) & =c_{1}(u)d_{2}(u+v)a_{2}(v)+b_{2}(u)b_{2}(u+v)d_{2}(v),\\
Y_{8,8}\text{:} &  & a_{2}(u)a_{2}(u+v)a_{2}(v)+d_{2}(u)c_{1}(u+v)d_{1}(v) & =a_{2}(u)a_{2}(u+v)a_{2}(v)+d_{1}(u)c_{2}(u+v)d_{2}(v).
\end{alignat*}

The two sets of differential equations derived from the\noun{ ybe}
(\ref{ybeA}) are presented below. The $E$ system corresponds to
equation (\ref{DYB1}) while the $F$ system corresponds to (\ref{DYB2}).
\begin{align*}
E_{1,1}\text{:} &  & 0 & =\delta_{2}c_{2}d_{1}-\delta_{1}c_{1}d_{2}, & F_{1,1}\text{:} &  & 0 & =\delta_{1}c_{2}d_{2}-\delta_{2}c_{1}d_{1},\\
E_{1,4}\text{:} &  & c_{2}d_{1}'-d_{1}c_{2}' & =\delta_{1}a_{1}^{2}-\delta_{1}b_{1}^{2}+(\alpha_{2}-\alpha_{1})c_{2}d_{1}, & F_{1,4}\text{:} &  & a_{1}d_{1}'-d_{1}a_{1}' & =\delta_{1}a_{2}c_{2}-\beta_{1}b_{1}d_{1}+(\alpha_{1}-\gamma_{2})a_{1}d_{1},\\
E_{1,6}\text{:} &  & b_{2}d_{1}'-d_{1}b_{2}' & =\beta_{1}a_{1}d_{1}-\delta_{1}b_{1}c_{1}+(\gamma_{2}-\alpha_{1})b_{2}d_{1}, & F_{1,6}\text{:} &  & d_{1}b_{1}'-b_{1}d_{1}' & =\delta_{1}b_{2}c_{2}-\beta_{2}a_{1}d_{1}+(\alpha_{1}-\gamma_{1})b_{1}d_{1},\\
E_{1,7}\text{:} &  & d_{1}a_{1}'-a_{1}d_{1}' & =\beta_{2}b_{2}d_{1}-\delta_{1}a_{2}c_{1}+(\gamma_{1}-\alpha_{1})a_{1}d_{1}, & F_{1,7}\text{:} &  & d_{1}c_{1}'-c_{1}d_{1}' & =\delta_{1}(b_{2}^{2}-a_{1}^{2})+(\alpha_{1}-\alpha_{2})c_{1}d_{1},\\
E_{2,2}\text{:} &  & d_{2}d_{1}'-d_{1}d_{2}' & =(\gamma_{2}-\gamma_{1})d_{1}d_{2}, & F_{2,2}\text{:} &  & 0 & =\delta_{1}c_{2}d_{2}-\delta_{2}c_{1}d_{1},\\
E_{2,3}\text{:} &  & b_{1}a_{1}'-a_{1}b_{1}' & =\beta_{2}d_{1}d_{2}-\beta_{1}c_{1}c_{2}, & F_{2,3}\text{:} &  & b_{1}c_{1}'-c_{1}b_{1}' & =\delta_{1}b_{2}d_{2}-a_{1}\beta_{1}c_{1}+(\alpha_{1}-\gamma_{2})b_{1}c_{1},\\
E_{2,5}\text{:} &  & c_{1}a_{1}'-a_{1}c_{1}' & =\delta_{2}a_{2}d_{1}-\beta_{1}b_{2}c_{1}+(\alpha_{1}-\gamma_{1})a_{1}c_{1}, & F_{2,5}\text{:} &  & c_{1}a_{1}'-a_{1}c_{1}' & =\delta_{1}a_{2}d_{2}-b_{1}\beta_{2}c_{1}+(\alpha_{1}-\gamma_{1})a_{1}c_{1},\\
E_{2,8}\text{:} &  & a_{2}d_{1}'-d_{1}a_{2}' & =\delta_{1}a_{1}c_{1}-\beta_{1}b_{1}d_{1}+(\alpha_{2}-\gamma_{1})a_{2}d_{1}, & F_{2,8}\text{:} &  & c_{1}d_{1}'-d_{1}c_{1}' & =\delta_{1}(a_{2}^{2}-b_{1}^{2})+(\alpha_{1}-\alpha_{2})c_{1}d_{1},\\
E_{3,2}\text{:} &  & a_{1}b_{1}'-b_{1}a_{1}' & =\beta_{1}c_{1}c_{2}-\beta_{2}d_{1}d_{2}, & F_{3,2}\text{:} &  & c_{2}b_{1}'-b_{1}c_{2}' & =\beta_{1}a_{1}c_{2}-\delta_{2}b_{2}d_{1}+(\gamma_{1}-\alpha_{1})b_{1}c_{2},\\
E_{3,3}\text{:} &  & c_{2}c_{1}'-c_{1}c_{2}' & =(\gamma_{1}-\gamma_{2})c_{1}c_{2}, & F_{3,3}\text{:} &  & c_{2}c_{1}'-c_{1}c_{2}' & =(\gamma_{1}-\gamma_{2})c_{1}c_{2},\\
E_{3,5}\text{:} &  & b_{2}c_{1}'-c_{1}b_{2}' & =\delta_{2}b_{1}d_{1}-\beta_{2}a_{1}c_{1}+(\alpha_{1}-\gamma_{2})b_{2}c_{1}, & F_{3,5}\text{:} &  & b_{2}a_{1}'-a_{1}b_{2}' & =\beta_{1}d_{1}d_{2}-\beta_{2}c_{1}c_{2},\\
E_{3,8}\text{:} &  & d_{1}b_{1}'-b_{1}d_{1}' & =\delta_{1}b_{2}c_{1}-\beta_{2}a_{2}d_{1}+(\alpha_{2}-\gamma_{2})b_{1}d_{1}, & F_{3,8}\text{:} &  & b_{2}d_{1}'-d_{1}b_{2}' & =\beta_{1}a_{2}d_{1}-\delta_{1}b_{1}c_{2}+(\gamma_{1}-\alpha_{2})b_{2}d_{1},\\
E_{4,1}\text{:} &  & d_{2}c_{1}'-c_{1}d_{2}' & =\delta_{2}(b_{1}^{2}-a_{1}^{2})+(\alpha_{1}-\alpha_{2})c_{1}d_{2}, & F_{4,1}\text{:} &  & d_{2}a_{1}'-a_{1}d_{2}' & =\beta_{1}b_{1}d_{2}-\delta_{2}a_{2}c_{1}+(\gamma_{1}-\alpha_{1})a_{1}d_{2},\\
E_{4,4}\text{:} &  & 0 & =\delta_{1}c_{1}d_{2}-\delta_{2}c_{2}d_{1}, & F_{4,4}\text{:} &  & d_{2}d_{1}'-d_{1}d_{2}' & =(\gamma_{1}-\gamma_{2})d_{1}d_{2},\\
E_{4,6}\text{:} &  & c_{1}b_{1}'-b_{1}c_{1}' & =\beta_{1}a_{2}c_{1}-\delta_{2}b_{2}d_{1}+(\gamma_{2}-\alpha_{2})b_{1}c_{1}, & F_{4,6}\text{:} &  & a_{2}b_{1}'-b_{1}a_{2}' & =\beta_{1}c_{1}c_{2}-\beta_{2}d_{1}d_{2},\\
E_{4,7}\text{:} &  & a_{2}c_{1}'-c_{1}a_{2}' & =\beta_{2}b_{1}c_{1}-\delta_{2}a_{1}d_{1}+(\gamma_{1}-\alpha_{2})a_{2}c_{1}, & F_{4,7}\text{:} &  & a_{2}c_{1}'-c_{1}a_{2}' & =\beta_{1}b_{2}c_{1}-\delta_{1}a_{1}d_{2}+(\gamma_{1}-\alpha_{2})a_{2}c_{1},\\
E_{5,2}\text{:} &  & a_{1}c_{2}'-c_{2}a_{1}' & =\beta_{1}b_{2}c_{2}-\delta_{1}a_{2}d_{2}+(\gamma_{2}-\alpha_{1})a_{1}c_{2}, & F_{5,2}\text{:} &  & a_{1}c_{2}'-c_{2}a_{1}' & =\beta_{2}b_{1}c_{2}-\delta_{2}a_{2}d_{1}+(\gamma_{2}-\alpha_{1})a_{1}c_{2}\\
E_{5,3}\text{:} &  & c_{2}b_{2}'-b_{2}c_{2}' & =\beta_{2}a_{1}c_{2}-b_{1}d_{2}\delta_{1}+(\gamma_{1}-\alpha_{1})b_{2}c_{2}, & F_{5,3}\text{:} &  & a_{1}b_{2}'-b_{2}a_{1}' & =\beta_{2}c_{1}c_{2}-\beta_{1}d_{1}d_{2},\\
E_{5,5}\text{:} &  & 0 & =\delta_{2}c_{2}d_{1}-\delta_{1}c_{1}d_{2}, & F_{5,5}\text{:} &  & d_{1}d_{2}'-d_{2}d_{1}' & =(\gamma_{2}-\gamma_{1})d_{1}d_{2},\\
E_{5,8}\text{:} &  & d_{1}c_{2}'-c_{2}d_{1}' & =\delta_{1}(b_{2}^{2}-a_{2}^{2})+(\alpha_{2}-\alpha_{1})c_{2}d_{1}, & F_{5,8}\text{:} &  & d_{1}a_{2}'-a_{2}d_{1}' & =\beta_{2}b_{2}d_{1}-\delta_{1}a_{1}c_{2}+(\gamma_{2}-\alpha_{2})a_{2}d_{1},\\
E_{6,1}\text{:} &  & d_{2}b_{2}'-b_{2}d_{2}' & =\delta_{2}b_{1}c_{2}-\beta_{1}a_{1}d_{2}+(\alpha_{1}-\gamma_{1})b_{2}d_{2}, & F_{6,1}\text{:} &  & b_{1}d_{2}'-d_{2}b_{1}' & =\beta_{2}a_{1}d_{2}-\delta_{2}b_{2}c_{1}+(\gamma_{2}-\alpha_{1})b_{1}d_{2},\\
E_{6,4}\text{:} &  & b_{1}c_{2}'-c_{2}b_{1}' & =\delta_{1}b_{2}d_{2}-\beta_{1}a_{2}c_{2}+(\alpha_{2}-\gamma_{1})b_{1}c_{2}, & F_{6,4}\text{:} &  & b_{1}a_{2}'-a_{2}b_{1}' & =\beta_{2}d_{1}d_{2}-\beta_{1}c_{1}c_{2},\\
E_{6,6}\text{:} &  & c_{1}c_{2}'-c_{2}c_{1}' & =(\gamma_{2}-\gamma_{1})c_{1}c_{2}, & F_{6,6}\text{:} &  & c_{1}c_{2}'-c_{2}c_{1}' & =(\gamma_{2}-\gamma_{1})c_{1}c_{2},\\
E_{6,7}\text{:} &  & a_{2}b_{2}'-b_{2}a_{2}' & =\beta_{2}c_{1}c_{2}-\beta_{1}d_{1}d_{2}, & F_{6,7}\text{:} &  & c_{1}b_{2}'-b_{2}c_{1}' & =\beta_{2}a_{2}c_{1}-\delta_{1}b_{1}d_{2}+(\gamma_{2}-\alpha_{2})b_{2}c_{1},\\
E_{7,1}\text{:} &  & a_{1}d_{2}'-d_{2}a_{1}' & =\delta_{2}a_{2}c_{2}-\beta_{2}b_{2}d_{2}+(\alpha_{1}-\gamma_{2})a_{1}d_{2}, & F_{7,1}\text{:} &  & c_{2}d_{2}'-d_{2}c_{2}' & =\delta_{2}(a_{1}^{2}-b_{2}^{2})+(\alpha_{2}-\alpha_{1})c_{2}d_{2},\\
E_{7,4}\text{:} &  & c_{2}a_{2}'-a_{2}c_{2}' & =\delta_{1}a_{1}d_{2}-\beta_{2}b_{1}c_{2}+(\alpha_{2}-\gamma_{2})a_{2}c_{2} & F_{7,4}\text{:} &  & c_{2}a_{2}'-a_{2}c_{2}' & =\delta_{2}a_{1}d_{1}-\beta_{1}b_{2}c_{2}+(\alpha_{2}-\gamma_{2})a_{2}c_{2},\\
E_{7,6}\text{:} &  & b_{2}a_{2}'-a_{2}b_{2}' & =\beta_{1}d_{1}d_{2}-\beta_{2}c_{1}c_{2}, & F_{7,6}\text{:} &  & b_{2}c_{2}'-c_{2}b_{2}' & =\delta_{2}b_{1}d_{1}-\beta_{2}a_{2}c_{2}+(\alpha_{2}-\gamma_{1})b_{2}c_{2},\\
E_{7,7}\text{:} &  & d_{1}d_{2}'-d_{2}d_{1}' & =(\gamma_{1}-\gamma_{2})d_{1}d_{2}, & F_{7,7}\text{:} &  & 0 & =\delta_{2}c_{1}d_{1}-\delta_{1}c_{2}d_{2},\\
E_{8,2}\text{:} &  & d_{2}a_{2}'-a_{2}d_{2}' & =\beta_{1}b_{1}d_{2}-\delta_{2}a_{1}c_{2}+(\gamma_{2}-\alpha_{2})a_{2}d_{2}, & F_{8,2}\text{:} &  & d_{2}c_{2}'-c_{2}d_{2}' & =\delta_{2}(b_{1}^{2}-a_{2}^{2})+(\alpha_{2}-\alpha_{1})c_{2}d_{2},\\
E_{8,3}\text{:} &  & b_{1}d_{2}'-d_{2}b_{1}' & =\beta_{2}a_{2}d_{2}-\delta_{2}b_{2}c_{2}+(\gamma_{1}-\alpha_{2})b_{1}d_{2}, & F_{8,3}\text{:} &  & d_{2}b_{2}'-b_{2}d_{2}' & =\delta_{2}b_{1}c_{1}-\beta_{1}a_{2}d_{2}+(\alpha_{2}-\gamma_{2})b_{2}d_{2},\\
E_{8,5}\text{:} &  & c_{1}d_{2}'-d_{2}c_{1}' & =\delta_{2}(a_{2}^{2}-b_{2}^{2})+(\alpha_{1}-\alpha_{2})c_{1}d_{2}, & F_{8,5}\text{:} &  & a_{2}d_{2}'-d_{2}a_{2}' & =\delta_{2}a_{1}c_{1}-\beta_{2}b_{2}d_{2}+(\alpha_{2}-\gamma_{1})a_{2}d_{2},\\
E_{8,8}\text{:} &  & 0 & =\delta_{1}c_{1}d_{2}-\delta_{2}c_{2}d_{1}, & F_{8,8}\text{:} &  & 0 & =\delta_{2}c_{1}d_{1}-\delta_{1}c_{2}d_{2}.
\end{align*}

\bibliographystyle{elsarticle-num}
\bibliography{DYBE}

\begin{thebibliography}{10}
\expandafter\ifx\csname url\endcsname\relax
  \def\url#1{\texttt{#1}}\fi
\expandafter\ifx\csname urlprefix\endcsname\relax\def\urlprefix{URL }\fi
\expandafter\ifx\csname href\endcsname\relax
  \def\href#1#2{#2} \def\path#1{#1}\fi

\bibitem{Yang1967}
C.~N. Yang, Some exact results for the many-body problem in one dimension with
  repulsive delta-function interaction, Physical Review Letters 19~(23) (1967)
  1312.
\newblock \href {https://doi.org/10.1103/PhysRevLett.19.1312}
  {\path{doi:10.1103/PhysRevLett.19.1312}}.

\bibitem{Yang1968}
C.~N. Yang, {$S$} matrix for the one-dimensional $n$-body problem with
  repulsive or attractive $\delta$-function interaction, Physical Review
  168~(5) (1968) 1920.
\newblock \href {https://doi.org/10.1103/PhysRev.168.1920}
  {\path{doi:10.1103/PhysRev.168.1920}}.

\bibitem{Zamolodchikov1979}
A.~B. Zamolodchikov, A.~B. Zamolodchikov, Factorized {$S$}-matrices in two
  dimensions as the exact solutions of certain relativistic quantum field
  theory models, Annals of Physics 120~(2) (1979) 253--291.
\newblock \href {https://doi.org/10.1142/9789812798336_0005}
  {\path{doi:10.1142/9789812798336_0005}}.

\bibitem{Baxter1972}
R.~J. Baxter, Partition function of the eight-vertex lattice model, Annals of
  Physics 70~(1) (1972) 193--228.
\newblock \href {https://doi.org/10.1016/0003-4916(72)90335-1}
  {\path{doi:10.1016/0003-4916(72)90335-1}}.

\bibitem{Baxter1978}
R.~J. Baxter, Solvable eight-vertex model on an arbitrary planar lattice,
  Philosophical Transactions of the Royal Society of London A: Mathematical,
  Physical and Engineering Sciences 289~(1359) (1978) 315--346.
\newblock \href {https://doi.org/10.1098/rsta.1978.0062}
  {\path{doi:10.1098/rsta.1978.0062}}.

\bibitem{KulishSklyanin1982}
P.~Kulish, E.~Sklyanin, Solutions of the {Yang-Baxter} equation, Journal of
  Mathematical Sciences 19~(5) (1982) 1596--1620.
\newblock \href {https://doi.org/10.1007/BF01091463}
  {\path{doi:10.1007/BF01091463}}.

\bibitem{Jimbo1990}
M.~Jimbo, {Yang-Baxter} equation in integrable systems, Vol.~10, World
  Scientific, 1990.

\bibitem{Kulish1996}
P.~P. Kulish, {Yang-Baxter} equation and reflection equations in integrable
  models, in: Low-dimensional models in statistical physics and quantum field
  theory, Springer, 1996, pp. 125--144.
\newblock \href {https://doi.org/10.1007/BFb0102555}
  {\path{doi:10.1007/BFb0102555}}.

\bibitem{SklyaninTakhtadzhyanFaddeev1979}
E.~K. Sklyanin, L.~A. Takhtadzhyan, L.~D. Faddeev, Quantum inverse problem
  method {I}, Theoretical and Mathematical Physics 40~(688-706) (1979) 86.
\newblock \href {https://doi.org/10.1007/BF01018718}
  {\path{doi:10.1007/BF01018718}}.

\bibitem{TakhtadzhanFaddeev1979}
L.~A. Takhtadzhan, L.~D. Faddeev, The quantum method of the inverse problem and
  the {Heisenberg} {XYZ} model, Russian Mathematical Surveys 34~(5) (1979)
  11--68.
\newblock \href {https://doi.org/10.1070/RM1979v034n05ABEH003909}
  {\path{doi:10.1070/RM1979v034n05ABEH003909}}.

\bibitem{Sklyanin1982B}
E.~K. Sklyanin, Quantum version of the method of inverse scattering problem,
  Journal of Mathematical Sciences 19~(5) (1982) 1546--1596.
\newblock \href {https://doi.org/10.1007/BF01091462}
  {\path{doi:10.1007/BF01091462}}.

\bibitem{Sklyanin1982A}
E.~K. Sklyanin, Some algebraic structures connected with the {Yang-Baxter}
  equation, Functional Analysis and its Applications 16~(4) (1982) 263--270.
\newblock \href {https://doi.org/10.1007/BF01077848}
  {\path{doi:10.1007/BF01077848}}.

\bibitem{Jimbo1985}
M.~Jimbo, A $q$-difference analogue of {$U(g)$} and the {Yang-Baxter} equation,
  Letters in Mathematical Physics 10~(1) (1985) 63--69.
\newblock \href {https://doi.org/10.1007/BF00400222}
  {\path{doi:10.1007/BF00400222}}.

\bibitem{Drinfeld1985}
V.~G. Drinfel'd, {Hopf} algebra and {Yang-Baxter} equation, Soviet Mathematics
  Doklady 32 (1985) 254--258.

\bibitem{Drinfeld1988B}
V.~G. Drinfel'd, Quantum groups, Journal of Soviet Mathematics 41~(2) (1988)
  898--915.
\newblock \href {https://doi.org/10.1007/BF01247086}
  {\path{doi:10.1007/BF01247086}}.

\bibitem{FaddeevReshetikhinTakhtajan1988}
L.~D. Faddeev, N.~Y. Reshetikhin, L.~Takhtajan, Quantization of {Lie} groups
  and {Lie} algebras, in: Algebraic Analysis, Volume 1, Elsevier, 1988, pp.
  129--139.
\newblock \href {https://doi.org/10.1016/B978-0-12-400465-8.50019-5}
  {\path{doi:10.1016/B978-0-12-400465-8.50019-5}}.

\bibitem{Turaev1988}
V.~G. Turaev, The {Yang-Baxter} equation and invariants of links, Inventiones
  mathematicae 92~(3) (1988) 527--553.
\newblock \href {https://doi.org/10.1007/BF01393746}
  {\path{doi:10.1007/BF01393746}}.

\bibitem{KauffmanLomonaco2004}
L.~H. Kauffman, S.~J. Lomonaco~Jr, Braiding operators are universal quantum
  gates, New Journal of Physics 6~(1) (2004) 134.
\newblock \href {https://doi.org/10.1088/1367-2630/6/1/134}
  {\path{doi:10.1088/1367-2630/6/1/134}}.

\bibitem{MinahanZarembo2003}
J.~A. Minahan, K.~Zarembo, The {Bethe-ansatz} for {$\mathcal{N}=4$} super
  {Yang-Mills}, Journal of High Energy Physics 2003~(03) (2003) 013.
\newblock \href {https://doi.org/10.1088/1126-6708/2003/03/013}
  {\path{doi:10.1088/1126-6708/2003/03/013}}.

\bibitem{BeisertEtal2012}
N.~Beisert, C.~Ahn, L.~F. Alday, Z.~Bajnok, J.~M. Drummond, L.~Freyhult,
  N.~Gromov, R.~A. Janik, V.~Kazakov, T.~Klose, et~al., Review of {AdS/CFT}
  integrability: an overview, Letters in Mathematical Physics 99~(1-3) (2012)
  3--32.
\newblock \href {https://doi.org/10.1007/s11005-011-0529-2}
  {\path{doi:10.1007/s11005-011-0529-2}}.

\bibitem{Witten1989}
E.~Witten, Gauge theories and integrable lattice models, Nuclear Physics B
  322~(3) (1989) 629--697.
\newblock \href {https://doi.org/10.1016/0550-3213(89)90232-0}
  {\path{doi:10.1016/0550-3213(89)90232-0}}.

\bibitem{CostelloWittenYamazaki2017}
K.~Costello, E.~Witten, M.~Yamazaki,
  \href{https://arxiv.org/abs/1709.09993}{Gauge theory and integrability, {I}}.
\newline\urlprefix\url{https://arxiv.org/abs/1709.09993}

\bibitem{CostelloWittenYamazaki2018}
K.~Costello, E.~Witten, M.~Yamazaki,
  \href{https://arxiv.org/abs/1802.01579}{Gauge theory and integrability,
  {II}}.
\newline\urlprefix\url{https://arxiv.org/abs/1802.01579}

\bibitem{Korepin1997}
V.~E. Korepin, N.~M. Bogoliubov, A.~G. Izergin, Quantum inverse scattering
  method and correlation functions, Vol.~3, Cambridge University Press, 1997.

\bibitem{Jones1990}
V.~F.~R. Jones, Baxterization, International Journal of Modern Physics B 4~(05)
  (1990) 701--713.
\newblock \href {https://doi.org/10.1142/S0217751X91001027}
  {\path{doi:10.1142/S0217751X91001027}}.

\bibitem{Jimbo1986A}
M.~Jimbo, Quantum {$R$} matrix for the generalized {Toda} system,
  Communications in Mathematical Physics 102~(4) (1986) 537--547.
\newblock \href {https://doi.org/10.1007/BF01221646}
  {\path{doi:10.1007/BF01221646}}.

\bibitem{Bazhanov1987}
V.~V. Bazhanov, Integrable quantum systems and classical {Lie} algebras,
  Communications in Mathematical Physics 113~(3) (1987) 471--503.
\newblock \href {https://doi.org/10.1007/BF01221256}
  {\path{doi:10.1007/BF01221256}}.

\bibitem{BazhanovShadrikov1987}
V.~V. Bazhanov, A.~G. Shadrikov, Trigonometric solutions of triangle equations.
  {Simple} {Lie} superalgebras, Theoretical and Mathematical Physics 73~(3)
  (1987) 1302--1312.
\newblock \href {https://doi.org/10.1007/BF01041913}
  {\path{doi:10.1007/BF01041913}}.

\bibitem{Krichever1981}
I.~M. Krichever, Baxter's equations and algebraic geometry, Functional Analysis
  and Its Applications 15~(2) (1981) 92--103.
\newblock \href {https://doi.org/10.1007/BF01082280}
  {\path{doi:10.1007/BF01082280}}.

\bibitem{CoxLittleOshea2015}
D.~A. Cox, J.~B. Little, D.~O'Shea, Ideals, varieties, and algorithms: {An}
  introduction to computational algebraic geometry and commutative algebra, 4th
  Edition, Springer, 2015.

\bibitem{Abel1823}
N.~H. Abel, M{\'e}thode g{\'e}n{\'e}rale pour trouver des fonctions d'une seule
  quantit{\'e} variable, lorsqu'une propri{\'e}t{\'e} de ces fonctions est
  exprim{\'e}e par une {\'e}quation entre deux variables, Magazin for
  Naturvidenskaberne 1~(2) (1823) 1--10.

\bibitem{Aczel1966}
J.~Acz{\'e}l, Lectures on functional equations and their applications, Vol.~19,
  Academic Press, 1966.

\bibitem{Sklyanin1988}
E.~K. Sklyanin, Boundary conditions for integrable quantum systems, Journal of
  Physics A: Mathematical and General 21~(10) (1988) 2375.
\newblock \href {https://doi.org/10.1088/0305-4470/21/10/015}
  {\path{doi:10.1088/0305-4470/21/10/015}}.

\bibitem{Mezincescu1991}
L.~Mezincescu, R.~I. Nepomechie, Integrable open spin chains with nonsymmetric
  {R}-matrices, Journal of Physics A: Mathematical and General 24~(1) (1991)
  L17.
\newblock \href {https://doi.org/10.1088/0305-4470/24/1/005}
  {\path{doi:10.1088/0305-4470/24/1/005}}.

\bibitem{MalaraLima2006}
R.~Malara, A.~Lima-Santos, {On, {$A_{n-1}^{(1)}, B_n^{(1)}, C_n^{(1)},
  D_n^{(1)}, A_{2n}^{(2)}, A_{2n-1}^{(2)}$ and $D_{n+1}^{(2)}$} and reflection
  {$K$}-matrices}, Journal of Statistical Mechanics: Theory and Experiment
  2006~(09) (2006) P09013.
\newblock \href {https://doi.org/10.1088/1742-5468/2006/09/P09013}
  {\path{doi:10.1088/1742-5468/2006/09/P09013}}.

\bibitem{Lima2009A}
A.~Lima-Santos, {Reflection matrices for the {$U_q[spo(2n|2m)]$} vertex model},
  Journal of Statistical Mechanics: Theory and Experiment 2009~(04) (2009)
  P04005.
\newblock \href {https://doi.org/10.1088/1742-5468/2009/04/P04005}
  {\path{doi:10.1088/1742-5468/2009/04/P04005}}.

\bibitem{Lima2009B}
A.~Lima-Santos, {Reflection matrices for the {$U_q [osp (r|2m)^{(1)}]$} vertex
  model}, Journal of Statistical Mechanics: Theory and Experiment 2009~(07)
  (2009) P07045.
\newblock \href {https://doi.org/10.1088/1742-5468/2009/07/P07045}
  {\path{doi:10.1088/1742-5468/2009/07/P07045}}.

\bibitem{Lima2009C}
A.~Lima-Santos, {Reflection matrices for the {$U_q [sl (m| n)^{(1)}]$} vertex
  model}, Journal of Statistical Mechanics: Theory and Experiment 2009~(08)
  (2009) P08006.
\newblock \href {https://doi.org/10.1088/1742-5468/2009/08/P08006}
  {\path{doi:10.1088/1742-5468/2009/08/P08006}}.

\bibitem{Lima2010}
A.~Lima-Santos, W.~Galleas, {Reflection matrices for the {$U_q [sl (r|
  2m)^{(2)}]$} vertex model}, Nuclear Physics B 833~(3) (2010) 271--297.
\newblock \href {https://doi.org/10.1016/j.nuclphysb.2010.02.009}
  {\path{doi:10.1016/j.nuclphysb.2010.02.009}}.

\bibitem{VieiraLima2013}
R.~S. Vieira, A.~Lima-Santos, On the multiparametric {$U_q[D_{n+1}^{(2)}]$}
  vertex model, Journal of Statistical Mechanics: Theory and Experiment
  2013~(02) (2013) P02011.
\newblock \href {https://doi.org/10.1088/1742-5468/2013/02/P02011}
  {\path{doi:10.1088/1742-5468/2013/02/P02011}}.

\bibitem{VieiraLima2017A}
R.~S. Vieira, A.~Lima-Santos, Reflection {$K$}-matrices for a nineteen vertex
  model with {$U_q[osp(2| 2)^{(2)}]$} symmetry, Physics Letters A 381~(36)
  (2017) 3015--3020.
\newblock \href {https://doi.org/10.1016/j.physleta.2017.07.032}
  {\path{doi:10.1016/j.physleta.2017.07.032}}.

\bibitem{VieiraLima2017C}
R.~S. Vieira, A.~Lima-Santos, Reflection matrices with {$U_q[osp^{(2)}(2|2m)]$}
  symmetry, Journal of Physics A: Mathematical and Theoretical 50~(37) (2017)
  375204.
\newblock \href {https://doi.org/10.1088/1751-8121/aa836c}
  {\path{doi:10.1088/1751-8121/aa836c}}.

\bibitem{SuzukiFisher1971}
M.~Suzuki, M.~E. Fisher, Zeros of the partition function for the {Heisenberg},
  ferroelectric, and general {Ising} models, Journal of Mathematical Physics
  12~(2) (1971) 235--246.
\newblock \href {https://doi.org/10.1063/1.1665583}
  {\path{doi:10.1063/1.1665583}}.

\bibitem{SogoEtal1982}
K.~Sogo, M.~Uchinami, Y.~Akutsu, M.~Wadati, Classification of exactly solvable
  two-component models, Progress of Theoretical Physics 68~(2) (1982) 508--526.
\newblock \href {https://doi.org/10.1143/PTP.68.508}
  {\path{doi:10.1143/PTP.68.508}}.

\bibitem{KhachatryanSedrakyan2013}
S.~Khachatryan, A.~Sedrakyan, On the solutions of the {Yang-Baxter} equations
  with general inhomogeneous eight-vertex {$R$}-matrix: relations with
  {Zamolodchikov's} tetrahedral algebra, Journal of Statistical Physics 150~(1)
  (2013) 130--155.
\newblock \href {https://doi.org/10.1007/s10955-012-0666-8}
  {\path{doi:10.1007/s10955-012-0666-8}}.

\bibitem{Baxter1985}
R.~J. Baxter, Exactly solved models in statistical mechanics, in: Integrable
  systems in statistical mechanics, World Scientific, 1985, pp. 5--63.

\bibitem{NIST2010}
F.~W. Olver, D.~W. Lozier, R.~F. Boisvert, C.~W. Clark, {NIST Handbook of
  Mathematical Functions}, 1st Edition, Cambridge University Press, New York,
  NY, USA, 2010.

\bibitem{Wang2003}
M.~Wang, Y.~Zhou, The periodic wave solutions for the
  {Klein--Gordon--Schr{\"o}dinger} equations, Physics Letters A 318~(1) (2003)
  84--92.
\newblock \href
  {https://doi.org/https://doi.org/10.1016/j.physleta.2003.07.026}
  {\path{doi:https://doi.org/10.1016/j.physleta.2003.07.026}}.

\bibitem{HuaiTangHongQing2003}
C.~Huai-Tang, Z.~Hong-Qing, New doubly periodic and multiple soliton solutions
  of the generalized (3+1)-dimensional {KP} equation with variable
  coefficients, Chinese Physics 12~(11) (2003) 1202.
\newblock \href {https://doi.org/10.1088/1009-1963/12/11/303}
  {\path{doi:10.1088/1009-1963/12/11/303}}.

\bibitem{HuaiTangHongQing2004}
C.~Huai-Tang, Z.~Hong-Qing, New double periodic and multiple soliton solutions
  of the generalized (2+1)-dimensional {Boussinesq} equation, Chaos, Solitons
  \& Fractals 20~(4) (2004) 765 -- 769.
\newblock \href {https://doi.org/10.1016/j.chaos.2005.08.148}
  {\path{doi:10.1016/j.chaos.2005.08.148}}.

\bibitem{Yan2004}
Z.~Yan, An improved algebra method and its applications in nonlinear wave
  equations, Chaos, Solitons \& Fractals 21~(4) (2004) 1013--1021.
\newblock \href {https://doi.org/10.1016/j.chaos.2003.12.042}
  {\path{doi:10.1016/j.chaos.2003.12.042}}.

\bibitem{Kudryashov2009}
N.~A. Kudryashov, On ``new travelling wave solutions'' of the {KdV} and the
  {KdV-Burgers} equations, Communications in Nonlinear Science and Numerical
  Simulation 14~(5) (2009) 1891--1900.
\newblock \href {https://doi.org/10.1016/j.cnsns.2008.09.020}
  {\path{doi:10.1016/j.cnsns.2008.09.020}}.

\bibitem{EbaidAly2012}
A.~Ebaid, E.~H. Aly, Exact solutions for the transformed reduced {Ostrovsky}
  equation via the {$F$}-expansion method in terms of {Weierstrass-elliptic}
  and {Jacobian-elliptic} functions, Wave Motion 49~(2) (2012) 296--308.
\newblock \href {https://doi.org/10.1016/j.wavemoti.2011.11.003}
  {\path{doi:10.1016/j.wavemoti.2011.11.003}}.

\bibitem{Zamolodchikov1979B}
A.~B. Zamolodchikov, {$Z_4$}-symmetric factorized {$S$}-matrix in two
  space-time dimensions, Communications in Mathematical Physics 69~(2) (1979)
  165--178.
\newblock \href {https://doi.org/10.1007/BF01221446}
  {\path{doi:10.1007/BF01221446}}.

\bibitem{Felderhof1973}
B.~U. Felderhof, Direct diagonalization of the transfer matrix of the
  zero-field free-fermion model, Physica 65~(3) (1973) 421--451.
\newblock \href {https://doi.org/10.1016/0031-8914(73)90059-1}
  {\path{doi:10.1016/0031-8914(73)90059-1}}.

\bibitem{BazhanovStroganov1985}
V.~V. Bazhanov, Y.~G. Stroganov, Hidden symmetry of free fermion model {I}.
  {Triangle} equation and symmetric parametrization, Theoretical and
  Mathematical Physics 62~(3) (1985) 253--260.
\newblock \href {https://doi.org/10.1007/BF01018266}
  {\path{doi:10.1007/BF01018266}}.

\bibitem{MacdonaldEtal1968}
C.~T. MacDonald, J.~H. Gibbs, A.~C. Pipkin, Kinetics of biopolymerization on
  nucleic acid templates, Biopolymers: Original Research on Biomolecules 6~(1)
  (1968) 1--25.
\newblock \href {https://doi.org/10.1002/bip.1968.360060102}
  {\path{doi:10.1002/bip.1968.360060102}}.

\bibitem{MotegiSakai2013}
K.~Motegi, K.~Sakai, Vertex models, {TASEP} and {Grothendieck} polynomials,
  Journal of Physics A: Mathematical and Theoretical 46~(35) (2013) 355201.
\newblock \href {https://doi.org/10.1088/1751-8113/46/35/355201}
  {\path{doi:10.1088/1751-8113/46/35/355201}}.

\bibitem{Cherednik1980}
I.~V. Cherednik, On a method of constructing factorized {S} matrices in
  elementary functions, Theoretical and Mathematical Physics 43~(1) (1980)
  356--358.
\newblock \href {https://doi.org/10.1007/BF01018470}
  {\path{doi:10.1007/BF01018470}}.

\bibitem{BelavinDrinfeld1982}
A.~A. Belavin, V.~G. Drinfel'd, Solutions of the classical {Yang-Baxter}
  equation for simple {Lie} algebras, Functional Analysis and Its Applications
  16~(3) (1982) 159--180.
\newblock \href {https://doi.org/10.1007/BF01081585}
  {\path{doi:10.1007/BF01081585}}.

\bibitem{FanWu1969}
C.~Fan, F.~Y. Wu, Ising model with second-neighbor interaction. {I}. {Some}
  exact results and an approximate solution, Physical Review 179~(2) (1969)
  560.
\newblock \href {https://doi.org/10.1103/PhysRev.179.560}
  {\path{doi:10.1103/PhysRev.179.560}}.

\bibitem{GalleasMartins2002}
W.~Galleas, M.~J. Martins, {Yang-Baxter} equation for the asymmetric
  eight-vertex model, Physical Review E 66~(4) (2002) 047103.
\newblock \href {https://doi.org/10.1103/PhysRevE.66.047103}
  {\path{doi:10.1103/PhysRevE.66.047103}}.

\bibitem{Sklyanin1979}
E.~K. Sklyanin, On complete integrability of the {Landau-Lifshitz}
  equation~(LOMI-E-3-1979) (1979) 1--32.

\end{thebibliography}

\end{document}